\newif\ifShowKeys
\ShowKeystrue

\newif\ifshowtikz
\showtikztrue
\showtikzfalse   

\documentclass[11pt]{article}
	\pdfoutput=1
\usepackage[T1]{fontenc}

\usepackage[T1]{fontenc}
\usepackage{textcomp}

\usepackage{listings}
\lstnewenvironment{arkady}  
{\lstset{language=C,frame=trbl,basicstyle = \footnotesize \ttfamily , breaklines = true,showstringspaces=false}}{}

\usepackage{datetime}
\usepackage{comment}					

\usepackage[usenames,dvipsnames]{xcolor}
\usepackage[setpagesize=false,pagebackref=false, 
linktocpage, bookmarksopen=true, colorlinks=true, 
linkcolor=Maroon,citecolor=Maroon,urlcolor=Maroon]{hyperref}

\usepackage[parsep]{collref}				

\usepackage{amsmath, amssymb,amsthm}
\usepackage{stackrel}
\numberwithin{equation}{section}
\usepackage{bm,environ,mathrsfs,array,arydshln}
\usepackage{booktabs,float,slashed}
\usepackage{appendix}
\usepackage[mathcal]{euscript}
\usepackage{tensor} 						
\usepackage{mathabx}
\usepackage[vcentermath]{youngtab}

\usepackage{graphicx,epsfig,epic}
\usepackage{tikz}
\usepackage{tikz-feynman} 

\allowdisplaybreaks

\usepackage{framed}						
\definecolor{shadecolor}{rgb}{0.9996078, 0.984314, 0.960784}
\definecolor{framecolor}{rgb}{0,0,0}
\definecolor{TFTitleColor}{RGB}{1,1,1}
\definecolor{TFFrameColor}{RGB}{249	218	181}		
\definecolor{TFFrameColor}{RGB}{230 230 230 }

\newenvironment{frshaded}{%
    \MakeFramed {\FrameRestore}}%
    {\endMakeFramed}

\newcommand{\red}[1]{\textcolor{red}{#1}}

\newcommand{\blue}[1]{\textcolor{blue}{#1}}

\definecolor{myred}{RGB}{233, 33, 45}

\newcommand{\bs}{\begin{frshaded}}			
\newcommand{\es}{\end{frshaded}\noindent}

\def\ba#1\ea{\begin{align}#1\end{align}}		        
\newcommand{\be}{\begin{equation}}
\newcommand{\ee}{\end{equation}}
\newcommand{\bea}{\begin{equation} \begin{aligned}} 
\newcommand{\eea}{\end{aligned} \end{equation}}
\newcommand{\mc}{\mathcal }
\newcommand{\wh}{\widehat}
\newcommand{\wt}{\widetilde}

\newcommand{\la}{\label}
\newcommand{\eps}{\varepsilon}

\newcommand{\lp}{\notag \\ & }

\DeclareMathOperator{\tr}{\text{tr}}

\newcommand{\cf}{\textit{cf.} }
\newcommand{\ie}{\textit{i.e.} }

\newcommand{\T}{{\sf T} }
\newcommand{\CounterTerm}{\kappa}
\newcommand{\hh}{{\rm h}}
\newcommand{\D}{{\rm D}}

\begin{document}
\begin{titlepage}

\vspace*{15mm}
\begin{center}
{\bf \large\sc  \bf  2-loop  scattering on superstring and supermembrane  in flat space}
\vspace*{10mm}

{ M. Beccaria$^{a}$, \ \ R. Roiban$^{b,c}$,\ \ \ A.A. Tseytlin$^{d,}$\footnote{Also at  ITMP and Lebedev Inst.
}} 

\vspace*{4mm}
{\small 	
${}^a$ Universit\`a del Salento, Dipartimento di Matematica e Fisica \textit{Ennio De Giorgi},\\ 
		and INFN - sezione di Lecce, Via Arnesano, I-73100 Lecce, Italy
			\vskip 0.01cm
${}^{b}$ Institute for Gravitation and the Cosmos, \\
		Pennsylvania State University, University Park, PA 16802, USA
			\vskip 0.01cm
${}^{c}$ Institute for Computational and Data Sciences, \\
		Pennsylvania State University, University Park, PA 16802, USA
			\vskip 0.01cm
${}^d$ Abdus Salam Centre for Theoretical Physics,\\ Imperial College London,  SW7 2AZ, U.K.
			\vskip 0.01cm
\vskip 0.2cm {\small E-mail: \texttt{matteo.beccaria@le.infn.it,\ radu@phys.psu.edu, \ tseytlin@ic.ac.uk}}
}
\vspace*{0.7cm}
\end{center}
\begin{abstract}  
\vskip 0.01cm
	{We consider the S-matrix of transverse scalar  
	excitations on an infinite   $D=10$  GS  superstring  
	and $D=11$  supermembrane 
	in flat  target space. 
	We compute the 4-particle scattering amplitude in the 2-loop approximation 
	and demonstrate  that, like in the Nambu string case,  
	 the $D=10$ GS string S-matrix does  not contain non-trivial 2d UV divergences 
	(UV pole not accompanied by  terms with logarithms of  momenta  is an  artefact
	of dimensional regularization).
	This  is  consistent with underlying integrability of this model   which is maintaining by  adding appropriate local counterterms. 
	In the supermembrane case  there are no 1-loop divergences  
	but  we find  a genuine 2-loop UV  pole. This demonstrates 
	 non-finiteness
	of the world-volume S-matrix of the M2 brane  theory. 
		}
	\end{abstract}
\end{titlepage}
\def \np {\newpage}
\def \ed {\np \small
\baselineskip 11pt
\bibliography{BT-Biblio}
\small
\bibliographystyle{JHEP-v2.9}
\end{document}
}
\def \iffa  {\iffalse}
\def \te {\textstyle}
\newcommand{\rf}[1]{(\ref{#1})}
\def\ov{\over}
\def \ci {\cite}
\def \foot {\footnote}
\def \bi{\bibitem}
\def\la{\label}\def \a {\alpha}
\def\foot{\footnote}
\def \adss {AdS$_5 \times S^5$\ }
\def \adsc {AdS$_4\times { \rm CP}^3$ } 
\def \tb {$T\bar T$ }
\def \s {\sigma} \def \del {\partial} 
\def \ha {\tfrac{1}{2}}
\def \DD {{\rm D}}

{\small 
\makeatletter
\newcommand*{\toccontents}{\@starttoc{toc}}
\makeatother
\toccontents
}

\def \bb {{\rm b}}\def \veps {\epsilon}
\def \OO {{\cal O}}
\def \cc {{\rm c}}
\def \no {\nonumber}
 \def \T {\nu} 
 \def \g {\gamma}
 \def \DB  {{\rm DB}}

\newpage 
\setcounter{footnote}{0}
\section{Introduction}

Even  most-symmetric  
 examples  of AdS/CFT   duality   have their 
AdS   side less well defined   than the CFT side. 
The  GS  superstring \ci{Green:1983wt}    and  BST  supermembrane \ci{Bergshoeff:1987cm} 
theories are formally  non-renormalizable   even  in the case of  flat  target space  and thus 
require special  definition of the corresponding quantum theory. 

One may view these 2d and 3d world-volume  models   just  as effective  theories 
 that may  receive higher-derivative corrections 
(like  what happens in the case of D-brane actions that  get $\a'$ corrections  controlled  by open-string dynamics, 
cf. \ci{Simon:2011rw}).
 This  point of view, however,  is not   satisfactory as it is not clear a priori
 which  should be a more fundamental   UV-consistent  world-volume 
  theory  that specifies  higher-order counterterms or  defines  an effective  cutoff
  as these  models   themselves are expected to describe fundamental degrees of freedom. 

 
Alternatively, one    may hope that a large amount of supersymmetry and possible  existence of  hidden symmetries 
 (in the case of special symmetric   backgrounds) may  provide   enough constraints to define the world-volume quantum theory
 unambiguously. That may   then  allow    to match its predictions (at least for some special observables) 
 to the corresponding  strong-coupling expansion  on the  CFT side. 
 In the case  of  the  2d  GS theory  in \adss \ci{Metsaev:1998it} 
 its integrability \ci{Bena:2003wd}   should provide such   stringent constraints.\foot{One can draw an analogy with  \tb deformation   \ci{Zamolodchikov:2004ce,Smirnov:2016lqw,Cavaglia:2016oda,Rosenhaus:2019utc}
  where the assumption of quantum integrability may allow to define a  formally non-renormalizable 2d 
 theory   by specifying  the required  counterterms  at each order in loop expansion.}
 Indeed, the  semiclassical 1-loop  computations in \adss  GS theory  gave   finite consistent results (see, e.g. \ci{Drukker:2000ep,Frolov:2002av}). What is  much more non-trivial is  that 
 there are  also  examples of 2-loop  GS string computations in \adss 
 \ci{Roiban:2007jf,Klose:2007rz,Roiban:2007dq,Giombi:2009gd,Giombi:2010fa}  and \adsc 
 \ci{Bianchi:2014ada}
 which  gave finite results in agreement  with the   integrability  predictions. 
  
 
 In  the case of the  M2  brane  theory in  AdS backgrounds the semiclassical expansion near  non-degenerate 
3d world surfaces is also well defined:  the 1-loop  correction has  no log UV  divergences
and matches   predictions on the  dual gauge  theory side 
\cite{Drukker:2020swu,Giombi:2023vzu,Beccaria:2023ujc,Beccaria:2023sph,
Beccaria:2025vdj,Gautason:2025per,Gautason:2025plx} (see also \ci{Drukker:2023jxp,
Drukker:2023bip,Astesiano:2024sgi,Giombi:2024itd,Beccaria:2025npl}).

 What happens  at  higher loops  is an important open problem.  
  On general grounds, one may  expect to  find logarithmic UV divergences in M2 brane theory  starting at 2-loop level 
    making the results   ambiguous. 
  This would be  problematic  for checks of AdS/CFT duality  as the dual  gauge theory  predictions for the corresponding 
   subleading
 (in inverse M2 brane tension) terms  in relevant observables are finite and  unambiguous (cf.  discussions in 
 \cite{Giombi:2023vzu,Beccaria:2023ujc}). 
 While there is no  analog of 2d integrability  in the case of 3d Lorentz-invariant field theories, 
 consistency with AdS/CFT  appears to imply that 2-loop UV  logs   may  actually cancel  in     special observables due to some hidden symmetry  of the M2 brane theory  yet to be uncovered.
 
 \
 
  With a motivation to shed light on  possible existence of such hidden symmetry 
  ref. \cite{Seibold:2024oyr}  considered  the M2 brane theory  in flat $D=11$  target space 
  and computed  the 1-loop  on-shell $2\to 2$  scattering amplitude  of the 8 transverse  massless scalars 
  in flat infinite membrane vacuum. The  
    resulting finite  expression for the amplitude   turned out to be   much simpler 
   than the corresponding one \cite{Seibold:2023zkz}  in the bosonic membrane case, 
   indicating  that the M2  brane theory may have some  special features.

 Our aim  here  will be  to extend the  $D=11$ M2 brane computation  of  \cite{Seibold:2024oyr} to the 2-loop level.
 We shall   also    find a    similar 2-loop correction to the 
 4-point amplitude  in  flat  $D=10$ GS string case. 
 
The S-matrix of massless excitations on a long  string in flat target space  is expected 
to have a   special (elastic, pure phase) form reflecting  the  integrability of the world-sheet theory~\ci{Dubovsky:2012sh,Dubovsky:2012wk,Cooper:2014noa,Conkey:2016qju}.
This    implies that the  direct   computation of the amplitude  should give the  expression 
 consistent with the pure-phase  form  provided   one makes  a specific choice of { local}   counterterms. 
 This means, in  particular, that the amplitude   cannot   contain   non-polynomial    logarithm of   momentum  ($\log s$, etc.) 
    terms 
  associated with  divergent  UV  behaviour. 
 
This is indeed what happens at the 1-loop level  in both   bosonic  
 string  case 
in any  dimension $D$ \ci{Dubovsky:2012sh}  and in the GS string  case in $D=10$ \cite{Seibold:2024oyr}.
The 2-loop amplitude in the bosonic case was computed in \ci{Conkey:2016qju} 
 and  found  also not to contain  $\log s$ terms.  Below   we will  confirm the result of \ci{Conkey:2016qju}  
 and extend it to  the case of the $D=10$   GS string  with a similar conclusion  about consistency with 
 integrability   under a specific choice of local counterterms. 
 Like  in the  Nambu string case the remaining  UV poles   in the 2-loop  GS amplitude 
 are effectively  unphysical (or artefacts of dimensional regularization, 
 originating from an evanescent 1-loop counterterm).

In the  M2 brane case we will  find  the  non-trivial 2-loop  log UV  divergences  that  are accompanied 
 by  the associated non-analytic $\log s$ terms.  This 
  demonstrates  non-renormalizability of the  world-volume S-matrix in     M2 brane theory in flat $D=11$ target space.
  This  does not a priori rule out a possibility that   some other  observables in the M2 brane  theory 
  (like   partition function near minimal surface for M2 brane in  AdS space   considered in \cite{Giombi:2023vzu,Beccaria:2023ujc}) 
  may  
  turn  out to be UV finite. 
 
\

Let us   first  review  the known  expressions  for the 
  1-loop  amplitudes    in  the (super) string  and   in  the  (super) membrane cases. 
  We shall   then  summarize the  2-loop  results of the present  paper. 
 
In  the bosonic string ($d=2$)  and  membrane ($d=3$) cases,  expanding  the standard  (Nambu-Goto or Dirac) $d$-dimensional
 induced volume action  in the static gauge 
one  may  compute  the scattering of  massless  scalar bosons  
represented by  the $\wh D=D-d$ transverse  coordinates $X^{i}$.
The amplitude for the  two incoming   bosons  with
$SO(\wh D)$ indices $i_{1},i_{2}$ and  momenta $p_{1},p_{2}$, and two outgoing bosons with  indices 
$i_{3}, i_{4}$ and  momenta $p_{3},p_{4}$
has the following general structure 
\ba
\la{2.8}
\mc M^{i_{1}i_{2}i_{3}i_{4}}(s,t,u) =& A(s,t,u)\, \delta^{i_{1}i_{2}}\delta^{i_{3}i_{4}}+B(s,t,u)\,\delta^{i_{1}i_{3}}\delta^{i_{2}i_{4}}+C(s,t,u)\,\delta^{i_{1}i_{4}}\delta^{i_{2}i_{3}}\ .
\ea
Here the annihilation $A$, transmission $B$  and reflection   $C$ amplitudes 
 are related by crossing\footnote{Note that in  \cite{Dubovsky:2012sh,Seibold:2023zkz,Seibold:2024oyr,Conkey:2016qju}
the directions of $p_{3}$, $p_{4}$  momenta  were outward,   so  that  $t$ and $u$ were defined with minus signs.
}
\ba
\la{1.2}
& B(s,t,u) =B(u,t,s)= A(t,s,u), \qquad\qquad  C(s,t,u) = A(u,t,s), \\
& s = -(p_{1}+p_{2})^{2}, \qquad t = -(p_{1}+p_{3})^{2}, \qquad u = -(p_{1}+p_{4})^{2}, \qquad s+t+u=0.\la{221}
\ea
In general, 
using   loop or inverse  tension $T$  expansion   one gets  for $A(s,t,u)$
\be 
A = T^{-1} A^{(0)} + T^{-2} A^{(1)}  + T^{-3} A^{(2)} + ... \ , \ \ \ \qquad  A^{(0)}= - \ha  u t  \ , \la{10} \ee
where $A^{(0)}$ is the tree-level  amplitude,  $A^{(1)}$ is the  1-loop one,   etc.

In the string case the massless $d=2$   kinematics imposes the constraint $stu=0$ that  can be solved by choosing 
  $t=0$ so that $u=-s$.     Then 
 $A,B,C$ become 
functions of the single variable  $s$ and  
crossing symmetry and the real analyticity requirements imply that
\be\la{14}
C(s) = A^*(-s)\ ,  \qquad \qquad B(s) = B^*(-s) \ ,  \  
\ee
where $s\to -s$   is the analytic continuation $s\to e^{i \pi } s$ through the upper half-plane.
At the tree level  
  \be \la{144}
  A^{(0)}=C^{(0)}=0, \qquad \quad   B^{(0)}=\ha s^2 \ . \ee
The exact S-matrix of the bosonic  string  expanded near long string vacuum 
 suggested  in \cite{Dubovsky:2012sh} corresponds to the pure transmission 
$A(s)=C(s)=0$ and is simply given by a  pure-phase   
factor,  thus 
representing an integrable theory
\be
\la{1.1}
\mathbb S(s) = 1+\tfrac{i}{2s}\,B(s) = e^{\frac{is}{4T}} =\te    1    +  {i  s\ov  4 T }   - {s^2\ov 32 T^2}  - {{i s^3\ov 2 \times 192 T^3} }   + ...\ .
\ee
{As was shown in \cite{Dubovsky:2012wk},  using this S-matrix in the thermodynamic Bethe Ansatz one reproduces the expected free bosonic string spectrum.}
The linear in $s$ term  in \rf{1.1}  corresponds to the   tree-level  contribution $B^{(0)}$ in \rf{144}. 
At the  1-loop order one finds \cite{Dubovsky:2012sh} 
\be \la{11} 
\te A^{(1)}=-C^{(1)} = -\frac{1}{192\pi}\, (D-26)\, s^{3}, \qquad \qquad  B^{(1)} = i\frac{1}{16}\, s^{3}\ . \ee
  $A^{(1)}$  and $C^{(1)}$ can be cancelled \cite{Dubovsky:2012sh,Aharony:2013ipa}  by adding a local real {\it finite} 
  Polyakov-Polchinski-Strominger (PPS) counterterm \ci{Polyakov:1981rd,Polchinski:1991ax}\foot{Here $R^{(2)} $ is the curvature of the induced metric in the static gauge, i.e.  
 $h_{ab}= \eta_{ab} + \del_a X^i \del_b X^i$. Cancellation of conformal anomaly  in any $D$ is required  for  preservation of 
 equivalence  to the free string spectrum found in conformal gauge. Note that  in the static gauge 
  $\eta^{ab} \del_a X^\mu \del_b X_\mu  =   2 + \del^a X^i \del_a X^i $. }
\ba  
\Delta S_{\rm PPS}=2 b \int  R^{(2)}  \nabla^{-2}  R^{(2)} 
= - 2 b    \int d^2 \s\, ( \del^a X^i \del_a \del_b X^i)^2  + ... \ , 
 \quad   b =  \te { 1\ov 192  \pi} \bb 
  \ , \quad  \bb =D-26 \ , \la{13}  \ea 
while the value of $B^{(1)}$ is  indeed  consistent with the $s^2$ term in \rf{1.1}.

A similar  result  for the 1-loop  massless boson  scattering amplitude is found  in the case  of  the  $D=10$  GS string   \cite{Seibold:2024oyr} (see also \ci{Tseytlin:2025dud})
\be\la{15} 
\te A^{(1)} =- C^{(1)}  =  {1\ov 16 \pi}   \, s^3  \ , \qquad \ \ \  B^{(1)} = i{1 \ov 16} s^3 \ . \ee 
Here  again   $A^{(1)}$ and $C^{(1)} $ can be cancelled  by a  local  counterterm  as in  \rf{13} 
and the resulting 1-loop S-matrix is the same as in \rf{1.1}.

The 2-loop  correction to  the  amplitude \rf{2.8}   in the bosonic string  was found 
 in \cite{Conkey:2016qju}  using dimensional regularization with $d=2 - 2\eps$  
  (including the  required 1-loop  evanescent counterterm $ \int \sqrt{-h}\, R^{(d)}$ that is a total derivative  in $d=2$). The resulting 2-loop    corrections to  $A(s),\, B(s),\, C(s)$  
   that we  will confirm below   contain the UV pole parts  as well as the finite parts  (cf. \rf{10},\rf{14})
\ba
\la{75}
&A^{(2)} = A^{(2)}_{1\ov \eps} + A^{(2)}_{\rm fin} \ , \qquad \qquad 
A^{(2)}_{1\ov \eps} = C^{(2)}_{1\ov\eps} = \tfrac{(D-12)(D-8)}{9216\pi^{2}\, \eps}\, s^{4}, \qquad
B^{(2)}_{1\ov\eps} =- \tfrac{D-8}{768\pi^{2}\, \eps}\, s^{4}\ , \\
 \la{1.5} 
&A^{(2)}_{\rm fin}(s)  = -i\,\tfrac{D-26}{768\pi}\,s^{4}-\tfrac{6D^{2}-143D+448}{13824\pi^{2}}\, s^{4}\ , \qquad 
C^{(2)}_{\rm fin}(s) =  i\,\tfrac{D-26}{768\pi}\,s^{4}-\tfrac{6D^{2}-143D+448}{13824\pi^{2}}\, s^{4}\ ,\\
 &\qquad \qquad B^{(2)}_{\rm fin} = 
{-\tfrac{1}{192}s^{4}}
+\tfrac{11(D+4)}{4608\pi^{2}}\, s^{4} \ .\la{76}
\ea
The imaginary parts of $A^{(2)}$  and $C^{(2)}$  in \rf{1.5}  can be cancelled  for any $D$ by including  the  contribution of the  1-loop counterterm \rf{13}  that was required for 1-loop  integrability. Then the pole parts  as well as the real ${1\ov \pi^2} s^4$ 
terms in the finite parts  in \rf{1.5},\rf{76}  can be cancelled  by a local counterterm  \cite{Conkey:2016qju}  of the form (here $K$ is the extrinsic curvature)\foot{Note that while 
$\tr(K^{i}K^{i})^{2}$ is the square of the Ricci scalar, the  first  term $\tr(K^{i}K^{j})^{2}$ cannot be written just  in terms 
of the curvature of the induced metric.}
\be
\la{80}
\int d^{2}\sigma\, \sqrt{-h}\, \big[c_{1} \tr(K^{i}K^{j})^{2}+c_{2}\tr(K^{i}K^{i})^{2}\big], \qquad 
\qquad  K^i_{ab} =\del_a \del_b X^i + ...\ . 
\ee
Then one ends up with  
\be \la{100}
A^{(2)}=C^{(2)}=0\ , \qquad \qquad     B^{(2)}= {-\tfrac{1}{192}s^{4}}   \ , \ee
which   is  consistent  with the pure-phase S-matrix in \rf{1.1}. 

This  special choice of the  counterterms or  ``scheme''  is thus required
  for preservation of integrability at the 2-loop level.\foot{The need to introduce special   counterterms
     may be related to dimensional regularization 
 not preserving higher  conserved charges manifestly (indeed, integrability is expected to be present only in $d=2$ theory). 
 One  may wonder
  if there may exist an alternative approach (e.g.   based on interpreting the Nambu   action in the static gauge as 
  a \tb  deformation of a free  scalar theory, cf.  \ci{Cavaglia:2016oda,Rosenhaus:2019utc} 
  and using  conformal perturbation theory) 
 in which preservation of integrability may be more manifest. We thank R. Tateo for a related comment.}
Its existence  relies on cancellation of  non-analytic $\log s$ terms.
 The presence of $1\ov \eps$   poles   not  accompanied  by $\log{s\ov \mu^2}$ terms suggests 
 that these  do not actually represent genuine  logarithmic UV divergences  and should be simply subtracted out.\foot{\la{Pol}In particular, the  cancellation  of $\log \mu$ terms   means  the absence of  ambiguous 
   scheme-dependent coefficients in the finite part which allows one to maintain the   underlying symmetry.  
   The remaining
 poles may be  viewed as  an 
 artefact of dimensional regularization that requires the introduction of an
  evanescent  
  counterterm
 which produces extra $1\ov \eps$  poles   not accompanied  by  $\log s$  terms. 
Examples    when    the coefficients of   $1\ov \eps$   poles  in dimensional regularization 
are   are not correlated with those of  $\log{s\ov \mu^2}$ terms  
were discussed in 4d case  in   \cite{Bern:2015xsa,Bern:2017puu}. 
The  model  considered in  \cite{Bern:2017puu} was the 
$\mc N=1$  supergravity with one matter multiplet 
 where   the 2-loop  UV poles  are present 
 but   do not have  any physical consequences since all momentum logarithms cancel  out.
 This  mismatch is again due to the effect of evanescent
operator  (here  the Gauss-Bonnet term). 
As was  shown in  \cite{Bern:2017puu},  the  absence   of  $\log\mu^{2}$  terms 
related to $\log s$ terms
can be seen  
directly  using 4d unitarity cuts ($\log\mu^{2}$ term itself    is not associated to a cut).
This approach avoids the need for  an  ultraviolet regularization and   thus for  the introduction of 
evanescent operators.
 }

 The 2-loop  computation  in  the $D=10$  GS string case that we shall  carry  out below   gives a similar  result:
 there  is $1\ov \eps$ pole but all 
 $\log s$ terms   cancel   out and  the expression for the amplitude is consistent with 2-particle unitarity. 
 Explicitly, we find\foot{Considering GS  model  in general dimension $D$  one finds that $\log s$ terms cancel only  
 for $D=10$.} 
 \ba
\la{73}
A^{(2)}_{1\ov \eps} = C^{(2)}_{1\ov\eps} = - \tfrac{1}{1536\pi^{2}\, \eps}\, s^{4}, \qquad&\qquad 
B^{(2)}_{1\ov\eps} =- \tfrac{1}{128\pi^{2}\, \eps}\, s^{4}\ , \\
 \la{115} 
A^{(2)}_{\rm fin}(s)  = i\,\tfrac{1}{64\pi}\,s^{4}+ \tfrac{145}{9216\pi^{2}}\, s^{4}\ , &\qquad 
C^{(2)}_{\rm fin}(s) = - i\,\tfrac{1}{64\pi}\,s^{4}+ \tfrac{145}{9216\pi^{2}}\, s^{4}\ ,\\
 B^{(2)}_{\rm fin} = & {-\tfrac{1}{192}s^{4}}   +\tfrac{25}{768\pi^{2}}\, s^{4} \ .\la{74}
\ea
Like in the bosonic string case, the   imaginary parts in $A^{(2)}$ and $C^{(2)}$  are again cancelled   
by the  contribution of the finite  1-loop  PPS counterterm   \rf{13}  that was needed  to remove the 
non-vanishing $A^{(1)}$ and $C^{(1)}$ in \rf{15}. 
Then the  remaining real $s^4$  terms  can be cancelled   by a local counterterm like \rf{80}   so  that  at the end  we get 
 the same integrability-consistent  expression  \rf{100} as in the bosonic string case.

\

Let us now turn  to the membrane case.   Here the  momentum invariants are  subject only to  $s+t+u=0$ but otherwise generic 
and   the amplitudes $B$ and $C$  are related to $A$ by \rf{1.2} so it is enough to present  just $A(s,t,u)$. 
For the   bosonic 
membrane  in general dimension $D$,  one finds that  the 1-loop correction to the massless scalar 
tree-level amplitude $A$ 
  in \rf{10}  is given by  the following finite expression 
 \cite{Seibold:2023zkz} 
 \be
A^{(1)}  = \te \frac{1}{256}\Big[(-s)^{3/2}\big(\frac{3 D-41}{32}s^{2}-\frac{D-3}{4}tu\big)+(-t)^{5/2}(3t+2s)+(-u)^{5/2}(3u+2s)\Big]\la{1.8} \ . \ee
As was found in  \cite{Seibold:2024oyr},   the M2 brane   analog of \rf{1.8} 
takes particularly simple form  for  $D=11$ 
\be\la{18} 
A^{(1)} = \tfrac{1}{32}\big[(-s)^{3/2}+(-t)^{3/2}+(-u)^{3/2}\big]\, A^{(0)} \ , \ \ \ \ \  \qquad  A^{(0)} =-\ha   tu \ . 
\ee
As we will show below the 2-loop  correction  to the $D=11$ M2  brane   amplitude 
 computed in  dimensional regularization with $d=3-2\eps$  is given by 
\ba
A^{(2)} =  
  \Big[  \big( -{\te \frac{1 }{256\pi^{2}\eps} } +  & {\te  \frac{3}{2048}} \big)     s t u   
+{\te \frac{1}{384\pi^{2}}} \te \big[s^{3}\log({-{s\ov \mu^2})}+t^{3}\log({-{t\ov \mu^2}})
+u^{3}\log({-{u\ov \mu^2})}\big]\Big]\ tu  \no \\
& +\tfrac{1}{80640\pi^{2}}\, s(6s^{4}+124 s^{3}t-1833 s^{2}t^{2}-3914 st^{3}-1957 t^{4}) 
 \ . \la{20}
\ea
This expression is simpler than the corresponding one  in  the bosonic membrane case (see \rf{4.4},\rf{230})
but still contains the  UV pole.\foot{Note that the double-pole terms cancel out which is a consequence of the 1-loop finiteness in $d=3$.} Here  it is  accompanied by  the  corresponding  $\log s$-like terms 
and thus represents  a genuine logarithmic   UV divergence. 
Some  coefficients in the second line in \rf{20}  are scheme-dependent.\foot{This scheme dependence is due to 
the  well-known ambiguity  in treating  products  of 3d Levi-Civita tensor  in dimensional regularization. 
An alternative approach  using  dimensional reduction  regularization \cite{Siegel:1979wq} is beyond the scope of the present   work.}
Note that  like  in  \rf{18} the scheme-independent part of 
\rf{20}  is again proportional to  the tree-level amplitude $A^{(0)} =-\ha   tu$. 

The  presence of the pole indicates non-renormalizability of  
the M2 brane theory.\foot{One could try to argue  that  supersymmetry and special structure of the   GS  and  M2 actions 
may imply  that despite  being power-counting non-renormalizable they  may  represent  UV  finite theories.  
Indeed, the  WZ terms in the actions   should not be renormalized  but then  
  $\kappa$-symmetry    may   relate  it to  the volume  part of the action. This argument  implies only  non-renormalization of the tension 
  but does not a priori exclude the presence of higher supersymmetric and $\kappa$-symmetric invariants
  ($\kappa$-symmetry may also be deformed at the quantum level). An attempt to 
  argue for finiteness of M2 brane theory  was in \cite{Paccanoni:1989hd}. 
  Note also that  possible higher derivative corrections to M2 brane action   were  discussed in 
   \cite{Howe:2001wc,Howe:2003sa}.} 
At the moment there is no  indication of some hidden  symmetry 
 that would  guide  the definition of this  theory. 
One may speculate (cf. \cite{Seibold:2024oyr}) 
that after adding some specific counterterms to remove the pole and part of rational  finite  part of \rf{20} 
to  make the   whole amplitude \rf{10} 
 look as   $A(s,t,u) = F(s,t,u) A^{(0)}$ where $F=1 + T^{-1}  F^{(1)} + T^{-2}  F^{(2)} +... $ is 
 a  totally symmetric function, the latter  may  ``exponentiate'' 
  by  analogy with the 2d   case in  \rf{1.1}.

\

In section 2 we shall  discuss the 2-loop S-matrix for the bosonic string and the membrane.
We shall reproduce  the string 2-loop  amplitude in \ci{Conkey:2016qju}  clarifying the origin of the remaining UV pole term
and find  the analogous  result  for the  bosonic membrane. 

In section 3 we shall   present the expansion of  the GS string and the M2 brane  actions
 in static gauge  to quartic order in bosons  
and fermions   fixing  a particular $\kappa$-symmetry gauge  in which  there is no cubic interaction vertices 
(generalizing  the expressions in \ci{Seibold:2024oyr} to quartic fermionic terms).  This  will be the starting point  for computing 4-scalar   2-loop amplitudes.

In section 4 and 5  we shall compute the 2-loop GS string and M2 brane amplitudes  treating both cases in parallel.
We shall  find   extra fermionic loop contributions  that should be added to the bosonic loop expressions in  section 2. 
In the GS case there will be extra 1-loop counterterm contributions  related to  two  fermion  propagators 
 in the loop or on the external lines. 
 
 In Appendix A  we will present   details  of the expansion of the  M2 brane and GS string actions.
Appendix  B will contain  some   useful 
 fermionic traces and 1-loop integrals.
  In Appendix C we will review  few   relations for the extrinsic curvature  and  the scalar curvature   for the induced  world-volume metric. Appendix D   will contain details of the computation of 2-loop  diagrams in the bosonic case. 

\newpage

\section{Massless   scattering  on bosonic string and membrane }

We shall start with the bosonic  case treating  the string and membrane case in parallel.
We  will  first review the   results   for  the  1-loop  string   \ci{Dubovsky:2012sh}
 and  membrane \ci{Seibold:2023zkz} 
  amplitudes. We will  then reproduce   the      2-loop   contribution    \cite{Conkey:2016qju}   to the string  amplitude
and generalize it to the membrane case.

\subsection{Classical  action}

A  brane  moving in a $D$-dimensional target space
has the induced volume  action 
\be
\la{2.1}
S = -T\int d^{d}\sigma\,\sqrt{-\det h_{ab}}, \qquad h_{ab}=\eta_{\mu\nu}\partial_{a}X^{\mu}\partial_{b}X^{\nu},
\ee
where $a,b=0, 1, \dots, d-1$ and $\mu,\nu=0,1,\dots,D-1$. 
We shall consider its expansion   near   an   infinite flat brane   vacuum   and  fix the static gauge $X^{a}=\sigma^{a}$  so that 
the induced metric  is 
\be
\la{2.2}
h_{ab}=\eta_{ab}+\partial_{a}X^{i}\partial_{b}X^{i}, \qquad \qquad \quad i= 1, \dots, \wh D, \qquad \wh D = D-d.
\ee
Expanding in powers of  derivatives gives (dropping  constant term)
\be
\la{2.3}
S = -T\int d^{d}\sigma\,\Big[\te \ 
 \ha \partial^{a}X^{i}\partial_{a}X^{i} + 
{1\ov 8} (  \partial^{a}X^{j}\partial_{a}X^{j})^2  - {1\ov 8} (  \partial_{a}X^{i}\partial_{b}X^{i})(  \partial^{a}X^{j}\partial^{b}X^{j}) + 
\OO\big((\del X)^6\big)\Big]\ . 
 \ee 
 We shall not need $(\del X)^6$ terms  as they will not contribute  non-trivially to 2-loop 4-point scattering amplitude
 (such terms   may be  relevant only for tadpole  contributions that  vanish in dimensional  regularization that we will use). 

After rescaling $X^{i}\to \frac{1}{\sqrt T}\, X^{i}$, the factors of  inverse tension $T^{-1} $ 
will appear in the  interaction vertices and thus in the corresponding scattering amplitudes.

\subsection{Tree  amplitude}
Starting with \rf{2.3} 
the expression for the  scattering amplitude $X^{i_{1}}(p_{1})\, X^{i_{2}}(p_{2})\to X^{i_{3}}(p_{3})\,X^{i_{4}}(p_{4})$
follows from the quartic vertex  
\be
\begin{tikzpicture}[thick,baseline=0,scale=0.75]
	\begin{feynman}
	\vertex (a1) at (-1,1);
	\vertex (a2) at (-1,-1);
	\vertex (a3) at (1,1);
	\vertex (a4) at (1,-1);
	\vertex (oo) at (0,0);
	\diagram* {
	(a1) -- (oo);
	(a2) -- (oo);
	(a3) -- (oo);
	(a4) -- (oo);
	};
	\end{feynman}

	\draw[->,thin] ($(a1)!0.25!(oo)+(0,0.15)$) -- ($(a1)!0.75!(oo)+(0,0.15)$);
	\draw[->,thin] ($(a2)!0.25!(oo)-(0,0.15)$) -- ($(a2)!0.75!(oo)-(0,0.15)$);
	\draw[->,thin] ($(a3)!0.25!(oo)+(0,0.15)$) -- ($(a3)!0.75!(oo)+(0,0.15)$);
	\draw[->,thin] ($(a4)!0.25!(oo)-(0,0.15)$) -- ($(a4)!0.75!(oo)-(0,0.15)$);
	
	\node[left] at (a1) {$(p_{1},i_{1})$};	
	\node[left] at (a2) {$(p_{2},i_{2})$};	
	\node[right] at (a3) {$(p_{3},i_{3})$};	
	\node[right] at (a4) {$(p_{4},i_{4})$};	
	\node at (-5.5,0) {$V^{i_{1}i_{2}i_{3}i_{4}}(p_{1},p_{2},p_{3},p_{4}) =\quad \qquad \qquad $};
\end{tikzpicture} \no 
\ee
\ba
&  =-i\Big[(p_1\cdot p_2\, p_3\cdot p_4-p_1\cdot p_4\,p_2\cdot p_3-p_1\cdot p_3\,p_2\cdot p_4)\, \delta^{i_{1}i_{2}}\delta^{i_{3}i_{4}}  \no \\ &\ \ \qquad 
+(p_1\cdot p_3\,p_2\cdot p_4-p_1\cdot p_4\,p_2\cdot p_3-p_1\cdot p_2\,p_3\cdot p_4)\, \delta^{i_{1}i_{3}}\delta^{i_{2}i_{4}}\no\\
&\ \ \qquad  +(p_1\cdot p_4\, p_2\cdot p_3-p_1\cdot p_3\,p_2\cdot p_4-p_1\cdot p_2\,p_3\cdot p_4)\,\delta^{i_{1}i_{4}}\delta^{i_{2}i_{3}}\Big].\la{240}
\ea
The resulting  tree amplitude is (cf. \rf{2.8},\rf{10}) 
\be\la{25}
\mc M^{(0)\, i_{1}i_{2}i_{3}i_{4}} = \tfrac{1}{i}V^{i_{1}i_{2}i_{3}i_{4}} = 
-\tfrac{1}{2}tu\,\delta^{i_{1}i_{2}}\delta^{i_{3}i_{4}}
-\tfrac{1}{2}su\,\delta^{i_{1}i_{3}}\delta^{i_{2}i_{4}}
-\tfrac{1}{2}st\delta^{i_{1}i_{4}}\delta^{i_{2}i_{3}}.
\ee
It  has the same universal form for  any $d$, simplifying  in the string case (under  the $t=0$  choice)  
to \rf{144}. 


\subsection{1-loop amplitude}

The non-trivial 1-loop  diagram is
\be
\begin{tikzpicture}[thick,baseline=0]
	\begin{feynman}
	\vertex (a1) at (-1,1);
	\vertex (a2) at (-1,-1);
	\vertex (a3) at (3,1);
	\vertex (a4) at (3,-1);
	\vertex (oo1) at (0,0);
	\vertex (oo2) at (2,0);
	\diagram* {
	(a1) -- (oo1);
	(a2) -- (oo1);
	(a3) -- (oo2);
	(a4) -- (oo2);
	(oo1) -- [half left] (oo2);
	(oo1) -- [half right] (oo2);
	};
	\end{feynman}

	\draw[->,thin] ($(a1)!0.25!(oo1)+(0,0.15)$) -- ($(a1)!0.75!(oo1)+(0,0.15)$);
	\draw[->,thin] ($(a2)!0.25!(oo1)-(0,0.15)$) -- ($(a2)!0.75!(oo1)-(0,0.15)$);
	\draw[->,thin] ($(a3)!0.25!(oo2)+(0,0.15)$) -- ($(a3)!0.75!(oo2)+(0,0.15)$);
	\draw[->,thin] ($(a4)!0.25!(oo2)-(0,0.15)$) -- ($(a4)!0.75!(oo2)-(0,0.15)$);
	
         \draw[->,thin]  ($ (oo1)!0.25!(oo2) + (0,0.6) $) to [out=15,in=165] ($ (oo1)!0.75!(oo2) + (0,0.6) $);
         \draw[->,thin]  ($ (oo2)!0.25!(oo1) - (0,0.6) $) to [out=180+15,in=-15] ($ (oo2)!0.75!(oo1) - (0,0.6) $);
	
	\node[left] at (a1) {$(p_{1},i_{1})$};	
	\node[left] at (a2) {$(p_{2},i_{2})$};	
	\node[right] at (a3) {$(p_{3},i_{3})$};	
	\node[right] at (a4) {$(p_{4},i_{4})$};	
	\node at (1,1.2) {\small $(k+p_{1}+p_{2}, j_{1})$};
	\node at (1,-1.2) {\small $(k, j_{2})$};
	\node at (-3.5,0) {$D^{i_{1}i_{2}i_{3}i_{4}}(p_{1}, p_{2}, p_{3}, p_{4})=$};
\end{tikzpicture}
\la{66}
\ee
The  corresponding  amplitude is given by  the sum over 
crossing\foot{We set  $  D^{i_{1}i_{2}i_{3}i_{4}}(s, t, u) \equiv D^{i_{1}i_{2}i_{3}i_{4}}(p_{1}, p_{2}, p_{3}, p_{4})$.}

\ba
\la{3.6}
& \mc M^{(1)\, i_{1}i_{2}i_{3}i_{4}}(s, t, u) 
= D^{i_{1}i_{2}i_{3}i_{4}}(s, t, u)
+D^{i_{1}i_{3}i_{4}i_{2}}(t, u, s)
+D^{i_{1}i_{4}i_{2}i_{3}}(u,s,t).
\ea
In   dimensional regularization  
\be
 D^{i_{1}i_{2}i_{3}i_{4}}(s,t,u) 
  = -
 \tfrac{1}{2i}\int \frac{d^{d}k}{(2\pi)^{d}}\
  \frac{V^{i_{1}i_{2}j_{1}j_{2}}(p_{1},p_{2},-k-p_{1}-p_{2}, k)\, 
 V^{j_{1}j_{2}i_{3}i_{4}}  (k+p_{1}+p_{2},-k, p_{3}, p_{4}) }{
  k^{2}\ (k+p_{1}+p_{2})^{2}}, 
\ee
where $V$ is given in \rf{240}. 
This integral  may be evaluated in the standard way  by introducing Feynman parameters and doing the 
loop integration (see Appendix \ref{app7})
or by reducing it to simpler integrals using the Integration By Parts (IBP) and tensor reduction \cite{Passarino:1978jh, Tkachov:1981wb, Chetyrkin:1981qh, Laporta:2000dsw} implemented in the automated Mathematica software {\sf FIRE}~\cite{Smirnov:2008iw,Smirnov:2013dia,Smirnov:2014hma,Smirnov:2019qkx,Smirnov:2023yhb} that we will use also below.
The result is ($s=-2 p_1\cdot p_2$)
\ba\la{39}
D^{i_{1}i_{2}i_{3}i_{4}}(s, t, u)  =& -\frac{i\, s^{2}}{16(d-1)}G_{1,1}(s)\Big[
\frac{(d-2)[d(\wh D-8)-8]\, s^{2}-8\wh D\, t u}{8(d+1)}\, \delta^{i_{1}i_{2}}\delta^{i_{3}i_{4}}\lp\qquad \qquad \qquad \qquad\ \ \ \  
+s(ds+2t)\delta^{i_{1}i_{3}}\delta^{i_{2}i_{4}}+s(ds+2u)\delta^{i_{1}i_{4}}\delta^{i_{2}i_{4}}\Big],
\\
\la{3.9}
G_{1,1}(s) \equiv  &\int\frac{d^{d}k}{(2\pi)^{d}}\frac{1}{k^{2}(k+p_{1}+p_{2})^{2}} = \frac{i}{(4\pi)^{d/2}}
\frac{\Gamma(2-\frac{d}{2})\big[\Gamma(\frac{d}{2}-1)\big]^{2}}{\Gamma(d-2)}\, (-s)^{\frac{d}{2}-2}.
\ea
In the 2d {string} case  we are to set  $d=2-2\eps$  and $\wh D= D-2 $.\foot{Note
that  while  we introduced  in general   $\wh D$ as $D-d$   we are not  continuing $d$ there  away from 2 or 3
as that would break the target-space supersymmetry (the number of  bosons will change, but the  number of fermionic components 
 is always the same). Thus  in the general   expressions we set 
$\wh D= D-2$  in  the  string case and $\wh D= D-3$ in  the membrane case.}
 Then  
\be
G_{1,1}(s) = \frac{i}{2\pi\, s}\, \frac{1}{\eps}+\dots \ . 
\ee
and  we get  the following UV pole contribution to   (\ref{3.6}) 
\be
\la{3.11}
\mc M^{(1)\, i_{1}i_{2}i_{3}i_{4}}_{\frac{1}{\eps}}(s,t,u) = - \tfrac{1}{96\pi\, \eps}\, (D-8)\, stu\, (\delta^{i_{1}i_{2}}\delta^{i_{3}i_{4}}
+\delta^{i_{1}i_{3}}\delta^{i_{2}i_{4}}+\delta^{i_{1}i_{4}}\delta^{i_{2}i_{3}})\ .
\ee
This pole  represents the standard UV  divergences  so that 
 the  finite part contains also the  corresponding $stu \log s$  term \ci{Dubovsky:2012sh,Seibold:2023zkz}.
These contributions   vanish  in $d=2$  due  to  the kinematic $stu=0$
relation  and then the finite  part is given by \rf{11}.

Keeping $d$  general in the numerator of \rf{3.11}  the pole term \rf{3.11} can be cancelled  by adding an 
``evanescent''  scalar curvature 
 counterterm which is trivial only in $d=2$
\be
\la{3.14}
\delta S = \CounterTerm
  \int d^{d}\sigma\,\sqrt{-h}\, R^{(d)} =  \CounterTerm 
    \int d^{d}\sigma\, \Big[(\partial_{a}\partial_{b} X^{i}  \,   \del_c X^i) \ (\partial^{a}\partial^{c} X^{j}\, \partial^{b} X^{j})+\dots\Big]  \ , \ \ \ \ \ \ \ \ \  \CounterTerm  =  -\frac{1}{48\pi\, \eps} \cc
    \ . 
\ee 
Here   we ignored the contribution $\sim \del^2 X$
  that vanishes on the free equation of motion (see
Appendix \ref{app:GB}).\foot{\la{Box}The term  in \rf{3.14}  proportional to $ \del^2 X$
may be ignored also when inserting it into 2-loop counterterm diagram (third in \rf{222}) 
as it gives   extra tadpole  contribution that  vanishes in dimensional regularization.
In general,  there may be contributions to higher point functions  that may be accounted for 
by replacing $\del^2 X$ by $X^2$ terms   using equations of motion (or doing field redefinitions). 
}
The  corresponding  vertex is  
\be\la{214}
\begin{tikzpicture}[thick,baseline=0,scale=0.75]
	\begin{feynman}
	\vertex (a1) at (-1,1);
	\vertex (a2) at (-1,-1);
	\vertex (a3) at (1,1);
	\vertex (a4) at (1,-1);
	\vertex (oo) at (0,0);
	\diagram* {
	(a1) -- (oo);
	(a2) -- (oo);
	(a3) -- (oo);
	(a4) -- (oo);
	};
	\end{feynman}

	\draw[fill=gray, thin] (oo) circle (0.15);

	\draw[->,thin] ($(a1)!0.25!(oo)+(0,0.15)$) -- ($(a1)!0.75!(oo)+(0,0.15)$);
	\draw[->,thin] ($(a2)!0.25!(oo)-(0,0.15)$) -- ($(a2)!0.75!(oo)-(0,0.15)$);
	\draw[->,thin] ($(a3)!0.25!(oo)+(0,0.15)$) -- ($(a3)!0.75!(oo)+(0,0.15)$);
	\draw[->,thin] ($(a4)!0.25!(oo)-(0,0.15)$) -- ($(a4)!0.75!(oo)-(0,0.15)$);
	
	\node[left] at (a1) {$(p_{1},i_{1})$};	
	\node[left] at (a2) {$(p_{2},i_{2})$};	
	\node[right] at (a3) {$(p_{3},i_{3})$};	
	\node[right] at (a4) {$(p_{4},i_{4})$};	
	\node at (-5,0) {$E^{i_{1}i_{2}i_{3}i_{4}}(p_{1},p_{2},p_{3},p_{4}) =$};
\end{tikzpicture}
\ee 
{\small \ba 
=&-2  i\CounterTerm (p_1\cdot p_3\,  p_1\cdot p_4\,  p_2\cdot p_3+p_1\cdot p_3\,  p_1\cdot p_4 
\, p_2\cdot p_4+p_1\cdot p_3\,  p_2\cdot p_3\,  p_2\cdot p_4+p_1\cdot p_4\,  p_2\cdot p_3\,  p_2\cdot p_4) \delta^{i_{1}i_{2}}\, \delta^{i_{3}i_{4}} 
\no\\
&-2  i\CounterTerm (p_1\cdot p_2\, p_1\cdot p_4\,  p_2\cdot p_3 +p_1\cdot p_2\,  p_1\cdot p_4\,  
p_3\cdot p_4+p_1\cdot p_2\,  p_2\cdot p_3\,  p_3\cdot p_4\, +p_1\cdot p_4\,  p_2\cdot p_3\,  p_3\cdot p_4) \delta^{i_{1}i_{3}} \,\delta^{i_{2}i_{4}}\no\\
&
-2  i\CounterTerm (p_1\cdot p_2 \,p_1\cdot p_3\, p_2\cdot p_4\,+p_1\cdot p_2\, p_1\cdot p_3\, p_3\cdot p_4+p_1\cdot p_2\, 
p_2\cdot p_4 \, p_3\cdot p_4+p_1\cdot p_3\, p_2\cdot p_4\, p_3\cdot p_4) \delta^{i_{1}i_{4}} \delta^{i_{2}i_{3}}.\no
\ea
}
Its contribution cancels  the pole in \rf{3.11}  if we choose (cf. \cite{Conkey:2016qju} and also footnote 5 in \ci{Seibold:2023zkz})
\be
\la{3.18} \cc = D-8 \ . 
\ee
In the {  membrane} case,  setting $d=3-2\eps$  and $\wh D= D-3$  in \rf{39},  
 we    automatically get a finite expression  \cite{Seibold:2024oyr}  given in \rf{1.8}, so that no  ``evanescent'' 
 counterterm is needed  here.

\subsection{2-loop amplitude: string}

Since the expansion in \rf{2.3}  starts with quartic vertex, non-trivial contributions to the 2-loop  amplitude  may come from the following diagrams\foot{\la{IR}As was  already mentioned,  we are ignoring  massless 
 tadpole diagrams  as they  vanish in dimensional regularization.} 
\be
\begin{tikzpicture}[line width=1 pt, scale=1]
	\draw (-0.7,0.7)--(0,0);
	\draw (-0.7,-0.7)--(0,0);
	\draw (0.5,0) circle (0.5);
	\draw (1.5,0) circle (0.5);
	\draw (2,0)--(2.7,0.7);
	\draw (2,0)--(2.7,-0.7);
	\node at (1,-1.2) {``Double-bubble''};
\end{tikzpicture}
\qquad\quad
\begin{tikzpicture}[line width=1 pt, scale=1]
	\draw (-2,0.7)--(2,0.7);
	\draw (-2,-0.7)--(2,-0.7);
	\draw (-1,0.7) to [out=55, in=180-55] (1,0.7);
	\draw (-1,0.7) -- (0,-0.7) -- (1,0.7);
	\node at (0,-1.2) {``Wine-glass''};
\end{tikzpicture}
\qquad\quad
\begin{tikzpicture}[line width=1 pt, scale=1]
	\draw (-0.7,0.7)--(0,0);
	\draw (-0.7,-0.7)--(0,0);
	\draw (0.5,0) circle (0.5);
	\draw (1,0)--(1.7,0.7);
	\draw (1,0)--(1.7,-0.7);
	\draw[fill=gray, thin] (1,0) circle (0.1); 
	\node at (0.5,-1.2) {``Counterterm''};
\end{tikzpicture} 
\la{222}
\ee
Detailed evaluation of these 2-loop diagrams is presented in Appendix \ref{app:NG-2-loop}.
The last diagram in \rf{222}  represent the  insertion of the 1-loop  counterterm \rf{3.14}, i.e. its contribution is proportional to $\cc$. 

The  $1\ov \eps^2$ double pole terms   that may come only from the  first two diagrams in \rf{222} 
cancel out,   reflecting the fact that there are no  $1\ov \eps $ divergences  at 1-loop level in $d=2$
(this is true even off shell as the  candidate $\int R^{(2)}$ counterterm \rf{3.14}  is  trivial directly in  $d=2$ so  that there are  no 
$1\ov \eps  $  sub-divergences).\foot{
In general, the  coefficient of $1\ov \eps^2$   is proportional   to $stu $   (like the 1-loop pole term  in \rf{3.11}) 
  and thus vanishes  due to 2d kinematics  \cite{Conkey:2016qju}.}

 The simple pole  contributions  to the amplitude  are   found to be as in \cite{Conkey:2016qju} (assuming  as usual $t=0$ 
   choice of 2d kinematics)
\ba
\la{3.20}
A^{(2)}_{ \frac{1}{\eps}} =& C^{(2)}_{ \frac{1}{\eps}} = -\tfrac{1}{9216\pi^{2}\, \eps}\, (D-12)(D-8-2\cc)\,  s^{4}, \qquad\qquad \ \ \ 
B^{(2)}_{ \frac{1}{\eps}} = \tfrac{1}{768\pi^{2}\,\eps}\, (D-8-2\cc)\, s^{4},
\ea
 For the finite parts  we find (we absorb $4\pi e^{-\gamma_E}$ into the  normalization factor $\mu^2$) 
\ba
A^{(2)}_{\rm  fin}(s) =& -\tfrac{1}{13824\pi^{2}} \big[1024-255D+11D^{2} +(72-5D)\cc\big]\ s^4 +i \tfrac{1}{768\pi}(34-2D+\cc) \, s^4 \lp
 +\tfrac{1}{4608\pi^{2}}(D-12)(D-8-\cc)\ s^4\, \log(-\tfrac{s}{\mu^2})\ , \la{321}\\
C^{(2)}_{\rm  fin}(s) =& -\tfrac{1}{13824\pi^{2}} \big[1024-255D+11D^{2} +(72-5D)\cc\big]
\, s^4 -   i\tfrac{1}{4608\pi}\big[108+ 8D-D^{2} + (D-6)\cc \big]\, s^4 \lp
+\tfrac{s^{4}}{4608\pi^{2}}(D-12)(D-8-\cc)\ s^4\, \log(-\tfrac{s}{\mu^2})\ , \la{322}\\
B^{(2)}_{\rm  fin}(s) =& -\tfrac{1}{192}\, s^4 -\tfrac{1}{4608\pi^{2}}(148-35D+24\cc)- i \tfrac{1}{768\pi}(D-8-\cc)\, s^4 \lp
-\tfrac{1}{384\pi^{2}}(D-8-\cc)\ s^4\, \log(-\tfrac{s}{\mu^2})\ .\la{32}
\ea
For $\cc=D-8$ all $\log s$ terms cancel out  but the pole terms survive and we  get the expressions in \rf{75}\rf{1.5},\rf{76} \cite{Conkey:2016qju}, i.e. 
 \ba
 \la{420}
&A^{(2)}_{ \frac{1}{\eps}} =C^{(2)}_{ \frac{1}{\eps}} = \tfrac{1}{9216\pi^{2}\, \eps}\, (D-12)(D-8)\,  s^{4}, \qquad \ \ \ 
B^{(2)}_{ \frac{1}{\eps}} =- \tfrac{1}{768\pi^{2}\,\eps}\, (D-8)\, s^{4},\\
\la{3.21}
&A^{(2)}_{\rm  fin}(s) = [C^{(2)}_{\rm  fin}(-s)]^{*} = -i\,\tfrac{1}{768\pi}\,(D-26)\, s^{4}-\tfrac{1}{13824\pi^{2}}\, (6D^{2}-143D+448)\, s^{4}, \\ 
&\qquad \quad \ \ \ B^{(2)}_{\rm  fin}  
= -\tfrac{1}{192}s^{4}+\tfrac{11}{4608\pi^{2}}\, (D+4)\,  s^{4}\ .  \la{223}
\ea
 Note that  for $\cc=0$  the coefficients of the pole terms in \rf{3.20}  are   the same as of the $\log s$ terms in \rf{321}--\rf{32}.
 That means that the pole  contributions of the genuine 2-loop diagrams (the first two in \rf{222})  represent 
  the standard log UV  divergences, 
 like that was  the  case for  the 1-loop pole terms in \rf{3.11}. At the same time, half 
  of the pole terms  among those proportional to $\cc$  which come from the  evanescent counterterm diagram  and 
   are not accompanied 
 by the $\log s$  terms  should   have a   different  nature  and thus their  presence 
  may be viewed as an artefact of   dimensional regularization (cf.  \cite{Bern:2015xsa,Bern:2017puu}). 
  They should  be just subtracted as they   have no  physical consequences.\foot{\la{M}In more detail, 
  the counterterm contribution  is given by  the ${1\ov \eps}$  factor in \rf{3.14} (which  has 1-loop UV origin)  times 
  the combination 
   $J=(d-2)(s G_{1,1} +G_{1,0}+G_{0,1})$   where 
    the first integral $G_{1,1}$ is the  contribution of the bubble diagram  \rf{3.9} and the other  two are the 
    tadpoles (we assume they are IR-regularized by introducing a mass $m$). 
     The $\log m$  IR divergence of the  bubble  diagram cancels
     against the $ \log m $ terms in the  tadpoles, and the remainder is a ${1\ov d-2}$  UV divergence of the tadpoles.
$J\sim \eps \times {1\ov \eps}$ is then finite, and we get  just a simple  UV pole  factor remaining. 
The reason why   part of the $1\ov \eps$ poles in the counterterm  contribution 
 are not accompanied by $\log s$ terms can be   understood also by tracing the dependence on the  dimensional  regularization scale  
 parameter $\mu$. The factor of  $\mu^{-2\eps}$ is introduced to  compensate for  the dimension of each $\int d^d\sigma=\int d^{2-2\eps}\sigma $  or each factor of $T$  and  thus its inverse  appears 
  in  the vertices that are multiplied by $T^{-1}$  factors. 
   In   general,   at $L$-loop order   we  then get  from a simple pole factor 
  ${1\ov \eps} \mu^{2L \eps} = {1\ov \eps} + L  \log \mu^2  + ...$.  
  Thus for $L=2$   there is a factor of 2 difference  between the coefficient  of $ {1\ov \eps}$ and of $\log \mu^2$, in agreement with  what one has for the $\cc$-independent terms in \rf{3.20}--\rf{32}. 
  At the same time, the contribution of the 1-loop counterterm \rf{3.14}  contains no $\mu^{2\eps}$ factor, i.e.
  there is just   one factor of $\mu^{-2\eps}$  coming from the   left 4-vertex \rf{240} in the last diagram in \rf{222}
    (the one from  the Nambu action \rf{2.3}).
    The dependence  on $\mu$ is correlated with the dependence on $s$  in \rf{321}--\rf{32}   and that explains why 
    $1\ov \eps$ factor  still survives after   all $\log s$ terms  are cancelled out    due to  choosing $\cc=D-8$.}

 
 
 The pole terms in \rf{3.20} or \rf{420} may be canceled  by   adding a  counterterm \rf{80} 
 whose  contribution to the  amplitudes are  given by  (as usual, here $t=0$)
\be
\la{3.26}
\Delta A^{(2)} = \Delta C^{(2)} = \tfrac{1}{4}(c_{1}+2c_{2})\,s^{4}, \qquad \qquad 
\Delta B^{(2)} = \tfrac{1}{2}c_{1}\,s^{4} \ . 
\ee
Thus   to cancel pole terms in \rf{420}   we need to choose the constants in \rf{80} as 
\be \la{327}
c_{1} = \tfrac{1}{384\pi^{2}\eps}(D-8)\ , \ \  \qquad \ \ c_{2} = -\tfrac{1}{4608\pi^{2}\eps}(D-6)(D-8) \ . \ee
  The cancellation of $\log s$   terms that is possible due to the  inclusion of the   contribution of the evanescent counterterm \rf{3.14}
  is crucial  for preservation of integrability. Indeed,  for  $D=26$  the imaginary part in \rf{3.21} vanishes 
  and  then all  the real $s^4$ terms  in \rf{420}-\rf{223}  apart from $-\tfrac{1}{192}s^{4}$  in   $B^{(2)}$ 
  required for consistency with \rf{1.1},\rf{100}
   may be removed  by adding a   specific real  local  counterterm  \cite{Conkey:2016qju} (see below). 
  
  To preserve integrability for  generic $D$ we   need also to take into account the {\it finite} 
   counterterm \rf{13}   that was required for    integrability  for  any $D$   at the 1-loop  order. 
   Its  contribution to 2-loop  diagram is again given  by the  third  diagram in \rf{222}  with the vertex coming from \rf{13}
   given by 
   \ba\la{400}
&{\rm V}^{i_{1}i_{2}i_{3}i_{4}} (p_{1},p_{2},p_{3},p_{4}) = 4ib\, \Big[
p_1\cdot p_2\,p_3\cdot p_4(p_{1}+p_{2})\cdot (p_{3}+p_{4})\delta^{i_{1}i_{2}}\delta^{i_{3}i_{4}}\lp
+p_1\cdot p_3\,p_2\cdot p_4(p_{1}+p_{3})\cdot (p_{2}+p_{4})\delta^{i_{1}i_{3}}\delta^{i_{2}i_{4}}
+p_1\cdot p_4\,p_2\cdot p_3(p_{1}+p_{4})\cdot (p_{2}+p_{3})\delta^{i_{1}i_{4}}\delta^{i_{2}i_{3}}\Big] \ . 
\ea
Its tree-level contribution  changes the 1-loop  amplitudes by 
\be
\la{675}
\Delta A^{(1)} = -\Delta C^{(1)}  = b\, s^{3}, \qquad\qquad  \Delta B^{(1)}  = 0 \ .
\ee
 Choosing  $b= {D-26\ov 192 \pi}$  we cancel   the $D-26$ terms in \rf{11}.  Its  contribution to  the 1-loop 
 counterterm diagram in 
 \rf{222}  is  finite  and changes the  
  finite part  of the 2-loop  amplitudes in \rf{3.21},\rf{223}  as (see Appendix \ref{ap1})
 \ba
\Delta A^{(2)} = \tfrac{1}{4}  i b\, s^4 +  \tfrac{1}{48\pi} (6D-37)\,  b\,  s^4  \ , &\qquad 
\qquad
 \Delta C^{(2)}  =  - \tfrac{1}{4}  i b\, s^4 +    \tfrac{1}{48\pi} (6D-37)\,  b\,  s^4     \ , \no \\ \la{6.75} 
  \qquad
&  \Delta B^{(2)}  =\tfrac{1}{6 \pi}  b\,s^4 \,  \ .
\ea
 These contributions  cancel the imaginary parts in  $ A^{(2)}$ and $ C^{(2)}$   and the remaining real $s^4$ terms  in them 
 can be  eliminated by adding the   local counterterm
 \rf{80} with  finite $c_1$   and $c_2$ (cf. \rf{3.26}).\foot{Note that the signs in \rf{675} and \rf{6.75}
are  consistent with \rf{14}. Also, the non-trivial part of $B$  that should be consistent  with \rf{1.1}  has imaginary  1-loop $s^3$ term 
 but  has  real  2-loop $s^4$  term.}

\subsection{2-loop amplitude:  membrane }

Starting with  the  general expressions for the  two  2-loop diagrams in \rf{222} 
we may set  $d=3-2\eps$ and $\wh D= D-3$  (for $d=3$   there is no  1-loop  counterterm  contribution). 
We find that  the first double-bubble diagram is finite, 
while  the wine-glass  diagram  gives the following  single UV pole  contribution  (double-pole terms cancel)
\ba
A^{(2)}_{\frac{1}{\eps}}(s,t,u) =  &
-\tfrac{1}{322560\pi^{2}\eps}\Big[(27 D+455 )\, st^{3}(2s+t)+3(D-3) \, s^{3}(4s^{2}+st+10t^{2})\no \\ \la{4.4}
& \qquad \qquad \qquad \ \ \ +8 s^{3}(-8s^{2}+4st+71t^{2})\Big], 
\ea
with  $B^{(2)}$ and $C^{(2)}$  related to $A^{(2)} $ as in \rf{1.2}.
The finite part   contains $\log$ momentum terms that accompany the pole terms in \rf{4.4}, 
i.e. the latter represent genuine log UV divergences. 
We find  explicitly 
\ba
\la{230}
&A^{(2)}_{\rm  fin}(s,t,u) =  \tfrac{1}{161280 \pi ^2}\Big\{ s^3 \big[3(7D-25)s^{2}-2(9D-17)(t^{2}+u^{2})\big]\, \log(-\tfrac{s}{\mu^{2}}) 
\lp 
- 
 t^4\, \big[s (250-6 D)+t (191+3 D)\big]\log(-\tfrac{t}{\mu^{2}}) 
+ 
 u^4\, \big[t (191+3 D)+s (-59+9 D)\big]\, \log(-\tfrac{u}{\mu^{2}}) \Big\} \lp
-\tfrac{1}{45158400 \pi 
^2} \, s\,  \Big\{\big[2(8869D-97807)tu+8(1027D-8745)s^{2}\big] (t^{2}+u^{2})+(57777D+74797)t^{2}u^{2}\Big\}\lp
+\tfrac{1}{4194304}  \, s\, \Big\{ \big[
2(-289-138D+7D^{2})tu+3(617-38D+D^{2})s^{2}\big](t^{2}+u^{2})\lp\qquad \qquad \qquad 
+4(1055-138D+7D^{2})t^{2}u^{2}
\Big\}.
\ea
One can check that the total sum of coefficients of $\log\mu^2$ here is   twice  that of the  $1\ov \eps$ term in \rf{4.4}
(cf. footnote \ref{M}).

The presence of  the UV divergence in the 2-loop S-matrix \rf{4.4}   demonstrates the  non-renormalizability 
of the bosonic membrane  theory. While  this divergence may be eliminated  by a local  $\del^{10} X^4$ counterterm  
there is no known  underlying principle (like  integrability in the $d=2$  string case) 
that would fix  the coefficients of  the remaining  finite terms  in the amplitude  order by order in loop expansion.


In general, starting with the  brane action  \rf{2.1}  that has  the   coupling  given by  (inverse) tension $T$
that has  mass dimension $d$
one may  classify possible counterterms  expected at $L$-loop order. 
They should be covariant, i.e. should be  built out the extrinsic curvature 
and its covariant derivatives (see, e.g.,  \cite{Aharony:2013ipa,Conkey:2016qju,Goon:2020myi} 
 and  Appendix \ref{app:ext})\footnote{Using the  Gauss-Codazzi relation in the case of  flat  target  space (\ref{B.7}), 
one  can express the Riemann tensor of the induced metric $h_{ab}$  in terms of the extrinsic curvature, see (\ref{B.7}).}
Assuming $X^i$   have the same  dimension (of length) as $\sigma^a$    so that $\partial_a X^{i}$  and $h_{ab}$ are dimensionless, 
a    local   logarithmically  divergent term at  the $L$ loop  order  contribution to  the  
 quantum effective action  of the theory \rf{2.1} 
should have the following   structure
\be
\la{2.7}
T^{1-L}\sum_n  \int d^d \sigma\     \partial^{ d\, L}   (\partial X)^{2n}\ .    
\ee
Here  $d\,L$ derivatives should be distributed  between $2n$ factors of $X^i$.
The  4-point scattering  amplitude  probes  the terms  with  $n=2$, i.e.  with the integrand $\partial^{d\, L+4}X^4$
 corresponding  to $s^{\frac{1}{2}d\,L+2}$, etc.,   in \rf{2.8}.
Such  terms may originate from   two covariant  structures  built out of the
 extrinsic curvature $K^i_{ab}$ and covariant derivative  (cf. \rf{80})
\be\la{72}
I_1 = \sqrt{-h}\,  \nabla^{dL-2} KK, \ \ \ \ \ \qquad   I_2=   \nabla^{dL-4} KKKK  \ .
\ee
In  $I_1$    we should  pick up the leading  $\partial^{2}X$  term from the expansion  of one factor of $K$ 
and  the subleading   $\partial^4 X^3$ term  from the second  factor  (cf. Appendix \ref{app:ext}).    In $I_2$ 
 we should  pick up just the  leading $\partial^{2}X$  term in  each of the 4 factors of $K$.

In the  string   case ($d=2$)  at the  1-loop order  the only possibility is $\sqrt{-h}\, KK$, or, equivalently  $\sqrt{-h} R$  (cf. on-shell relation (\ref{B.8}))
which is trivial being  a total derivative. 
At the  2-loop order  we may  have $\sqrt{-h}\,\nabla^{2}KK$  and  $\sqrt{-h}\, KKKK$.
The first invariant reduces  to the second one at  the $X^4$ order\foot{From  
 (\ref{B.13}) we  see that we  only need to consider $\nabla^{a}K^{i}_{bc}\nabla_{a}(K^{i})^{bc}$. Using 
(\ref{B.12}) this is same as $\nabla^{c} K^{i}_{ab} \nabla^{a} (K^{i})\indices{_{c}^{b}} $+ $ KKKK$ terms
where the first term  vanishes upon integration by parts.}
so that we end up with the same counterterm \rf{80} as 
in \cite{Conkey:2016qju}. 

In the  membrane case ($d=3$)  there is no  candidate   counterterm \rf{2.7}  at $L=1$ order 
(in general, for  odd $d$  there are no  relevant invariants  at  odd number of loops).\foot{We are assuming  that parity invariance is 
preserved, i.e. that one cannot use  odd number of $\epsilon^{abc}$  to  contract the  indices.}
At $L=2$  we have the two  candidate invariants  in \rf{72}, i.e.    $\sqrt{-h}\, \nabla^{4}KK$ or 
$\sqrt{-h}\, \nabla^{2}KKKK$.
Again, we can reduce the first one to the second (restricting to 
 on-shell $X^4$ vertex).\foot{One may  
commute derivatives until they hit  $K$   getting  $\nabla K \nabla K\,  KK$   term
(commutator also gives such term).
Equivalently,  we may  first  assume that   all derivatives
are  flat ones and  use that  $\del^a\del_a  X\to 0$ on-shell.
Then replacing derivatives  by  the  covariant ones makes them non-commuting but difference is just  extra $KK $  or curvature terms.
}
The divergent part    in \rf{4.4}   is reproduced by  the 
 following counterterm (here $\del_{ab...c} \equiv \del_a\del_b...\del_c$  and repeated 3d indices  are contracted by $\eta_{ab}$) 
\ba
&\  -\tfrac{1}{80640 \pi ^2\, \eps} (27 D + 455) 
\partial_{ab}X^{i}\partial_{ac}X^{j}\partial_{de}X^{j}\partial_{debc}X^{i} 
-\tfrac{1}{10080 \pi ^2\, \eps} (3D -25) 
\partial_{ab}X^{i}\partial_{be}X^{j}\partial_{acd}X^{i}\partial_{ecd}X^{j}\lp\qquad \qquad 
+\tfrac{1}{161280 \pi ^2\, \eps} (3 D - 191) 
\partial_{ab}X^{i}\partial_{ac} X^{i}\partial_{de}X^{j}\partial_{debc}X^{j} \ , \la{233}
\ea
where we  ignore terms with $\del^2 X$ that vanish on-shell. 
 A terms like 
$\partial_{ab}X^{i}$  come from  the leading term in the    expansion of $K^{i}_{ab}$,
while 
 terms like $\partial_{debc}X^{i}$ may come from  both $\nabla_{d}\nabla_{e}K^{i}_{bc}$ and 
$\nabla_{b}\nabla_{c}K^{i}_{de}$, etc. 
Note that the  corresponding  covariant $\nabla   K \nabla  K K K$
 counterterm in the  effective action will 
  cancel also  log divergences   present  in higher-point  2-loop scattering amplitudes.

\section{Superstring and  supermembrane: classical action}

Let us   now    consider  similar scattering    of massless scalars 
for $D=10$ GS string and $D=11$ M2 brane. The tree-level amplitude is  the same as in the bosonic  case \rf{25}
while the 1-loop correction   was computed in  \ci{Seibold:2024oyr}. 

Our aim will be to find the  corresponding 2-loop amplitude. 
The total amplitude  will be  given by the bosonic  contribution from the previous section plus the contribution from the loops 
involving fermions. The latter    will be our main focus in what follows. 

Like in   \ci{Seibold:2024oyr} it will be    convenient   to treat  the  GS  string   and  the M2  brane   cases in parallel  given that they are  directly related  by the double dimensional reduction  \ci{Duff:1987bx}. 
The supermembrane action in flat target space is given by \cite{Bergshoeff:1987cm,Bergshoeff:1987qx}\foot{We assume that  $\Gamma^{0}(\Gamma^{\mu})^{\dagger}\Gamma^{0}=\Gamma^{\mu}$  and also that 
 $\eps^{012}=-1$ and $ \Gamma^{\mu_{1}\dots \mu_{n}} = \Gamma^{[\mu_{1}\dots \mu_{n}]} = \frac{1}{n!}(\Gamma^{\mu_{1}}\dots \Gamma^{\mu_{n}}+\dots).$}
\ba
\la{5.1}
S =& S_{1}+S_{2}, \qquad \qquad S_{1}=-T\int d^{3}\sigma\,\sqrt{-\det g}\ , \\
\la{5.2}
S_{2} =& -T\int d^{3}\sigma\, \tfrac{i}{2}\eps^{abc}\, \bar\theta\Gamma_{\mu\nu}\, \partial_{a}\theta\, \big(\Pi_{b}^{\mu}\Pi_{c}^{\nu}+
i\,\Pi^{\mu}_{b}\, \bar\theta\Gamma^{\nu}\partial_{c}\theta-\tfrac{1}{3}\bar\theta\Gamma^{\mu}\partial_{b}\theta\, \bar\theta\Gamma^{\nu}\partial_{c}\theta\big)\ , \\ \la{333}
g_{ab} =& \eta_{\mu\nu}\Pi^{\mu}_{a}\Pi^{\nu}_{b}, \qquad \Pi^{\mu}_{a} = \partial_{a}X^{\mu}-i\bar\theta\Gamma^{\mu}\partial_{a}\theta, \qquad
\bar\theta=\theta^{\dagger}\Gamma^{0}, \qquad \{\Gamma^{\mu}, \Gamma^{\nu}\} = 2\eta^{\mu\nu}.
\ea
 The first term in the action 
$S_{1}$ is  manifestly invariant under 
the global supersymmetry  
$
\delta X^{\mu}=i\bar\eps\Gamma^{\mu}\theta, \  \ \delta\theta = \eps. 
$
The WZ  part  $S_{2}$ can be made supersymmetric in $D=4,5,7,11$ with an appropriate choice of spinors 
\ci{Achucarro:1987nc}.
In $D=11$ spinors are Majorana $\bar\theta=\theta^{t}C$, where $C$ satisfies  
$
C\Gamma^{\mu}C=(\Gamma^{\mu})^{t}, \  C^{t}=-C, \ C^{2}=-1, 
$  so that 
 $(\bar\theta\Gamma^{\mu_{1}\dots \mu_{n}}\partial_{\mu}\theta)^{t} = (-1)^{n}\partial_{\mu}\bar\theta\Gamma^{\mu_{1}\dots\mu_{n}}\theta$.
In $D=4,5,7,11$  the action has  local fermionic $\kappa$-symmetry
\be
\delta X^{\mu} = i\bar\theta\Gamma^{\mu}(1+\mathbf{\Gamma})\kappa, \qquad \delta\theta = (1+\mathbf{\Gamma})\kappa,
\  \qquad
\mathbf{\Gamma} \equiv  \tfrac{1}{6\sqrt{-g}}\eps^{abc}\Pi^{\mu}_{a}\Pi^{\nu}_{b}\Pi^{\rho}_{c}\Gamma_{\mu}\Gamma_{\nu}\Gamma_{\rho}, \qquad \mathbf{\Gamma}^{2}=1 \ .\la{344}
\ee
Fixing the static   gauge   as in the bosonic case (so that $h_{ab} = \eta_{ab} + \del_a X^i \del_b X^i$) 
and the $\kappa$-symmetry  gauge as in  \ci{Kallosh:1997ky,Kallosh:1997sw,Seibold:2024oyr} 
\ba
&
 P_{+}\theta=0, \qquad\qquad  P_{-}\theta=\theta , \qquad \qquad P_{\pm}\equiv \tfrac{1}{2}(1\pm \Gamma^{\star}) \ , \la{3.34}\\
\la{3.16}
&\Gamma^{\star}\equiv \Gamma^{012}, \qquad (\Gamma^{\star})^{2}=1, \qquad \eps^{abc}\Gamma_{bc}=2\Gamma^{\star}\Gamma^{a}, \qquad \eps^{abc}\Gamma_{ic}=\Gamma^{\star}\Gamma_{i}\Gamma^{ab} \ , 
\ea
 and expanding the  Lagrangian  in \rf{5.1},\rf{5.2} to quartic order  in both bosons and fermions  
 we find that  the cubic $X\theta \theta $  term  vanishes. The remaining terms to quartic order in fermions  may be 
 written as (see Appendix \ref{app:M2-expansion}) 
 \ba\la{5.10}
 L= & L_{B}  +  L_{\theta^2}  + L_{X^2\theta^2}  +  L_{\theta^4} + ... \ , \ \ \ \ \ \ \ 
\ \ \ \ \ \ 
L_{B} = -\sqrt{-h}\, , \qquad L_{\theta^2} = i \bar\theta\slashed{\partial}\theta\, ,\\
L_{X^2\theta^2} =& \tfrac{i}{4}\partial_{a}X^{i}\partial^{a}X^{i}\, \bar\theta\slashed{\partial}\theta
-\tfrac{i}{2}\partial_{a}X^{i}\partial_{b}X^{i}\bar\theta\Gamma^{a}\partial^{b}\theta
-\tfrac{i}{4}\eps^{abc}\, \partial_{a}X^{i}\,\partial_{b}X^{j} \, \bar\theta\Gamma_{ij}\, \partial_{c}\theta\ ,\la{510} \\
 L_{\theta^4} =& -\tfrac{1}{4}\bar\theta\Gamma_{a}\partial_{b}\theta\ \bar\theta\Gamma^{b}\partial^{a}\theta
 +\tfrac{1}{4}\bar\theta\slashed{\partial}\theta\ \bar\theta\slashed{\partial}\theta\ .\la{511}
\ea
Here $ \slashed{\partial} = \Gamma^a \del_a$  and we rescaled the fermions 
$\theta\to \frac{1}{\sqrt 2}\theta$ to  get  the canonical  normalization of their
 kinetic term. 

 The interaction  terms   with 
$\bar \theta \slashed{\partial}\theta $ factors  in \rf{510} and \rf{511} can be, in principle, 
eliminated by a field redefinition at the expense of introducing  higher-point  vertices that should not contribute  to the 4-scalar amplitude at 2-loop   
level consider here.\foot{\la{RED}
Explicitly, we may redefine 
$\theta \to  \big(1 +  a_1 \partial_{a}X^{i}\partial^{a}X^{i}  + a_2 \bar\theta\slashed{\partial}\theta\big)\theta + ... $. Then instead of $\bar\theta\slashed{\partial}\theta$ terms in \rf{510},\rf{511} we will get  6-point 
terms like   $(\del X)^4  (\theta \del \theta) + (\del X)^2  (\theta \del \theta)^2    +   (\theta \del \theta)^3 $.
These will  not give non-trivial contributions  to the 2-loop 4-scalar amplitude.}
Thus 
the $\bar \theta \slashed{\partial}\theta $ terms in \rf{510},\rf{511}  
 can be simply omitted in the present computation.


The double dimensional reduction of the classical $D= 11$ supermembrane action
 gives the type IIA GS superstring action \cite{Duff:1987bx}. 
 Setting  $X^{2}=\sigma^{2}$  we get from \rf{5.1},\rf{5.2} 
\ba\la{3100}
S =& S_{1}+S_{2}, \quad\qquad \qquad  S_{1}=-T\int d^{2}\sigma\,\sqrt{-\det g}, \\
S_{2} =& -T\int d^{2}\sigma\, i\, \eps^{ab}\, \bar\theta\Gamma_{\mu}\Gamma_{2}\, \partial_{a}\theta\, (
\Pi_{b}^{\mu}+\tfrac{i}{2}\,\bar\theta\Gamma^{\mu}\partial_{b}\theta)\ . \la{001}
\ea
Here $g_{ab} $ is defined as in \rf{333},  $\eps^{ab}=\eps^{ab2}$ and $\theta$ is the same  as in supermembrane action. 
Here 
$
a,b=0,1, \ \mu=0,1,3,\dots, \D$, \  $  \D \equiv  D-1$, 
so that  we will have  the same number $\D-2 = D-3$  of physical ``transverse'' coordinates. 
This action is supersymmetric and $\kappa$-symmetric in dimension $\D = 3,4,6,10$.\foot{In what follows in discussing superstring we will redefine the notation D $\to D$.}

 Fixing the  static gauge and the $\kappa$-symmetry gauge  
\be\la{444}
X^{\mu} = (\sigma^{a}, X^{i}), \ i=3,\dots,\D, \qquad \ \ P_{+}\theta=0, \qquad P_{\pm}=\tfrac{1}{2}(1\pm\Gamma^{\star}), \quad \Gamma^{\star}=\Gamma^{01}\Gamma^{2} \ , 
\ee
we find that  the expansion of the GS string Lagrangian to  quartic order in fermions  
is the same as  in the M2 brane case in \rf{5.10}--\rf{511}  with the $\eps^{abc}$ term dropped 
 (see Appendix A)
\ba\la{5210}
 L= & L_{B}  +  L_{\theta^2}  + L_{X^2\theta^2}  +  L_{\theta^4} + ... \ , \ \ \ \ \ \ \ 
\ \ \ \ \ \ 
L_{B} = -\sqrt{-h}\, , \qquad L_{\theta^2} = i \bar\theta\slashed{\partial}\theta\, ,\\
L_{X^2\theta^2} =& \tfrac{i}{4}\partial_{a}X^{i}\partial^{a}X^{i}\, \bar\theta\slashed{\partial}\theta
-\tfrac{i}{2}\partial_{a}X^{i}\partial_{b}X^{i}\bar\theta\Gamma^{a}\partial^{b}\theta
\ ,\qquad 
 L_{\theta^4} = -\tfrac{1}{4}\bar\theta\Gamma_{a}\partial_{b}\theta\ \bar\theta\Gamma^{b}\partial^{a}\theta
 +\tfrac{1}{4}\bar\theta\slashed{\partial}\theta\ \bar\theta\slashed{\partial}\theta\ .\la{5211}
\ea
Note that in contrast to \rf{510} in the membrane case   this  Lagrangian contains only
 $\Gamma_a$  matrices with 2d indices
which allows to relate it to  the corresponding expansion of the NSR action \ci{Tseytlin:2025dud}. 
Like in the membrane case, here the terms with $\bar\theta\slashed{\partial}\theta$   can be redefined away.

\section{2-loop  amplitude   on  superstring}

In addition to  the bosonic  contributions we need to  include also the fermionic loop ones. 
The fermion propagator  following from \rf{5210} in the gauge \rf{444}  is given by 
$
\frac{i P_{-}\slashed{p}}{p^{2}}.$\foot{Since 
$\Gamma_a$  matrices  satisfy the 2d   Clifford algebra
 they can be identified with 2d  Dirac matrices in 32$\times 32$ representation. 
}
The $X^2 \theta^2$ vertex   following from \rf{510} or \rf{5211}  may be represented as 
\be\la{4110}
\begin{tikzpicture}[thick,baseline=0,scale=0.75]
	\begin{feynman}
	\vertex (a1) at (-1,1);
	\vertex (a2) at (-1,-1);
	\vertex (a3) at (1,1);
	\vertex (a4) at (1,-1);
	\vertex (oo) at (0,0);
	\diagram* {
	(a1) -- [fermion] (oo);
	(oo) -- [fermion] (a2);
	(a3) -- (oo);
	(a4) -- (oo);
	};
	\end{feynman}

	\draw[->, thin] ($(a1)!0.25!(oo)+(0,0.2)$) -- ($(a1)!0.75!(oo)+(0,0.2)$);
	\draw[->, thin] ($(a2)!0.25!(oo)-(0,0.2)$) -- ($(a2)!0.75!(oo)-(0,0.2)$);
	\draw[->,thin] ($(a3)!0.25!(oo)+(0,0.15)$) -- ($(a3)!0.75!(oo)+(0,0.15)$);
	\draw[->,thin] ($(a4)!0.25!(oo)-(0,0.2)$) -- ($(a4)!0.75!(oo)-(0,0.2)$);
	
	\node[left] at (a1) {$(p_{1},\alpha_{1})$};	
	\node[left] at (a2) {$(p_{2},\alpha_{2})$};	
	\node[right] at (a3) {$(p_{3},i_{3})$};	
	\node[right] at (a4) {$(p_{4},i_{4})$};	
	\node at (-5,0) {$(V^{i_{3}i_{4}})\indices{^{\alpha_{1}}_{\alpha_{2}}}(p_{1}, p_{2},p_{3},p_{4}) =\qquad \qquad $};
\end{tikzpicture}
\ee
Explicitly, in  the  spinor matrix  notation   it is given by 
\ba
\la{5.23}
 V^{i_{3}i_{4}}&(p_{1}, p_{2},p_{3},p_{4}) =\tfrac{i}{2}\, P_{-}\big[\delta^{i_{3}i_{4}}(p_{3}\cdot p_{4}\, \slashed{p}_{1}
-p_{3}\cdot p_{1}\, \slashed{p}_{4}-p_{4}\cdot p_{1}\, \slashed{p}_{3})-\eps^{abc}p_{1,a}p_{3,b}p_{4,c}\Gamma^{i_{3}i_{4}}\big] \ , 
\ea
where in the superstring case we should omit  the last $\eps^{abc}$ term. 

The quartic fermion vertex   from \rf{5211}  may be represented as 
\be
 \begin{tikzpicture}[thick,baseline=0,scale=0.75]
	\begin{feynman}
	\vertex (a1) at (-1,1);
	\vertex (a2) at (-1,-1);
	\vertex (a3) at (1,1);
	\vertex (a4) at (1,-1);
	\vertex (oo) at (0,0);
	\diagram* {
	(a1) -- [fermion] (oo);
	(a2) -- [fermion] (oo);
	(oo) -- [fermion] (a3);
	(oo) -- [fermion] (a4);
	};
	\end{feynman}

	\draw[->, thin] ($(a1)!0.25!(oo)+(0,0.2)$) -- ($(a1)!0.75!(oo)+(0,0.2)$);
	\draw[->, thin] ($(a2)!0.25!(oo)-(0,0.2)$) -- ($(a2)!0.75!(oo)-(0,0.2)$);
	\draw[->,thin] ($(a3)!0.25!(oo)+(0,0.15)$) -- ($(a3)!0.75!(oo)+(0,0.15)$);
	\draw[->,thin] ($(a4)!0.25!(oo)-(0,0.2)$) -- ($(a4)!0.75!(oo)-(0,0.2)$);
	
	\node[left] at (a1) {$(p_{1},\alpha_{1})$};	
	\node[left] at (a2) {$(p_{2},\alpha_{2})$};	
	\node[right] at (a3) {$(p_{3},\alpha_{3})$};	
	\node[right] at (a4) {$(p_{4},\alpha_{4})$};	
	\node at (-5,0) {$V\indices{_{\alpha_{3}\alpha_{4}}^{\alpha_{1}\alpha_{2}}}(p_{1}, p_{2},p_{3},p_{4}) =\qquad \qquad $};
\end{tikzpicture}\no 
\ee
\be
\la{5.25}
= \tfrac{i}{2}\big[
(\slashed{p}_{1})\indices{_{\alpha_{4}}^{\alpha_{2}}}(\slashed{p}_{2})\indices{_{\alpha_{3}}^{\alpha_{1}}}
+(\slashed{p}_{1})\indices{_{\alpha_{4}}^{\alpha_{1}}}(\slashed{p}_{2})\indices{_{\alpha_{3}}^{\alpha_{2}}}
-(\slashed{p}_{1})\indices{_{\alpha_{3}}^{\alpha_{2}}}(\slashed{p}_{2})\indices{_{\alpha_{4}}^{\alpha_{1}}}
-(\slashed{p}_{1})\indices{_{\alpha_{3}}^{\alpha_{1}}}(\slashed{p}_{2})\indices{_{\alpha_{4}}^{\alpha_{2}}}\big],
\ee
where  it is understood that all $\slashed{p}$ factors  contain  also the projector $P_{-}$.

\subsection{1-loop  order }
\la{sec:fer-1loop}

The fermionic loop contribution to 4-scalar scattering amplitude  is given  by 
\be
\begin{tikzpicture}[thick,baseline=0]
	\begin{feynman}
	\vertex (a1) at (-1,1);
	\vertex (a2) at (-1,-1);
	\vertex (a3) at (3,1);
	\vertex (a4) at (3,-1);
	\vertex (oo1) at (0,0);
	\vertex (oo2) at (2,0);
	\diagram* {
	(a1) -- (oo1);
	(a2) -- (oo1);
	(a3) -- (oo2);
	(a4) -- (oo2);
	(oo1) -- [half left, fermion] (oo2);
	(oo2) -- [half left, fermion] (oo1);
	};
	\end{feynman}

	\draw[->,thin] ($(a1)!0.25!(oo1)+(0,0.15)$) -- ($(a1)!0.75!(oo1)+(0,0.15)$);
	\draw[->,thin] ($(a2)!0.25!(oo1)-(0,0.15)$) -- ($(a2)!0.75!(oo1)-(0,0.15)$);
	\draw[->,thin] ($(a3)!0.25!(oo2)+(0,0.15)$) -- ($(a3)!0.75!(oo2)+(0,0.15)$);
	\draw[->,thin] ($(a4)!0.25!(oo2)-(0,0.15)$) -- ($(a4)!0.75!(oo2)-(0,0.15)$);
	
         \draw[->,thin]  ($ (oo1)!0.25!(oo2) + (0,0.6) $) to [out=15,in=165] ($ (oo1)!0.75!(oo2) + (0,0.6) $);
         \draw[->,thin]  ($ (oo2)!0.25!(oo1) - (0,0.6) $) to [out=180+15,in=-15] ($ (oo2)!0.75!(oo1) - (0,0.6) $);
	
	\node[left] at (a1) {$(p_{1},i_{1})$};	
	\node[left] at (a2) {$(p_{2},i_{2})$};	
	\node[right] at (a3) {$(p_{3},i_{3})$};	
	\node[right] at (a4) {$(p_{4},i_{4})$};	
	\node at (1,1.2) {\small $k+p_{1}+p_{2}$};
	\node at (1,-1.2) {\small $k$};
	\node at (-2.5,0) {$\DD_{\rm  F}(1,2,3,4)=\qquad \qquad $};
\end{tikzpicture} \la{445}
\ee
\be
 = 
 \frac{1}{i}\int \frac{d^{d}k}{(2\pi)^{d}}\tr\Big[V^{i_{3}i_{4}}(k+p_{1}+p_{2}, -k,p_{3},p_{4})\frac{-iP_{-}(\slashed{k}+\slashed{p}_{1}+\slashed{p}_{2})}{(k+p_{1}+p_{2})^{2}} 
 V^{i_{1}i_{2}}(k,-k-p_{1}-p_{2},p_{1},p_{2})\frac{-iP_{-}\slashed{k}}{k^{2}}\Big],\no 
\ee
where  $ V^{i_{1}i_{2}}$ was  given in \rf{5.23}.
The trace is proportional to $\tr P_{-}=n_F=16$; we also include  the factor $\frac{1}{2}$ as spinors  are  Majorana, and $(-1)$  to account for the  Fermi statistics  so that effectively 
$\tr P_{-}\to -\frac{1}{2}n_{F}=-8$. 
 Expanding  the above expression  for  $d=2-2\eps$ 
 we find that  resulting 1-loop contribution to the $D=10$ superstring  scattering amplitude has a divergent part
\be
\la{6.3}
\mc M^{(1)\, i_{1}i_{2}i_{3}i_{4}}_{{\rm F}, \frac{1}{\eps}} = -\tfrac{1}{24\pi\, \eps}\, 
\, stu\,  (\delta^{i_{1}i_{2}}\delta^{i_{3}i_{4}}
+\delta^{i_{1}i_{3}}\delta^{i_{2}i_{4}}+\delta^{i_{1}i_{4}}\delta^{i_{2}i_{3}}) \ . 
\ee
This vanishes  directly in $d=2$  where $stu=0$.  Combined with the bosonic loop contribution \rf{3.11} evaluated at $D=10$
that  corresponds to the  coefficient of the evanescent counterterm in  \rf{3.14}   being  changed to 
\be 
\cc=  D-8 + 4  \stackrel{D\to 10}{=}   6 \ . \la{6.7} \ee
 The finite part of the 
fermion loop  contribution  for $t=0$    is given  by 
\be\la{77}
A^{(1)}_{\rm F, fin}(s)  =- C^{(1)}_{\rm F, fin}(s)   = -\tfrac{1}{48\pi}s^{3}\ , \qquad 
\ \  \qquad B^{(1)}_{\rm F, fin} = 0 \ . 
\ee
Added to  the bosonic loop contribution 
 (\ref{11}) with $D=10$, the total 1-loop amplitude  in the  superstring  case is  given
   by \rf{15} \cite{Seibold:2024oyr}, i.e.
\ba
\la{6.5}
A^{(1)}_{\rm  fin} = -C^{(1)}_{\rm  fin} =  -\tfrac{1}{192\pi} (D-26 + 4)\, s^{3} \stackrel{D\to 10}{=} \tfrac{1}{16\pi}\,s^{3}\, ,\ \ \ \ \  \qquad \ \ \ \ 
B^{(1)}_{\rm  fin} = i\, \tfrac{1}{16}\, s^{3} \ . 
\ea
As in the bosonic case, one  can  satisfy the integrability requirement 
  $A^{(1)}=C^{(1)}=0$   by adding the  counterterm \rf{13}  with   \cite{Seibold:2024oyr}\foot{As was noticed   in  \cite{Seibold:2024oyr}, if one defines  the GS scattering amplitude as a ``double-dimensional'' \ci{Duff:1987bx}  limit of the amplitude on   $S^1$-compactified  M2 brane, 
   then one   gets directly  that $A^{(1)} =C^{(1)}=0$, 
implying that the definition of the GS  path integral   via  the  M2 brane  one  automatically  provides the  required 
measure factors or local counterterms, at least to the  1-loop order.}
  \be \bb= D-26 +4 \stackrel{D\to 10}{=}  -12 \ . \la{78}
\ee 

\subsection{2-loop order}

We get the same 2-loop diagram topologies as in \rf{222}  where 
now there are additional  contributions  from diagrams with  internal fermionic lines. 
We will discuss these   in turn. 

\subsubsection{Double-bubble diagrams}

There are 3 such  diagrams -- two with one bosonic and  one fermionic loop (related in an obvious way  by 
  exchanging $12\leftrightarrow 34$), and  the one with two fermionic loops:
{\small 
\be
\la{6.9}
\begin{tikzpicture}[thick,baseline]
	\begin{feynman}
	\vertex (a1) at (-1,1);
	\vertex (a2) at (-1,-1);
	\vertex (a3) at (5,1);
	\vertex (a4) at (5,-1);
	\vertex (oo1) at (0,0);
	\vertex (oo2) at (2,0);
	\vertex (oo3) at (4,0);
	\diagram* {
	(a1) -- (oo1);
	(a2) -- (oo1);
	(a3) -- (oo3);
	(a4) -- (oo3);
	(oo1) -- [half left, fermion] (oo2);
	(oo2) -- [half left, fermion] (oo1);
	(oo2) -- [half left] (oo3);
	(oo2) -- [half right] (oo3);
	};
	\end{feynman}

	\draw[->,thin] ($(a1)!0.25!(oo1)+(0,0.15)$) -- ($(a1)!0.75!(oo1)+(0,0.15)$);
	\draw[->,thin] ($(a2)!0.25!(oo1)-(0,0.15)$) -- ($(a2)!0.75!(oo1)-(0,0.15)$);
	\draw[->,thin] ($(a3)!0.25!(oo3)+(0,0.15)$) -- ($(a3)!0.75!(oo3)+(0,0.15)$);
	\draw[->,thin] ($(a4)!0.25!(oo3)-(0,0.15)$) -- ($(a4)!0.75!(oo3)-(0,0.15)$);
	
         \draw[->,thin]  ($ (oo1)!0.25!(oo2) + (0,0.6) $) to [out=15,in=165] ($ (oo1)!0.75!(oo2) + (0,0.6) $);
         \draw[->,thin]  ($ (oo2)!0.25!(oo1) - (0,0.6) $) to [out=180+15,in=-15] ($ (oo2)!0.75!(oo1) - (0,0.6) $);
         \draw[->,thin]  ($ (oo2)!0.25!(oo3) + (0,0.6) $) to [out=15,in=165] ($ (oo2)!0.75!(oo3) + (0,0.6) $);
         \draw[->,thin]  ($ (oo3)!0.25!(oo2) - (0,0.6) $) to [out=180+15,in=-15] ($ (oo3)!0.75!(oo2) - (0,0.6) $);
         	
	\node[left] at (a1) {$(p_{1},i_{1})$};	
	\node[left] at (a2) {$(p_{2},i_{2})$};	
	\node[right] at (a3) {$(p_{3},i_{3})$};	
	\node[right] at (a4) {$(p_{4},i_{4})$};	
	\node at (1-0.3,1.2) {\small $k_{1}+p_{1}+p_{2}$};
	\node at (1,-1.2) {\small $k_{1}$};
	\node at (3+0.3,1.2) {\small $(k_{2}+p_{1}+p_{2}, j_{1})$};
	\node at (3,-1.2) {\small $(k_{2}, j_{2})$};
	\node at (-3,0) {$({\rm DB}_{a, \rm F})^{i_{1}i_{2}i_{3}i_{4}}_{p_{1},p_{2},p_{3},p_{4}} =\qquad \qquad  $};
\end{tikzpicture}
\ee
\be 
\la{6.10}
\begin{tikzpicture}[thick,baseline=0]
	\begin{feynman}
	\vertex (a1) at (-1,1);
	\vertex (a2) at (-1,-1);
	\vertex (a3) at (5,1);
	\vertex (a4) at (5,-1);
	\vertex (oo1) at (0,0);
	\vertex (oo2) at (2,0);
	\vertex (oo3) at (4,0);
	\diagram* {
	(a1) -- (oo1);
	(a2) -- (oo1);
	(a3) -- (oo3);
	(a4) -- (oo3);
	(oo1) -- [half left] (oo2);
	(oo2) -- [half left] (oo1);
	(oo2) -- [half left, fermion] (oo3);
	(oo3) -- [half left, fermion] (oo2);
	};
	\end{feynman}

	\draw[->,thin] ($(a1)!0.25!(oo1)+(0,0.15)$) -- ($(a1)!0.75!(oo1)+(0,0.15)$);
	\draw[->,thin] ($(a2)!0.25!(oo1)-(0,0.15)$) -- ($(a2)!0.75!(oo1)-(0,0.15)$);
	\draw[->,thin] ($(a3)!0.25!(oo3)+(0,0.15)$) -- ($(a3)!0.75!(oo3)+(0,0.15)$);
	\draw[->,thin] ($(a4)!0.25!(oo3)-(0,0.15)$) -- ($(a4)!0.75!(oo3)-(0,0.15)$);
	
         \draw[->,thin]  ($ (oo1)!0.25!(oo2) + (0,0.6) $) to [out=15,in=165] ($ (oo1)!0.75!(oo2) + (0,0.6) $);
         \draw[->,thin]  ($ (oo2)!0.25!(oo1) - (0,0.6) $) to [out=180+15,in=-15] ($ (oo2)!0.75!(oo1) - (0,0.6) $);
         \draw[->,thin]  ($ (oo2)!0.25!(oo3) + (0,0.6) $) to [out=15,in=165] ($ (oo2)!0.75!(oo3) + (0,0.6) $);
         \draw[->,thin]  ($ (oo3)!0.25!(oo2) - (0,0.6) $) to [out=180+15,in=-15] ($ (oo3)!0.75!(oo2) - (0,0.6) $);
         	
	\node[left] at (a1) {$(p_{1},i_{1})$};	
	\node[left] at (a2) {$(p_{2},i_{2})$};	
	\node[right] at (a3) {$(p_{3},i_{3})$};	
	\node[right] at (a4) {$(p_{4},i_{4})$};	
	\node at (1-0.3,1.2) {\small $(k_{1}+p_{1}+p_{2}, j_{1})$};
	\node at (1,-1.2) {\small $(k_{1}, j_{2})$};
	\node at (3+0.3,1.2) {\small $k_{2}+p_{1}+p_{2})$};
	\node at (3,-1.2) {\small $k_{2}$};
	\node at (-3,0) {$({\rm DB}_{b, \rm F})^{i_{1}i_{2}i_{3}i_{4}}_{p_{1},p_{2},p_{3},p_{4}} =\qquad \qquad $};
\end{tikzpicture}
\ee
\be
\la{6.11}
\begin{tikzpicture}[thick,baseline]
	\begin{feynman}
	\vertex (a1) at (-1,1);
	\vertex (a2) at (-1,-1);
	\vertex (a3) at (5,1);
	\vertex (a4) at (5,-1);
	\vertex (oo1) at (0,0);
	\vertex (oo2) at (2,0);
	\vertex (oo3) at (4,0);
	\diagram* {
	(a1) -- (oo1);
	(a2) -- (oo1);
	(a3) -- (oo3);
	(a4) -- (oo3);
	(oo1) -- [half left, fermion] (oo2);
	(oo2) -- [half left, fermion] (oo1);
	(oo2) -- [half left, fermion] (oo3);
	(oo3) -- [half left, fermion] (oo2);
	};
	\end{feynman}

	\draw[->,thin] ($(a1)!0.25!(oo1)+(0,0.15)$) -- ($(a1)!0.75!(oo1)+(0,0.15)$);
	\draw[->,thin] ($(a2)!0.25!(oo1)-(0,0.15)$) -- ($(a2)!0.75!(oo1)-(0,0.15)$);
	\draw[->,thin] ($(a3)!0.25!(oo3)+(0,0.15)$) -- ($(a3)!0.75!(oo3)+(0,0.15)$);
	\draw[->,thin] ($(a4)!0.25!(oo3)-(0,0.15)$) -- ($(a4)!0.75!(oo3)-(0,0.15)$);
	
         \draw[->,thin]  ($ (oo1)!0.25!(oo2) + (0,0.6) $) to [out=15,in=165] ($ (oo1)!0.75!(oo2) + (0,0.6) $);
         \draw[->,thin]  ($ (oo2)!0.25!(oo1) - (0,0.6) $) to [out=180+15,in=-15] ($ (oo2)!0.75!(oo1) - (0,0.6) $);
         \draw[->,thin]  ($ (oo2)!0.25!(oo3) + (0,0.6) $) to [out=15,in=165] ($ (oo2)!0.75!(oo3) + (0,0.6) $);
         \draw[->,thin]  ($ (oo3)!0.25!(oo2) - (0,0.6) $) to [out=180+15,in=-15] ($ (oo3)!0.75!(oo2) - (0,0.6) $);
         	
	\node[left] at (a1) {$(p_{1},i_{1})$};	
	\node[left] at (a2) {$(p_{2},i_{2})$};	
	\node[right] at (a3) {$(p_{3},i_{3})$};	
	\node[right] at (a4) {$(p_{4},i_{4})$};	
	\node at (1-0.3,1.2) {\small $k_{1}+p_{1}+p_{2}$};
	\node at (1,-1.2) {\small $k_{1}$};
	\node at (3+0.3,1.2) {\small $k_{2}+p_{1}+p_{2}$};
	\node at (3,-1.2) {\small $k_{2}$};
	\node at (-3,0) {$({\rm DB}_{c, \rm F})^{i_{1}i_{2}i_{3}i_{4}}_{p_{1},p_{2},p_{3},p_{4}} =\qquad \qquad $};
\end{tikzpicture}
\ee
}
Explicitly, they are given by 
\ba
&({\rm DB}_{a, \rm F}) ^{i_{1}i_{2}i_{3}i_{4}}_{p_{1},p_{2},p_{3},p_{4}} =\tfrac{1}{2i}\int \wt{dk_{1}}\, \wt{dk_{2}}
\frac{1}{k_{2}^{2}(k_{2}+p_{1}+p_{2})^{2}}
V^{j_{1}j_{2}i_{3}i_{4}}_{p_{1}+p_{2}+k_{2}, -k_{2}, p_{3}, p_{4}}\lp\qquad \qquad 
\tr\Big[V^{i_{1}i_{2}}_{k_{1},-k_{1}-p_{1}-p_{2},p_{1},p_{2}}\frac{P_{-}(\slashed{k}_{1})}{k_{1}^{2}}
 V^{j_{1}j_{2}}_{k_{1}+p_{1}+p_{2}, -k_{1}, -k_{2}-p_{1}-p_{2}, k_{2}}\frac{P_{-}(\slashed{k}_{1}+\slashed{p}_{1}+\slashed{p}_{2})}{(k_{1}+p_{1}+p_{2})^{2}}\Big], \la{114}
\\ &
({\rm DB}_{b, \rm F}) ^{i_{1}i_{2}i_{3}i_{4}}_{p_{1},p_{2},p_{3},p_{4}} =\tfrac{1}{2i}\int \wt{dk_{1}}\, \wt{dk_{2}}
\frac{1}{k_{1}^{2}(k_{1}+p_{1}+p_{2})^{2}}
V^{i_{1}i_{2}j_{1}j_{2}}_{p_{1},p_{2},-k_{1}-p_{1}-p_{2},k_{1}}\lp\qquad \qquad 
\tr\Big[V^{i_{3}i_{4}}_{k_{2}+p_{1}+p_{2}, -k_{2}, p_{3}, p_{4}}\frac{P_{-}(\slashed{k}_{2}+\slashed{p}_{1}+\slashed{p}_{2})}{(k_{2}+p_{1}+p_{2})^{2}}
 V^{j_{1}j_{2}}_{k_{2},-k_{2}-p_{1}-p_{2},k_{1}+p_{1}+p_{2},-k_{1}}\frac{P_{-}(\slashed{k}_{2})}{k_{2}^{2}}\Big], \la{411}\\
&({\rm DB}_{c, \rm F}) ^{i_{1}i_{2}i_{3}i_{4}}_{p_{1},p_{2},p_{3},p_{4}} =\tfrac{1}{i}\int \wt{dk_{1}}\, \wt{dk_{2}}
\Big[S(k_{1}+p_{1}+p_{2})V^{i_{1}i_{2}}_{k_{1},-k_{1}-p_{1}-p_{2},p_{1},p_{2}}S(k_{1})\Big]_{\alpha_{2}}^{\  \alpha_{1}}
\lp\qquad \qquad 
\Big[S(k_{2})V^{i_{3}i_{4}}_{k_{2}+p_{1}+p_{2}, -k_{2}, p_{3}, p_{4}}S(k_{2}+p_{1}+p_{2})\Big]_{\alpha_{4}}^{\ \alpha_{3}}
V\indices{_{\alpha_{1}\alpha_{3}}^{\alpha_{2}\alpha_{4}}}(k_{1}+p_{1}+p_{2}, k_{2}), \la{415}
\ea
where  the vertices $V$ were given in \rf{5.23},\rf{5.25}  and we used the notation
$\wt{dk}= {d^d k\ov (2\pi)^d}$ and $S(p) = \frac{P_{-}\slashed{p}}{p^{2}}$. 

Introducing  $q=k_{1}+p_{1}+p_{2}$  and 
\ba
U\equiv S(k_{1}+p_{1}+p_{2})V^{i_{1}i_{2}}_{k_{1},-k_{1}-p_{1}-p_{2},p_{1},p_{2}}S(k_{1}), \qquad 
W\equiv S(k_{2})V^{i_{3}i_{4}}_{k_{2}+p_{1}+p_{2}, -k_{2}, p_{3}, p_{4}}S(k_{2}+p_{1}+p_{2}),
\ea
we have for the  integrand in \rf{415} (\cf (\ref{5.25}))
\ba
\la{6.19}
\tfrac{i}{2}\, &U\indices{_{\alpha_{2}}^{\alpha_{1}}} W\indices{_{\alpha_{4}}^{\alpha_{3}}}[
(\slashed{q})\indices{_{\alpha_{3}}^{\alpha_{4}}}(\slashed{k}_{2})\indices{_{\alpha_{1}}^{\alpha_{2}}}
+(\slashed{q})\indices{_{\alpha_{3}}^{\alpha_{2}}}(\slashed{k}_{2})\indices{_{\alpha_{1}}^{\alpha_{4}}}
-(\slashed{q})\indices{_{\alpha_{1}}^{\alpha_{4}}}(\slashed{k}_{2})\indices{_{\alpha_{3}}^{\alpha_{2}}}
-(\slashed{q})\indices{_{\alpha_{1}}^{\alpha_{2}}}(\slashed{k}_{2})\indices{_{\alpha_{3}}^{\alpha_{4}}}]\lp
= \tfrac{i}{2}[\tr(U\slashed{k}_{2})\tr(W \slashed{q})+\tr(U\slashed{q}W\slashed{k}_{2})-\tr(U\slashed{k}_{2}W\slashed{q})
-\tr(U\slashed{q})\tr(W\slashed{k}_{2})].
\ea

\subsubsection{Wine-glass diagrams}

There are 3 diagrams  of this topology: one with 2 fermionic propagators and two   with 3 fermionic propagators
(these are  related by $(p_{1}, i_{1})\leftrightarrow (p_{3},i_{3})$ exchange):
{\small
\be
\la{6.20}
\begin{tikzpicture}[thick, scale=0.8,baseline=0]
	\begin{feynman}
	\vertex (a1) at (-4,1.4);
	\vertex (a2) at (-4,-1.4);
	\vertex (a3) at (4,1.4);
	\vertex (a4) at (4,-1.4);
	\vertex (p1) at (-2,1.4);
	\vertex (p2) at (2,1.4);
	\vertex (p3) at (0,-1.4);
	\diagram* {
	(a1) -- (p1);
	(a2) -- (p3);
	(a3) -- (p2);
	(a4) -- (p3);
	(p2) -- [fermion] (p1);
	(p1) -- (p3) -- (p2);
	(p1) -- [fermion, bend left=45] (p2) ;
	};
	\end{feynman}

	\draw[->,thin] ($(a1)!0.25!(p1)+(0,0.15)$) -- ($(a1)!0.75!(p1)+(0,0.15)$);
	\draw[->,thin] ($(a3)!0.25!(p2)+(0,0.15)$) -- ($(a3)!0.75!(p2)+(0,0.15)$);
	\draw[->,thin] ($(a2)!0.25!(p3)+(0,0.15)$) -- ($(a2)!0.75!(p3)+(0,0.15)$);
	\draw[->,thin] ($(a4)!0.25!(p3)+(0,0.15)$) -- ($(a4)!0.75!(p3)+(0,0.15)$);
	\draw[->,thin] ($(p2)!0.25!(p1)-(0,0.15)$) -- ($(p2)!0.75!(p1)-(0,0.15)$);
	\draw[->,thin] ($(p1)!0.25!(p2)+(0,0.75)$) to [out=20, in=180-20]  ($(p1)!0.75!(p2)+(0,0.75)$);
	\draw[->,thin] ($(p3)!0.25!(p1)-(0.15,0)$) -- ($(p3)!0.75!(p1)-(0.15,0)$);
	\draw[->,thin] ($(p2)!0.25!(p3)+(0.15,0)$) -- ($(p2)!0.75!(p3)+(0.15,0)$);
         	
	\node[left] at (a1) {$(p_{1},i_{1})$};	
	\node[left] at (a2) {$(p_{2},i_{2})$};	
	\node[right] at (a3) {$(p_{3},i_{3})$};	
	\node[right] at (a4) {$(p_{4},i_{4})$};	
	\node at ($(p1)!0.5!(p2)+(0,1.5)$)  {$k_{1}$};
	\node at ($(p1)!0.5!(p2)-(0,0.6)$)  {$k_{1}-k_{2}$};
	\node at ($(p3)!0.5!(p1)-(1.5,0)$) {$(k_{2}-p_{1},j_{1})$};
	\node at ($(p3)!0.5!(p2)+(1.5,0)$) {$(k_{2}+p_{3},j_{2})$};
	\node at (-6,0) {$({\rm W}_{a, \rm F})^{i_{1}i_{2}i_{3}i_{4}}_{p_{1},p_{2},p_{3},p_{4}} = \qquad \qquad\qquad \qquad $};
\end{tikzpicture}
\ee
\be
\la{6.21}
\begin{tikzpicture}[thick,scale=0.8,baseline=0]
	\begin{feynman}
	\vertex (a1) at (-4,1.4);
	\vertex (a2) at (-4,-1.4);
	\vertex (a3) at (4,1.4);
	\vertex (a4) at (4,-1.4);
	\vertex (p1) at (-2,1.4);
	\vertex (p2) at (2,1.4);
	\vertex (p3) at (0,-1.4);
	\diagram* {
	(a1) -- (p1);
	(a2) -- (p3);
	(a3) -- (p2);
	(a4) -- (p3);
	(p2) -- [fermion] (p1);
	(p1) -- [fermion] (p3) -- [fermion] (p2);
	(p1) -- [bend left=45] (p2) ;
	};
	\end{feynman}

	\draw[->,thin] ($(a1)!0.25!(p1)+(0,0.15)$) -- ($(a1)!0.75!(p1)+(0,0.15)$);
	\draw[->,thin] ($(a3)!0.25!(p2)+(0,0.15)$) -- ($(a3)!0.75!(p2)+(0,0.15)$);
	\draw[->,thin] ($(a2)!0.25!(p3)+(0,0.15)$) -- ($(a2)!0.75!(p3)+(0,0.15)$);
	\draw[->,thin] ($(a4)!0.25!(p3)+(0,0.15)$) -- ($(a4)!0.75!(p3)+(0,0.15)$);
	\draw[->,thin] ($(p2)!0.25!(p1)-(0,0.15)$) -- ($(p2)!0.75!(p1)-(0,0.15)$);
	\draw[->,thin] ($(p1)!0.25!(p2)+(0,0.75)$) to [out=20, in=180-20]  ($(p1)!0.75!(p2)+(0,0.75)$);
	\draw[->,thin] ($(p1)!0.25!(p3)-(0.15,0)$) -- ($(p1)!0.75!(p3)-(0.15,0)$);
	\draw[->,thin] ($(p3)!0.25!(p2)+(0.15,0)$) -- ($(p3)!0.75!(p2)+(0.15,0)$);
         	
	\node[left] at (a1) {$(p_{1},i_{1})$};	
	\node[left] at (a2) {$(p_{2},i_{2})$};	
	\node[right] at (a3) {$(p_{3},i_{3})$};	
	\node[right] at (a4) {$(p_{4},i_{4})$};	
	\node at ($(p1)!0.5!(p2)+(0,1.5)$)  {$(k_{1}, j_{1})$};
	\node at ($(p1)!0.5!(p2)-(0,0.6)$)  {$k_{1}-k_{2}$};
	\node at ($(p3)!0.5!(p1)-(1.5,0)$) {$p_{1}-k_{2}$};
	\node at ($(p3)!0.5!(p2)+(1.5,0)$) {$-k_{2}-p_{3}$};
	\node at (-6,0) {$({\rm W}_{b, \rm F})^{i_{1}i_{2}i_{3}i_{4}}_{p_{1},p_{2},p_{3},p_{4}} =\qquad \qquad \qquad \qquad $};
\end{tikzpicture}
\ee
\be
\la{6.22}
\begin{tikzpicture}[thick,scale=0.8,baseline=0]
	\begin{feynman}
	\vertex (a1) at (-4,1.4);
	\vertex (a2) at (-4,-1.4);
	\vertex (a3) at (4,1.4);
	\vertex (a4) at (4,-1.4);
	\vertex (p1) at (-2,1.4);
	\vertex (p2) at (2,1.4);
	\vertex (p3) at (0,-1.4);
	\diagram* {
	(a1) -- (p1);
	(a2) -- (p3);
	(a3) -- (p2);
	(a4) -- (p3);
	(p1) -- [fermion] (p2);
	(p2) -- [fermion] (p3) -- [fermion] (p1);
	(p1) -- [bend left=45] (p2) ;
	};
	\end{feynman}

	\draw[->,thin] ($(a1)!0.25!(p1)+(0,0.15)$) -- ($(a1)!0.75!(p1)+(0,0.15)$);
	\draw[->,thin] ($(a3)!0.25!(p2)+(0,0.15)$) -- ($(a3)!0.75!(p2)+(0,0.15)$);
	\draw[->,thin] ($(a2)!0.25!(p3)+(0,0.15)$) -- ($(a2)!0.75!(p3)+(0,0.15)$);
	\draw[->,thin] ($(a4)!0.25!(p3)+(0,0.15)$) -- ($(a4)!0.75!(p3)+(0,0.15)$);
	\draw[->,thin] ($(p1)!0.25!(p2)-(0,0.15)$) -- ($(p1)!0.75!(p2)-(0,0.15)$);
	\draw[->,thin] ($(p1)!0.25!(p2)+(0,0.75)$) to [out=20, in=180-20]  ($(p1)!0.75!(p2)+(0,0.75)$);
	\draw[->,thin] ($(p3)!0.25!(p1)-(0.15,0)$) -- ($(p3)!0.75!(p1)-(0.15,0)$);
	\draw[->,thin] ($(p2)!0.25!(p3)+(0.15,0)$) -- ($(p2)!0.75!(p3)+(0.15,0)$);
         	
	\node[left] at (a1) {$(p_{1},i_{1})$};	
	\node[left] at (a2) {$(p_{2},i_{2})$};	
	\node[right] at (a3) {$(p_{3},i_{3})$};	
	\node[right] at (a4) {$(p_{4},i_{4})$};	
	\node at ($(p1)!0.5!(p2)+(0,1.5)$)  {$(k_{1}, j_{1})$};
	\node at ($(p1)!0.5!(p2)-(0,0.6)$)  {$k_{2}-k_{1}$};
	\node at ($(p3)!0.5!(p1)-(1.5,0)$) {$k_{2}-p_{1}$};
	\node at ($(p3)!0.5!(p2)+(1.5,0)$) {$k_{2}+p_{3}$};
	\node at (-6,0) {$({\rm W}_{c, \rm F})^{i_{1}i_{2}i_{3}i_{4}}_{p_{1},p_{2},p_{3},p_{4}} = \qquad \qquad\qquad \qquad$};
\end{tikzpicture}
\ee
}
The  corresponding momentum integrals are 
 \ba
&({\rm W}_{a, \rm F})^{i_{1}i_{2}i_{3}i_{4}}_{p_{1},p_{2},p_{3},p_{4}} =\tfrac{1}{i}\int \wt{dk_{1}}\, \wt{dk_{2}}
\frac{
V^{i_{2}j_{1}j_{2}i_{4}}_{p_{2},-k_{2}+p_{1},k_{2}+p_{3},p_{4}}
}{(k_{2}+p_{3})^{2}(k_{2}-p_{1})^{2}}\no \\ &\qquad \qquad \qquad \qquad\qquad \qquad\times 
\tr\Big[S(k_{1}) V^{i_{1}j_{1}}_{k_{1}-k_{2},-k_{1},p_{1},k_{2}-p_{1}}S(k_{1}-k_{2}) V^{i_{3}j_{2}}_{k_{1},-k_{1}+k_{2},p_{3},-k_{2}-p_{3}}\Big],\la{421} \\
&
({\rm W}_{b, \rm F})^{i_{1}i_{2}i_{3}i_{4}}_{p_{1},p_{2},p_{3},p_{4}} =\tfrac{1}{i}\int \wt{dk_{1}}\, \wt{dk_{2}}
\frac{1}{k_{1}^{2}}\
\tr\Big[S(k_{1}-k_{2})V^{i_{3}j_{1}}_{-k_{2}-p_{3},k_{2}-k_{1},p_{3},k_{1}}\no  \\ & \qquad\qquad \qquad \qquad\qquad \qquad  \times S(-k_{2}-p_{3})
V^{i_{2}i_{4}}_{p_{1}-k_{2},k_{2}+p_{3},p_{2},p_{4}}S(p_{1}-k_{2})V^{i_{1}j_{1}}_{k_{1}-k_{2},k_{2}-p_{1},p_{1},-k_{1}}\Big], \la{422}\\
&
({\rm W}_{c,\rm F})^{i_{1}i_{2}i_{3}i_{4}}_{p_{1},p_{2},p_{3},p_{4}} =\tfrac{1}{i}\int \wt{dk_{1}}\, \wt{dk_{2}}
\frac{1}{k_{1}^{2}}
\tr\Big[S(k_{2}-k_{1})V^{i_{1}j_{1}}_{k_{2}-p_{1},k_{1}-k_{2},p_{1},-k_{1}}S(k_{2}-p_{1})\no \\ & 
 \qquad\qquad \qquad \qquad\qquad \qquad \qquad \qquad \times V^{i_{2}i_{4}}_{k_{2}+p_{3},-k_{2}+p_{1},p_{2},p_{4}}
S(k_{2}+p_{3})V^{i_{3}j_{1}}_{k_{2}-k_{1},-k_{2}-p_{3},p_{3},k_{1}}\Big].\la{423}
\ea

\subsubsection{Counterterm diagram with  $X^4$ vertex  from  fermionic loop}

Next, let us consider possible  counterterm  diagrams
 corresponding to the third topology  in \rf{222}. 
In the case when the  loop  in that diagram is bosonic,  the   counterterm $X^4$  vertex 
may  correspond to the 1-loop diagram   from  either  bosonic or fermionic loop.

The case of the bosonic 1-loop counterterm \rf{3.14} was already discussed  in section 2. 
Let us now consider the  case of the 1-loop counterterm vertex  originating from the fermionic loop diagram \rf{445}. 
The  corresponding 2-loop order contribution is represented by
\be
\la{6.49}
\begin{tikzpicture}[thick,baseline=0]
	\begin{feynman}
	\vertex (a1) at (-1,1);
	\vertex (a2) at (-1,-1);
	\vertex (a3) at (3,1);
	\vertex (a4) at (3,-1);
	\vertex (oo1) at (0,0);
	\vertex (oo2) at (2,0);
	\diagram* {
	(a1) -- (oo1);
	(a2) -- (oo1);
	(a3) -- (oo2);
	(a4) -- (oo2);
	(oo1) -- [half left] (oo2);
	(oo2) -- [half left] (oo1);
	};
	\end{feynman}

	\draw[->,thin] ($(a1)!0.25!(oo1)+(0,0.15)$) -- ($(a1)!0.75!(oo1)+(0,0.15)$);
	\draw[->,thin] ($(a2)!0.25!(oo1)-(0,0.15)$) -- ($(a2)!0.75!(oo1)-(0,0.15)$);
	\draw[->,thin] ($(a3)!0.25!(oo2)+(0,0.15)$) -- ($(a3)!0.75!(oo2)+(0,0.15)$);
	\draw[->,thin] ($(a4)!0.25!(oo2)-(0,0.15)$) -- ($(a4)!0.75!(oo2)-(0,0.15)$);
	
         \draw[->,thin]  ($ (oo1)!0.25!(oo2) + (0,0.6) $) to [out=15,in=165] ($ (oo1)!0.75!(oo2) + (0,0.6) $);
         \draw[->,thin]  ($ (oo2)!0.25!(oo1) - (0,0.6) $) to [out=180+15,in=-15] ($ (oo2)!0.75!(oo1) - (0,0.6) $);
	
	 \draw[fill=gray, thin] (oo1) circle (0.15); 
		
	\node[left] at (a1) {$(p_{1},i_{1})$};	
	\node[left] at (a2) {$(p_{2},i_{2})$};	
	\node[right] at (a3) {$(p_{3},i_{3})$};	
	\node[right] at (a4) {$(p_{4},i_{4})$};	
	\node at (1,1.2) {\small $k+p_{1}+p_{2}$};
	\node at (1,-1.2) {\small $k$};
\end{tikzpicture} \ 
\ee
where   the dashed circle corresponds to the counterterm canceling the pole in (\ref{6.3}) due to a fermionic 
 loop, \ie the evanescent
operator (\ref{3.14}) with $\cc\equiv \cc_{_{\rm F}}=4$  (i.e. the  contribution  of 4 to  total $\cc$ in (\ref{6.7}). 
This gives\foot{Here we specify to the string case and set $\wh D=D-2$.} 
\ba
&\Delta {\cal M}^{i_1i_2i_3i_4}=  \Big\{  \tfrac{i}{1536(d^{2}-1)\pi} s^3
 \Big[(d-2)[-4+d (D-4)
 ]\,( t^{2} + u^2)\la{524} \\ &+2[d^{2} (D-4) 
 \, -2(d+2) (D-2) ]tu\Big]\delta^{i_{1}i_{2}}\delta^{i_{3}i_{4}} 
-\tfrac{i(d-2)}{384(d-1)\pi}s^5\, \big(\delta^{i_{1}i_{4}}\delta^{i_{2}i_{3}}+\delta^{i_{1}i_{3}}\delta^{i_{2}i_{4}}\big)\Big\}G_{1,1}(s).\no 
\ea
There is also  a similar  contribution from the diagram like \rf{6.49} 
 with the  counterterm vertex on the right.

\subsubsection{Counterterm diagram  with  $X^2 \theta^2$ vertex }

When expanding around $d=2$, 
the above 2-loop diagrams with internal fermions (both of double-bubble and wine-glass type) have divergent 
sub-diagrams corresponding also to the 
1-loop  $XX\to \bar\theta\theta$ process. We thus 
need  to 
determine the associated counterterms   to be inserted into the 
 1-loop diagrams  of  third type in \rf{222} with the loop being the fermionic one. 
 This is needed to cancel various 1-loop sub-divergences in diagrams \rf{6.9}--\rf{6.11} and \rf{6.20}--\rf{6.22} associated with 
 sub-diagrams with two bosonic and two fermionic legs.
 
 
 Let us first  find the counterterm that is required to cancel the pole   in the 
1-loop $XX\to \bar\theta \theta$   amplitude. This  amplitude may receive several contributions. 
 One is from the bosonic loop diagram\footnote{Diagram (\ref{6.30}) 
cancels the bosonic bubble sub-divergence of (\ref{6.9}) and its left-right flipped version cancels the sub-divergence of  (\ref{6.10}).
}
\be
\begin{tikzpicture}[thick,baseline=0]
	\begin{feynman}
	\vertex (a1) at (-1,1);
	\vertex (a2) at (-1,-1);
	\vertex (a3) at (3,1);
	\vertex (a4) at (3,-1);
	\vertex (oo1) at (0,0);
	\vertex (oo2) at (2,0);
	\diagram* {
	(a1) -- (oo1);
	(a2) -- (oo1);
	(a3) -- [fermion] (oo2);
	(oo2) -- [fermion] (a4);
	(oo1) -- [half left] (oo2);
	(oo2) -- [half left] (oo1);
	};
	\end{feynman}

	\draw[->,thin] ($(a1)!0.25!(oo1)+(0,0.15)$) -- ($(a1)!0.75!(oo1)+(0,0.15)$);
	\draw[->,thin] ($(a2)!0.25!(oo1)-(0,0.15)$) -- ($(a2)!0.75!(oo1)-(0,0.15)$);
	\draw[->,thin] ($(a3)!0.25!(oo2)+(0,0.15)$) -- ($(a3)!0.75!(oo2)+(0,0.15)$);
	\draw[->,thin] ($(a4)!0.25!(oo2)-(0,0.15)$) -- ($(a4)!0.75!(oo2)-(0,0.15)$);
	
         \draw[->,thin]  ($ (oo1)!0.25!(oo2) + (0,0.6) $) to [out=15,in=165] ($ (oo1)!0.75!(oo2) + (0,0.6) $);
         \draw[->,thin]  ($ (oo2)!0.25!(oo1) - (0,0.6) $) to [out=180+15,in=-15] ($ (oo2)!0.75!(oo1) - (0,0.6) $);
	
	\node[left] at (a1) {$(p_{1},i_{1})$};	
	\node[left] at (a2) {$(p_{2},i_{2})$};	
	\node[right] at (a3) {$\bar\theta^{\beta}(p_{3})$};	
	\node[right] at (a4) {$\theta_{\alpha}(p_{4})$};	
	\node at (1,1.2) {\small $(k+p_{1}+p_{2},j_{1})$};
	\node at (1,-1.2) {\small $(k,j_{2})$};
	\node at (-2.5,0) {$(\DD_{\rm  BB})\indices{_{\alpha}^{\beta}}=\qquad \qquad $};
\end{tikzpicture} \la{6.30}
\ee
\ba
\no 
\qquad \qquad  = 
 \tfrac{1}{2i}\int \frac{d^{d}k}{(2\pi)^{d}}\frac{V^{i_{1}i_{2}j_{1}j_{2}}_{p_{1},p_{2},-k-p_{1}-p_{2},k}
\ (V^{j_{1}j_{2}})\indices{_{\alpha}^{\beta}}(p_{3}, p_{4}, k+p_{1}+p_{2},-k)}{k^{2}(k+p_{1}+p_{2})^{2}}\ .
\ea
For notational simplicity we  will suppress  the trivial $\delta^{i_1i_2}$ factor in the expressions for these  amplitudes in what follows. 

%
The two-fermion counterterm is a matrix  with spinor indices, which we may decompose in a basis of Dirac matrices. From the 
perspective of the 2d worldsheet, this basis contains $\{I, \Gamma^a\}$. On dimensional and locality grounds, or 
simply because the two-fermion vertices are linear in the 2d Dirac matrices, the identity matrix cannot appear in the 
decomposition of this matrix. Thus, the counterterm must be 
of the form $\slashed{v}\equiv v_a \Gamma^a$ 
 where $v_a$ is some 2d vector. To
 simplify the calculations we take a  trace  of its product with  an auxiliary matrix $p_5^a \Gamma_a$, and extract the vector $v_a$
from the resulting scalar expression.

Computing $\tr(D_{\rm BB}\, \slashed{p}_{5})$ we get (cf.  (\ref{3.9}))
\ba
& \tr(\DD_{\rm  BB}\, \slashed{p}_{5}) = \tfrac{i}{8\,(d^{2}-1)}\, s^2 \Big[
-4\wh D[t(p_{1}\cdot p_{5}-p_{2}\cdot p_{5})+s(p_{1}\cdot p_{5}+p_{3}\cdot p_{5})]+2s(\wh D-6)(d-2)(p_{1}\cdot p_{5} \lp
\qquad  \qquad +p_{2}\cdot p_{5}+2p_{3}\cdot p_{5})
+s(\wh D-4)(d-2)^{2}(p_{1}\cdot p_{5}+p_{2}\cdot p_{5}+2p_{3}\cdot p_{5})
\Big]G_{1,1}(s) 
\ .\la{426}
\ea
Another contribution is from  the  diagram with the fermionic loop\footnote{The diagram (\ref{427})
 and its left-right flipped version cancel the sub-divergences of (\ref{6.11}).
}
\be
\begin{tikzpicture}[thick,baseline=0]
	\begin{feynman}
	\vertex (a1) at (-1,1);
	\vertex (a2) at (-1,-1);
	\vertex (a3) at (3,1);
	\vertex (a4) at (3,-1);
	\vertex (oo1) at (0,0);
	\vertex (oo2) at (2,0);
	\diagram* {
	(a1) -- (oo1);
	(a2) -- (oo1);
	(a3) -- [fermion] (oo2);
	(oo2) -- [fermion] (a4);
	(oo1) -- [half left, fermion] (oo2);
	(oo2) -- [half left, fermion] (oo1);
	};
	\end{feynman}

	\draw[->,thin] ($(a1)!0.25!(oo1)+(0,0.15)$) -- ($(a1)!0.75!(oo1)+(0,0.15)$);
	\draw[->,thin] ($(a2)!0.25!(oo1)-(0,0.15)$) -- ($(a2)!0.75!(oo1)-(0,0.15)$);
	\draw[->,thin] ($(a3)!0.25!(oo2)+(0,0.15)$) -- ($(a3)!0.75!(oo2)+(0,0.15)$);
	\draw[->,thin] ($(a4)!0.25!(oo2)-(0,0.15)$) -- ($(a4)!0.75!(oo2)-(0,0.15)$);
	
         \draw[->,thin]  ($ (oo1)!0.25!(oo2) + (0,0.6) $) to [out=15,in=165] ($ (oo1)!0.75!(oo2) + (0,0.6) $);
         \draw[->,thin]  ($ (oo2)!0.25!(oo1) - (0,0.6) $) to [out=180+15,in=-15] ($ (oo2)!0.75!(oo1) - (0,0.6) $);
	
	\node[left] at (a1) {$(p_{1},i_{1})$};	
	\node[left] at (a2) {$(p_{2},i_{2})$};	
	\node[right] at (a3) {$\bar\theta^{\beta}(p_{3})$};	
	\node[right] at (a4) {$\theta_{\alpha}(p_{4})$};	
	\node at (1,1.2) {\small $(k+p_{1}+p_{2},\alpha')$};
	\node at (1,-1.2) {\small $(k,\beta')$};
	\node at (-2.5,0) {$ (\DD_{\rm  FF})\indices{_{\alpha}^{\beta}}=\qquad \qquad $};
\end{tikzpicture} \la{427}
\ee
\ba
&
= 
 \tfrac{1}{i}\int \frac{d^{d}k}{(2\pi)^{d}}\frac{[S(k+p_{1}+p_{2})V^{i_{1}i_{2}}(k,-k-p_{1}-p_{2},p_{1},p_{2})S(k)]\indices{_{\alpha'}^{\beta'}}
 V\indices{_{\alpha\beta'}^{\beta\alpha'}}(p_{3},k+p_{1}+p_{2})
 }{k^{2}(k+p_{1}+p_{2})^{2}}.\no 
\ea
Let us define 
\be
q=k+p_{1}+p_{2}, \qquad\qquad  \mc V^{i_{1}i_{2}}=S(k+p_{1}+p_{2})V^{i_{1}i_{2}}(k,-k-p_{1}-p_{2},p_{1},p_{2})S(k).
\ee
Considering  like in \rf{426}  the combination $\DD_{\rm  FF} \slashed{p}_{5}$ and using (\ref{5.25})  we have 
\ba
& (\mc V^{i_{1}i_{2}})\indices{_{\alpha'}^{\beta'}}
 V\indices{_{\alpha\beta'}^{\beta\alpha'}}(p_{3},q) (\slashed{p}_{5})\indices{_{\beta}^{\alpha}}
 \lp = \tfrac{i}{2}(\mc V^{i_{1}i_{2}})\indices{_{\alpha'}^{\beta'}}(\slashed{p}_{5})\indices{_{\beta}^{\alpha}}
 \big[
(\slashed{p}_{3})\indices{_{\beta'}^{\alpha'}}(\slashed{q})\indices{_{\alpha}^{\beta}}
+(\slashed{p}_{3})\indices{_{\beta'}^{\beta}}(\slashed{q})\indices{_{\alpha}^{\alpha'}}
-(\slashed{p}_{3})\indices{_{\alpha}^{\alpha'}}(\slashed{q})\indices{_{\beta'}^{\beta}}
-(\slashed{p}_{3})\indices{_{\alpha}^{\beta}}(\slashed{q})\indices{_{\beta'}^{\alpha'}}\big]\lp
= \tfrac{i}{2}\Big(\tr[\mc V^{i_{1}i_{2}}\slashed{p}_{3}]\, \tr[\slashed{p}_{5}\slashed{q}]+\tr[\mc V^{i_{1}i_{2}}\slashed{p}_{3}\slashed{p}_{5}\slashed{q}]
-\tr[\mc V^{i_{1}i_{2}}\slashed{q}\slashed{p}_{5}\slashed{p}_{3}]-\tr[\mc V^{i_{1}i_{2}}\slashed{q}] \tr[\slashed{p}_{3}\slashed{p}_{5}]\Big)\ .
\ea
 Thus after the loop integration 
 \ba \tr(\DD_{\rm  FF}\slashed{p}_{5})= &-\tfrac{i}{d^{2}-1} \, s^2\, \Big[
2t(p_{1}\cdot p_{5}-p_{2}\cdot p_{5})+2s(p_{1}\cdot p_{5}+p_{3}\cdot p_{5})\no \\
&\qquad \qquad  \ \ \ +(s+2t)(p_{1}\cdot p_{5}-p_{2}\cdot p_{5})(d-2)
\Big]G_{1,1}(s) 
\ .\la{430}
\ea
The third  possible  diagram  has   one  bosonic and one fermionic propagator in  the loop\footnote{The diagram (\ref{431}) 
cancels the sub-divergence of the ``Wine-glass'' diagrams (\ref{6.20}, \ref{6.21}).
}
\be
\begin{tikzpicture}[thick,baseline=0]
	\begin{feynman}
	\vertex (a1) at (-1,1);
	\vertex (a2) at (-1,-1);
	\vertex (a3) at (3,1);
	\vertex (a4) at (3,-1);
	\vertex (oo1) at (0,0);
	\vertex (oo2) at (2,0);
	\diagram* {
	(a1) -- [fermion] (oo1);
	(a2) -- (oo1);
	(oo2) -- [fermion] (a3);
	(oo2) --  (a4);
	(oo1) -- [half left, fermion] (oo2);
	(oo2) -- [half left] (oo1);
	};
	\end{feynman}

	\draw[->,thin] ($(a1)!0.25!(oo1)+(0,0.15)$) -- ($(a1)!0.75!(oo1)+(0,0.15)$);
	\draw[->,thin] ($(a2)!0.25!(oo1)-(0,0.15)$) -- ($(a2)!0.75!(oo1)-(0,0.15)$);
	\draw[->,thin] ($(a3)!0.25!(oo2)+(0,0.15)$) -- ($(a3)!0.75!(oo2)+(0,0.15)$);
	\draw[->,thin] ($(a4)!0.25!(oo2)-(0,0.15)$) -- ($(a4)!0.75!(oo2)-(0,0.15)$);
	
         \draw[->,thin]  ($ (oo1)!0.25!(oo2) + (0,0.6) $) to [out=15,in=165] ($ (oo1)!0.75!(oo2) + (0,0.6) $);
         \draw[->,thin]  ($ (oo2)!0.25!(oo1) - (0,0.6) $) to [out=180+15,in=-15] ($ (oo2)!0.75!(oo1) - (0,0.6) $);
	
	\node[left] at (a1) {$\bar\theta^{\beta}(p_{3})$};	
	\node[left] at (a2) {$(p_{1},i_{1})$};	
	\node[right] at (a3) {$\theta_{\alpha}(p_{4})$};	
	\node[right] at (a4) {$(p_{2},i_{2})$};	
	\node at (1,1.2) {\small $(k+p_{1}+p_{3},\alpha')$};
	\node at (1,-1.2) {\small $(k,j)$};
	\node at (-2.5,0) {$(\DD_{\rm  FB})\indices{_{\alpha}^{\beta}}=\qquad \qquad $};
\end{tikzpicture} \la{431}
\ee
\ba
& 
 = 
 \frac{1}{i}\int \frac{d^{d}k}{(2\pi)^{d}}\frac{
(V^{i_{1},j})\indices{_{\alpha'}^{\beta}}(p_{3},-k-p_{1}-p_{3},p_{1},k)S\indices{_{\beta'}^{\alpha'}}(k+p_{1}+p_{3})
(V^{i_{2},j})\indices{_{\alpha}^{\beta'}}(k+p_{1}+p_{3},p_{4},p_{2},-k)}{k^{2}(k+p_{1}+p_{3})^{2}}.\no 
\ea
In this case 
\ba
& \tr(\DD_{\rm  FB}\slashed{p}_{5}) = 
 \tfrac{1}{i}\int \frac{d^{d}k}{(2\pi)^{d}}\frac{1}{k^{2}(k+p_{1}+p_{3})^{2}}\lp\qquad \ \ \ 
\times \tr\Big[\slashed{p}_{5}\,V^{i_{2},j}(k+p_{1}+p_{3},p_{4},p_{2},-k)\,
S(k+p_{1}+p_{2})\,
V^{i_{1},j}(p_{3},-k-p_{1}-p_{3},p_{1},k)
\Big]\la{234}\\
& 
=  -\tfrac{i}{8(d-1)}\, t^2 \, \Big[
(2(d+2)s+(5d-2)t)p_{1}\cdot p_{5}-4dt p_{2}\cdot p_{5}
+(2(d+2)s+(d-2)t)p_{3}\cdot p_{5}
\Big]\, G_{1,1}(t)\, .
\no 
\ea
There is  also a similar  diagram obtained by swapping $p_{1}$ and $p_{2}$ or $t\rightarrow u$ (we will denote its   contribution  as  $\DD_{\rm  BF})$. 

Summing up the above expressions  and setting 
$d=2-2\eps$ we get for the  pole part  in the  total $XX\to \bar \theta \theta $ amplitude 
\ba
\la{6.41}
\tr[(\DD_{\rm  BB} +\DD_{\rm  FF}+ \DD_{\rm  FB}+ \DD_{\rm  BF})\slashed{p}_{5}] 
= -\tfrac{1}{12\pi\eps}(D-4)\, s\, \big(u\, p_{1}\cdot p_{5}+t\, p_{2}\cdot p_{5}-s\, p_{3}\cdot p_{5}\big)
+\mc O(\eps^{0}).
\ea
The $\DD_{\rm  BB}$, etc.,  matrices  contain 
  an implicit  factor $P_{-}$.   Using that $\tr P_{-}=16$ 
    we  then get  for the  pole part 
\be
\la{6.42}
\DD_{\rm  tot, {1\ov \eps} }
=\tfrac{1}{192\pi\eps}\, (D-4) \, s\, \big(-u \slashed{p}_{1}-t \slashed{p}_{2}+s \slashed{p}_{3}\big)\, .  
\ee
This   is a matrix with spinor indices (it  is also proportional to $\delta^{i_{1}i_{2}}$
which we suppress  here).

In general,  at the $L$-loop order,   in addition to the  bosonic   counterterms \rf{2.7} we  may 
have also  the ones  containing fermions, e.g., 
$T^{1-L} \sum_{n,k} \int d^d \sigma\     \partial^{dL}   (\partial X)^{2n} \, (\bar\theta\partial \theta)^k$.\foot{Here  we  use  that 
$\del X$ and $\bar\theta\partial \theta$  that appear  in \rf{5.1},\rf{3100}  are dimensionless.}
In particular,  for $L=1$ and $n=k=1$ we  may have a counterterm 
of the following symbolic structure in   $d=2$
\be \la{uuu}
{1\ov \eps} \int d^2 \sigma\,     \partial^{2}    \partial X^i \del X^i   \,  \bar\theta\partial \theta  \ . \ee
A   covariant counterterm  that  contains \rf{uuu} in its expansion 
 is (cf. (\ref{B.8}) and (\ref{B.9}))\be \la{437}
R_{ab}\, \bar\theta\Gamma^{a}\partial^{b}\theta = -(K^{i}K^{i})_{ab}\, \bar\theta\Gamma^{a}\partial^{b}\theta = -\partial_{a}\del^c X^{i}\partial_{b}\del_c X^{i}\, \bar\theta\Gamma^{a}\partial^{b}\theta + ...\ , 
\ee
where we used  the on-shell conditions $\slashed{\partial}\theta= 0, \ \del^2 X =0$.\foot{The 
  tree level  contribution corresponding   to \rf{uuu} is
$
 -p_{1}\cdot p_{2}(p_{2}\cdot p_{3}\ \slashed{p}_{{1}}
+p_{1}\cdot p_{3}\, \slashed{p}_{{2}}) 
= \tfrac{1}{4}s(-u \slashed{p}_{{1}}-t \slashed{p}_{{2}}),
$
which  has the same structure as  \rf{6.42} after using that   $\slashed{p}_{{3}}$ term does not  contribute on-shell. 
}


We are now ready  to compute  the  analog of  the 4-scalar  counterterm diagram \rf{6.49}  
  containing the above $X^2 \theta^2$ counterterm vertex that cancels the pole in (\ref{6.42})
  \be
\begin{tikzpicture}[thick,baseline=0,scale=0.75]
	\begin{feynman}
	\vertex (a1) at (-1,1);
	\vertex (a2) at (-1,-1);
	\vertex (a3) at (1,1);
	\vertex (a4) at (1,-1);
	\vertex (oo1) at (0,0);
	\diagram* {
	(a1) -- (oo1);
	(a2) -- (oo1);
	(oo1) -- [fermion] (a4);
	(a3) -- [fermion] (oo1);
	};
	\end{feynman}

	\draw[->,thin] ($(a1)!0.25!(oo1)+(0,0.15)$) -- ($(a1)!0.75!(oo1)+(0,0.15)$);
	\draw[->,thin] ($(a2)!0.25!(oo1)-(0,0.15)$) -- ($(a2)!0.75!(oo1)-(0,0.15)$);
	\draw[->,thin] ($(a3)!0.25!(oo1)+(0,0.15)$) -- ($(a3)!0.75!(oo1)+(0,0.15)$);
	\draw[->,thin] ($(a4)!0.25!(oo1)-(0,0.15)$) -- ($(a4)!0.75!(oo1)-(0,0.15)$);
	
	 \draw[fill=gray, thin] (oo1) circle (0.15); 
		
	\node[left] at (a1) {$(p_{1},i_{1})$};	
	\node[left] at (a2) {$(p_{2},i_{2})$};	
	\node[right] at (a3) {$p_{3}$};	
	\node[right] at (a4) {$p_{4}$};	
	\node at (6,0) {$\qquad \qquad =-
	\tfrac{i}{192\pi\eps} (D-4) \,p_{1}\cdot p_{2}\, (p_{2}\cdot p_{3}\,\slashed{p}_{1}+p_{1}\cdot p_{3}\,\slashed{p}_{2})\, \delta^{i_{1}i_{2}}$};
\end{tikzpicture}, \la{439}
\ee
and the fermionic loop 
\be
\la{6.46}
\begin{tikzpicture}[thick,baseline=0]
	\begin{feynman}
	\vertex (a1) at (-1,1);
	\vertex (a2) at (-1,-1);
	\vertex (a3) at (3,1);
	\vertex (a4) at (3,-1);
	\vertex (oo1) at (0,0);
	\vertex (oo2) at (2,0);
	\diagram* {
	(a1) -- (oo1);
	(a2) -- (oo1);
	(a3) -- (oo2);
	(a4) -- (oo2);
	(oo1) -- [half left, fermion] (oo2);
	(oo2) -- [half left, fermion] (oo1);
	};
	\end{feynman}

	\draw[->,thin] ($(a1)!0.25!(oo1)+(0,0.15)$) -- ($(a1)!0.75!(oo1)+(0,0.15)$);
	\draw[->,thin] ($(a2)!0.25!(oo1)-(0,0.15)$) -- ($(a2)!0.75!(oo1)-(0,0.15)$);
	\draw[->,thin] ($(a3)!0.25!(oo2)+(0,0.15)$) -- ($(a3)!0.75!(oo2)+(0,0.15)$);
	\draw[->,thin] ($(a4)!0.25!(oo2)-(0,0.15)$) -- ($(a4)!0.75!(oo2)-(0,0.15)$);
	
         \draw[->,thin]  ($ (oo1)!0.25!(oo2) + (0,0.6) $) to [out=15,in=165] ($ (oo1)!0.75!(oo2) + (0,0.6) $);
         \draw[->,thin]  ($ (oo2)!0.25!(oo1) - (0,0.6) $) to [out=180+15,in=-15] ($ (oo2)!0.75!(oo1) - (0,0.6) $);
	
	 \draw[fill=gray, thin] (oo1) circle (0.15); 
		
	\node[left] at (a1) {$(p_{1},i_{1})$};	
	\node[left] at (a2) {$(p_{2},i_{2})$};	
	\node[right] at (a3) {$(p_{3},i_{3})$};	
	\node[right] at (a4) {$(p_{4},i_{4})$};	
	\node at (1,1.2) {\small $k+p_{1}+p_{2}$};
	\node at (1,-1.2) {\small $k$};
\end{tikzpicture}\ . 
\ee
Evaluation of the diagram  \rf{6.46}  gives  (cf. \rf{524})
\be\la{646}
\Delta {\cal M}^{i_1i_2i_3i_4}= \tfrac{i}{3072\pi(d^{2}-1)}\, (D-4) \, s^{3}\, \Big[ (d-2) (t^{2}+u^{2})-2d\,  tu\Big]\, G_{1,1}(s)\, \delta^{i_{1}i_{2}}\delta^{i_{3}i_{4}}.
\ee
A similar diagram  with the  counterterm vertex on the right 
obtained by  interchanging  $1234\leftrightarrow 3412$
gives  the equivalent  contribution.

\subsubsection{Total 2-loop contribution} 

Let us now sum up  the contributions  of the above diagrams with fermions. Let us denote by DB  the  sum of  the double-bubble 
 expressions in (\ref{6.9}),(\ref{6.10}), (\ref{6.11}), by W -- the sum of wine-glass  diagram expressions 
(\ref{6.20}),(\ref{6.21}),(\ref{6.22})  and  by  CT and CT$' $ the contributions of the counterterm diagrams in (\ref{6.46}) and (\ref{6.49})
plus  their  reflected versions. We should also sum over appropriate crossed versions of these diagrams. 

Keeping first $s,t,u$ generic  it is enough to find the $A$-amplitude as $B$ and $C$ 
can be found from it  by crossing as in \rf{1.2}. 
We find, in particular,  that the   double-pole contributions are given by\foot{As already mentioned,  this is due to the absence  of 1-loop divergences in $d=2$ and thus  absence of sub-divergences in 2-loop diagrams.} 
\be
\te 
A^{(2)}_{{\rm DB}, \frac{1}{\eps^{2}}} = -\frac{D}{576\pi^{2}\eps^2}\, s^{2}tu, \ \ 
A^{(2)}_{{\rm W}, \frac{1}{\eps^{2}}}      = \frac{17}{1152\pi^{2}\eps^2}\, s^{2}tu, \ \
A^{(2)}_{{\rm CT}, \frac{1}{\eps^{2}}} = \frac{D-4}{2304\pi^{2}\eps^2}\, s^{2}tu, \ \
A^{(2)}_{{\rm CT'}, \frac{1}{\eps^{2}}}        = \frac{D}{576\pi^{2}\eps^2}\, s^{2}tu, \la{441}
\ee
so that  each of them vanishes  due to  the kinematic condition $stu=0$ in $d=2$.

For the $t=0$ choice of the 2d kinematics we get  for the  single pole and  finite contributions of these diagrams (we again include $4\pi e^{-\gamma_E}$ into $\mu^2$) 
\ba\te 
&\te A^{(2)}_{{\rm DB}, \frac{1}{\eps}} = -\frac{D}{1152\pi^{2}\eps}\, s^{4}, \qquad \te 
A^{(2)}_{{\rm W}, \frac{1}{\eps}}      = \frac{41}{4608\pi^{2}\eps}\, s^{4}, \qquad 
A^{(2)}_{{\rm CT}, \frac{1}{\eps}} = \frac{D-4}{4608\pi^{2}\eps}\, s^{4}, \qquad
A^{(2)}_{{\rm CT'}, \frac{1}{\eps}} = \frac{D-12}{1152\pi^{2}\eps}\, s^{4}, \no\\ 
&\te A^{(2)}_{\rm DB, fin} = \frac{D}{576\pi^{2}}\, s^{4}\, \log(-\frac{s}{\mu^2})\, 
- \frac{8D-6}{1728\pi^{2}}s^4 , \no  \\
 \no 
&\te A^{(2)}_{\rm W, fin} = -\frac{1}{96\pi^{2}} \, s^4\, \log(\frac{s}{\mu^2})+\frac{17}{2304\pi^{2}}\,s^4\, 
\log(-\frac{s}{\mu^2})+\frac{935}{27648\pi^{2}} s^4, \\
\no 
&\te A^{(2)}_{\rm CT, fin} = -\frac{D-4}{4608\pi^{2}}\, s^4\log(-\frac{s}{\mu^2})+ \frac{D-4}{1728\pi^{2}}\, , \\
\la{6.62}
&\te A^{(2)}_{\rm CT', fin} = \frac{1}{192\pi^{2}}\, s^4\, \log(\frac{s}{\mu^2})-\frac{D-6}{1152\pi^{2}}\, s^{4} \log(-\frac{s}{\mu^2})   +\frac{5D-72}{3456\pi^{2}} s^4\ .
\ea
%
The  sum of 
these expressions 
 gives  the following result  for the total fermionic contribution to the 2-loop  amplitude 
\ba
 A^{(2)}_{\rm F}  (s)
=&  \tfrac{1}{4608\pi^{2}\, \eps}\, (D-11)\, s^4  + 
 \tfrac{1}{1536\pi^{2}}\, (D-10)\, s^4 \, \log(-\tfrac{s}{\mu^2})     \, \no \\  & \qquad
 - \tfrac{1}{27648\pi^{2}}\, (72D-391)\, s^4 - i\tfrac{1}{192\pi} s^4 \ , \la{999}\\
 C^{(2)}_{\rm F} (s)=&  \tfrac{1}{4608\pi^{2}\, \eps}\, (D-11)\, s^4  + 
 \tfrac{1}{1536\pi^{2}}\, (D-10)\, s^4 \, \log(-\tfrac{s}{\mu^2}) \, ,\no \\
& \qquad - \tfrac{1}{27648\pi^{2}}\, (72D-391)\, s^4 +  i\big[ \tfrac{1}{192\pi}  +  \tfrac{1}{1536\pi}\, (D-10) \big] s^4 \ ,\la{4458}  \\
B^{(2)}_{\rm F} (s)  =& -\tfrac{1}{192\pi^{2}\eps}\,  s^4  - \tfrac{1}{1152\pi^{2}}\, s^4 \ , \la{4486}
\ea
where we used that $\log s = \log (-s)+i\pi$.
 Note  that while in the bosonic case  (\ref{420}),(\ref{3.21}) the  $\log s$ terms cancelled for any $D$, 
here   in the superstring case that happens only in the special  case of $D=10$.
  Also, the  condition \rf{14}, i.e. $A(s) =  C^*(-s)$  holds only for $D=10$.

The total 2-loop  amplitude in the  superstring case  is found   by summing together the expressions in (\ref{420}),(\ref{3.21})
and \rf{999}--\rf{4486} evaluated  for  $D=10$. The result was given in \rf{73}--\rf{74}, i.e.  
\ba
 A^{(2)} (s) = [C^{(2)}(-s)]^*= &
 - \tfrac{1}{1536\pi^{2}\, \eps}\, s^4  
 + \tfrac{145}{9216\pi^{2}}\, s^4  + i\tfrac{1}{64\pi} s^4 \ , \la{990}\\
  B^{(2)} (s)  =& -\tfrac{1}{128\pi^{2}\eps}\,  s^4   + \tfrac{25}{768\pi^{2}}\, s^4   - \tfrac{1}{192} s^4 \ .\la{648}
\ea 
As in the bosonic case \ci{Conkey:2016qju},  one can check that these  2-loop GS string  contributions  are consistent 
  with  2-particle unitarity.

Like in the bosonic case,  the poles  here that   are not accompanied by $\log s$   terms 
  come from the evanescent counterterm contributions, so they 
 are artefacts of the dimensional regularization  and should be subtracted. 
 They   can be  cancelled out  by  adding    the  counterterm (\cf (\ref{80}),(\ref{3.26}),\rf{327})
 \be\la{778}
\tfrac{1}{64\pi^{2}\, \eps}\, \int d^{2}\sigma\, \sqrt{-h}\,\big[\tr(K^{i}K^{j})^{2}-\tfrac{5}{12}\tr(K^{i}K^{i})^{2}\big].
\ee
In addition,  like in the $D\neq 26$ bosonic case, we need 
to add  the contribution  of the PPS
1-loop  counterterm \rf{13} with the coefficient  given by \rf{78}
that was required for  the integrability of the $D=10$ GS string  at the 1-loop level \ci{Seibold:2024oyr}. 
The corresponding 2-loop  contribution  containing 4-vertex from this counterterm    cancels the imaginary  terms in $A^{(2)} $ and
 $C ^{(2)} $  (see  Appendix \ref{ap1}). Then the remaining  real $s^4$ terms   can be cancelled by a real  counterterm  of the 
 same form as in  \rf{80},\rf{778}. As a result,    we end  up with the  amplitudes in  \rf{100},
  i.e. $A^{(2)} = C^{(2)} =0, \ B^{(2)} = - \tfrac{1}{192} s^4$, 
 as required for  the 2-loop integrability of the $D=10$ GS  string.


\section{2-loop  amplitude   on  supermembrane} 

Like in  the bosonic membrane  case,  in the M2 brane  case  there are no  1-loop divergences   \cite{Seibold:2024oyr}
so the only 2-loop contributions  come  from the first two genuine  2-loop diagrams  in \rf{222}. 
This makes the computation effectively more straightforward  than in the GS  string  case. 

\subsection{1-loop order }

To get the 1-loop amplitude  we need   to add to the bosonic loop contribution \rf{1.8}  the contribution of the fermionic loop 
with the  vertices \rf{4110} including the  $\eps^{abc}$ term in  (\ref{5.23}). 
After summing over crossings the  fermionic loop result is  found  to be 
\ba
\la{7.1}
A^{(1)}_{\rm F}(s,t,u) =& \tfrac{1}{1024}\,\big[(-s)^{3/2}\, (s^{2}-8tu)+4 (-t)^{3/2}\, t(t+2s)+4(-u)^{3/2}\, u(u+2s)\big], \\
B^{(1)}_{\rm F}(s,t,u) =& A^{(1)}_{\rm F}(t,s,u), \qquad \quad 
C^{(1)}_{\rm F}(s,t,u) = A^{(1)}_{\rm F}(u,t,s),
\ea
Adding this to  \rf{1.8}   we get for the total $A$  amplitude   the following   expression \cite{Seibold:2024oyr} 
\ba
A^{(1)} (s,t,u) = -\tfrac{1}{64}tu\, \big[(-s)^{3/2}+(-t)^{3/2}+(-u)^{3/2}\big]\  .
\ea
Thus   the  total 1-loop amplitude has a simple form    proportional to the tree level one\foot{As
  in any odd number of dimensions
the 1-loop amplitude  is finite. 
A similar   computation of 4-particle scattering in the case of the supersymmetric D3-brane action ($d=4$) 
gives  1-loop   divergence $\sim  {1\ov \eps} (s^2 +t^2 + u^2) $ times tree-level amplitude   \ci{Shmakova:1999ai}.}
\ba\la{54}
&\mc M^{(1)}(s,t,u) = \tfrac{1}{32} \big[(-s)^{3/2}+(-t)^{3/2}+(-u)^{3/2}\big] \, \mc M^{(0)}(s,t,u) \ ,\\
 &\la{55}
\mc M^{(0)}{}^{i_1i_2i_3i_4}(s,t,u) 
=-\tfrac{1}{2}\big( tu\, \delta^{i_{1}i_{2}}\delta^{i_{3}i_{4}}+
su\, \delta^{i_{1}i_{3}}\delta^{i_{2}i_{4}}+
st\, \delta^{i_{1}i_{4}}\delta^{i_{2}i_{3}}\big) \ . 
\ea

\subsection{2-loop order}

The fermionic double-bubble diagrams \rf{6.9}, \rf{6.10}, \rf{6.11}  give (after sum over crossing)
a finite contribution which  for  $D=11$ is given by 
\be
\la{7.6}
A^{(2)}_{\rm DB_{F}} =-\tfrac{3 }{65536}\,  s \big(5 s^4+20 s^3 t+16 s^2 t^2-8 s t^3-4 t^4\big)\ . 
\ee
Each of the three  wine-glass  diagrams \rf{6.20},\rf{6.21},\rf{6.22}  should be summed over 6 crossed 
configurations.
 As a result, we get an expression containing a UV pole  part  (accompanied  by  the corresponding $\log s $, etc., terms).
In dimension $D=11$ it  reads 
\ba
\la{7.8}
A^{(2)}_{\rm W_{F}, \frac{1}{\eps}} =\tfrac{1}{80640\pi^{2}\, \eps }\, s\big(8s^{4}+14 s^{3}t-113 s^{2}t^{2}-254  st^{3}-127 t^{4}\big)\ . 
\ea
Adding this  to  the bosonic membrane  pole  contribution in  (\ref{4.4})   we conclude   that the total 
     UV pole part  of the  $D=11$ 
M2 brane  2-loop amplitude  is given by 
\ba
\la{7.9}
A^{(2)}_{\frac{1}{\eps}} =  - \tfrac{1}{256\pi^{2}\,\eps}\, s\,t^{2}u^{2} \ . 
\ea
Thus despite  its   maximal   target space supersymmetry the S-matrix of M2 brane  theory is not UV finite
starting from  2-loop order. 

The total expression for the finite part   of the M2 brane amplitude (including  the wine-glass, double-bubble  
(\ref{7.6})  and also  the bosonic  finite part in  (\ref{230}) evaluated at $D=11$) is given by 
\ba
A^{(2)}_{\rm fin} = &
 \tfrac{1}{384} tu\Big( \tfrac{1}{\pi^2} \big[s^{3}\log(-\tfrac{s}{\mu^2})+t^{3} \log(-\tfrac{t}{\mu^2}) + u^{3} \log(-\tfrac{u}{\mu^2}) \big] + \tfrac{9}{16} 
 stu \Big)
 \lp
+\tfrac{1}{80640\pi^{2}}\, s\big(6s^{4}+124 s^{3}t-1833 s^{2}t^{2}-3914 st^{3}-1957 t^{4}\big) 
\la{710}\ , 
\ea
where we absorbed $\pi$ and $e^{-\gamma_E}$ factors into $\mu^2$.\foot{Note that 
since 
$s^{3}+t^{3}+u^{3} =3 stu$
 the total  coefficient of the $\log \mu^2$ term  differs from the  coefficient of the pole term in \rf{7.9}  by factor of 2, as 
 expected in the 2-loop contribution (see footnote \ref{M}).}

Let us note that  the  coefficients  in the   finite   ${1\ov \pi^2}$  term in the second line  of \rf{710} 
are potentially scheme-dependent. 
They are sensitive to how one treats  contractions of the  four  $\eps^{abc}$ 
symbols in 2-loop diagrams in dimensional regularization. 
We may   first group them in pairs    and then use  the 
relation (\ref{5.17}) for each pair. Particular  choices of  ordering in pairs may   differ  by  order $\eps=-\ha(d-3)$ terms
and this may lead  to different finite parts  when multiplied by $1\ov \eps$ pole. 
The expression in \rf{710} was found  by  using the 
symmetric average over the three pairing  orders.\foot{In more detail, in the wine-glass diagrams \rf{6.21},\rf{6.22} 
there is just one    four $\eps^{abc}$   contraction.  Introducing the notation
 $\eps(p,q,r) \equiv \eps^{abc}p_{a}q_{b}r_{c}$, we find  such quartic  contraction where 
 one factor comes from the Dirac trace (\ref{5.18}) (we will indicate  this  it by underlining it)  and the other three  from the last term in (\ref{5.23}). This gives 
\ba
 \underline{\eps}(k_{1} - k_{2}, -k_{2} - p_{3}, -k_{2} + p_{1})\ \eps(k_{1} - k_{2}, p_{1}, -k_{1})\ \eps(-k_{2} + p_{1}, p_{2}, -p_{1} - p_{2} - p_{3})\ \eps(-k_{2} - p_{3}, p_{3}, k_{1})
\equiv\underline{\eps}\ \eps'\ \eps''\ \eps''' . \no 
\ea
We may group  these factors in pairs 
with  arbitrary coefficients $\a,\beta$  as 
 $ \a\, [\underline{\eps}\eps'][\eps''\eps''']+\beta\,[\underline{\eps}\eps''][\eps'\eps''']
+\g\,[ \underline{\eps}\eps'''][\eps'\eps'']$  where $\g=1- \a-\beta. $
We  find that   the  dependence on $\a$  in $A^{(2)}$ drops out,  so that there is just one 
 scheme-dependence parameter $\beta$. Then 
 the analog of  
  the   second  line in  (\ref{710}) is found to be 
\ba\no
\tfrac{1}{40320\pi^{2}} (7-12\beta)\, s^{5}
+\tfrac{1}{80640\pi^{2}} stu \big[
4s^{2}(-37+18\beta)-tu(1965-24\beta)
\big].
\ea
 The ``symmetric'' choice  of $\beta={1\ov 3} $  leads to the expression in the second line of  (\ref{710}). 
}

\section*{Acknowledgements}
We thank S. Dubovsky for sharing some
details  about the results  in \cite{Conkey:2016qju}.  
Part of this work was done while AAT was attending the  workshop ``Higher-d Integrability''
in Favignana, Italy (02-13.06.25)  and he thanks the organizers for the hospitality. 
 MB is supported by the INFN grant GAST. 
RR is  supported by the U.S. Department
of Energy (DOE) under award number DE-SC00019066.
The work of AAT   is supported by the STFC grant ST/T000791/1.

\newpage

\appendix

\section{Expansion of  M2 brane and GS   string actions to  fourth order} 
\la{app:M2-expansion}

Here we will present details of the derivation of the expansions  in \rf{5.10}--\rf{511}  and \rf{5210},\rf{5211}. 
The starting point is  the  covariant  M2 brane action in \rf{5.1},\rf{5.2}. 
In the  static gauge we fix $X^a=\sigma^a, \ a=0,1,2$ and   expand in powers of  $D-3$ fields  $X^{i}$. 
The expansion  of $\sqrt { - h}$ was  given in \rf{2.3}   and we are  interested in  similar terms  involving $\theta$. 
We have (see \rf{333}) 
\ba
\la{E.3}
L_1= -\sqrt{-g} 
= -\sqrt{-h}\Big[1 &-i \partial_{a}X_{\mu}\bar\theta\Gamma^{\mu}\partial^{a}\theta
-\tfrac{1}{2}\bar\theta\Gamma_{\mu}\partial_{a}\theta\, \bar\theta\Gamma^{\mu}\partial^{a}\theta
-{\tfrac{1}{2}}(\partial_{a}X_{\mu}\bar\theta\Gamma^{\mu}\partial^{a}\theta)^{2}\lp 
+{\tfrac{1}{2}}\partial_{a}X_{\mu}\bar\theta\Gamma^{\mu}\partial_{b}\theta\,
\partial^{b}X_{\nu}\bar\theta\Gamma^{\nu}\partial^{a}\theta
+{\tfrac{1}{2}\partial_{a}X_{\mu}\bar\theta\Gamma^{\mu}\partial_{b}\theta\,
\partial^{a}X_{\nu}\bar\theta\Gamma^{\nu}\partial^{b}\theta}
+\dots\Big], 
\ea
where $\mu=(a,i)$ and $\partial_{a}X^{b}=\delta^{b}_{a}$. 
Hence  ($\slashed{\partial} \equiv \Gamma^a \del_a$) 
\ba
\la{E.6}
L_{1} =& -\sqrt{-h}+i \bar\theta\slashed{\partial}\theta +i\,\partial_{a}X_{i}\bar\theta\Gamma^{i}\partial^{a}\theta 
+i (\tfrac{1}{2}\eta^{ab}\partial^{c}X^{i}\partial_{c}X^{i}-\partial^{a}X^{i}\partial^{b}X^{i})
\bar\theta\Gamma_{a}\partial_{b}\theta \\ &
+\tfrac{1}{2}\bar\theta\Gamma_{b}\partial_{a}\theta\, \bar\theta\Gamma^{b}\partial^{a}\theta
+ \tfrac{1}{2}\bar\theta\Gamma_{i}\partial_{a}\theta\, \bar\theta\Gamma^{i}\partial^{a}\theta
+{\tfrac{1}{2}}(\bar\theta\slashed{\partial}\theta)^{2}
-{\tfrac{1}{2}}\bar\theta\Gamma_{a}\partial_{b}\theta\, 
\bar\theta\Gamma^{b}\partial^{a}\theta
{-\tfrac{1}{2}\bar\theta\Gamma_{a}\partial_{b}\theta\, 
\bar\theta\Gamma^{a}\partial^{b}\theta}
+\dots. \no 
\ea
Similarly, expanding the integrand  $L_2= L_{2,1} + L_{2,2}$ of the WZ term   in \rf{5.2} we get (using \rf{3.16})
\ba
L_{2,1} =& -\tfrac{i}{2}\eps^{abc}\, \bar\theta\Gamma_{\mu\nu}\, \partial_{a}\theta\, \Pi_{b}^{\mu}\Pi_{c}^{\nu}
 =  -\tfrac{i}{2}\eps^{abc}\, \bar\theta\Gamma_{bc}\, \partial_{a}\theta 
-i\eps^{abc}\, \bar\theta\Gamma_{bi}\, \partial_{a}\theta\, \partial_{c}X^{i}
-\eps^{abc}\, \bar\theta\Gamma_{bi}\, \partial_{a}\theta\, \bar\theta\Gamma^{i}\partial_{c}\theta\lp
 -\tfrac{1}{2}\eps^{abc}(\bar\theta\Gamma_{bd}\partial_{a}\theta\, \bar\theta\Gamma^{d}\partial_{c}\theta
 -\bar\theta\Gamma_{cd}\partial_{a}\theta\, \bar\theta\Gamma^{d}\partial_{b}\theta )
 -\tfrac{i}{2}\eps^{abc}\, \bar\theta\Gamma_{ij}\, \partial_{a}\theta\, 
\partial_{b}X^{i}\,\partial_{c}X^{j}+\dots
\no\\
=&  -i\bar\theta\Gamma^{\star}\slashed{\partial}\theta -i \bar\theta\Gamma^{\star}\Gamma_{i}\Gamma^{ac}\, \partial_{a}\theta\, \partial_{c}X^{i}
 -\tfrac{i}{2}\eps^{abc}\, \bar\theta\Gamma_{ij}\, \partial_{a}\theta\, 
\partial_{b}X^{i}\,\partial_{c}X^{j}
\lp
 -\eps^{abc}\, \bar\theta\Gamma_{bi}\, \partial_{a}\theta\, \bar\theta\Gamma^{i}\partial_{c}\theta
 -\tfrac{1}{2}\eps^{abc}(\bar\theta\Gamma_{bd}\partial_{a}\theta\, \bar\theta\Gamma^{d}\partial_{c}\theta
 -\bar\theta\Gamma_{cd}\partial_{a}\theta\, \bar\theta\Gamma^{d}\partial_{b}\theta )
+\dots, 
\\
L_{2,b} =& \tfrac{1}{2}\eps^{abc}\, \bar\theta\Gamma_{\mu\nu}\, \partial_{a}\theta\, 
(\partial_{b}X^{\mu}-i\bar\theta\Gamma^{\mu}\partial_{b}\theta)\, \bar\theta\Gamma^{\nu}\partial_{c}\theta 
=  \tfrac{1}{2}\eps^{abc}\, \bar\theta\Gamma_{bd}\, \partial_{a}\theta\, 
\, \bar\theta\Gamma^{d}\partial_{c}\theta+\tfrac{1}{2}\eps^{abc}\, \bar\theta\Gamma_{bi}\, \partial_{a}\theta\, 
\, \bar\theta\Gamma^{i}\partial_{c}\theta  + \dots\no 
\ea
Thus  (see \rf{3.16})  
\ba
L =&L_1+ L_2 =  L_{B}+L_{\theta^2 }+L_{X\theta^2}+L_{X^2\theta^2}+L_{\theta^{4}}+\dots, 
\\
L_{B} =& -\sqrt{-h}, \qquad L_{\theta^2} = i \bar\theta(1-\Gamma^{\star})\slashed{\partial}\theta, \qquad 
L_{X\theta^2} = i\,\partial_{a}X_{i}\bar\theta\Gamma^{i}\partial^{a}\theta +i \partial_{a}X^{i}\bar\theta\Gamma^{\star}\Gamma_{i}\Gamma^{ab}\, \partial_{b}\theta,\la{a55} \\
L_{X^2\theta^2} =& \tfrac{i}{2}\partial_{a}X^{i}\partial^{a}X^{i}\, \bar\theta\slashed{\partial}\theta
-i\partial_{a}X^{i}\partial_{b}X^{i}\bar\theta\Gamma^{a}\partial^{b}\theta
-\tfrac{i}{2}\partial_{a}X^{i}\,\partial_{b}X^{j} \eps^{abc}\, \bar\theta\Gamma_{ij}\, \partial_{c}\theta, \la{a56}\\
 L_{\theta^{4}} =& -\eps^{abc}\, \bar\theta\Gamma_{bi}\, \partial_{a}\theta\, \bar\theta\Gamma^{i}\partial_{c}\theta
 -\tfrac{1}{2}\eps^{abc}(\bar\theta\Gamma_{bd}\partial_{a}\theta\, \bar\theta\Gamma^{d}\partial_{c}\theta
 -\bar\theta\Gamma_{cd}\partial_{a}\theta\, \bar\theta\Gamma^{d}\partial_{b}\theta ) 
  + \tfrac{1}{2}\eps^{abc}\, \bar\theta\Gamma_{b\nu}\, \partial_{a}\theta\, 
\, \bar\theta\Gamma^{\nu}\partial_{c}\theta\lp
+\tfrac{1}{2}\bar\theta\Gamma_{i}\partial_{a}\theta\, \bar\theta\Gamma^{i}\partial^{a}\theta
+{\tfrac{1}{2}}(\bar\theta\slashed{\partial}\theta)^{2}
{-\tfrac{1}{2}}\bar\theta\Gamma_{a}\partial_{b}\theta\, 
\bar\theta\Gamma^{b}\partial^{a}\theta 
\bar\theta\Gamma^{a}\partial^{b}\theta
\ .
\ea
Here \rf{a55},\rf{a56} are the same as in 
  \cite{Seibold:2024oyr}   where $\theta^4$ terms were not included as they do not contribute to the  bosonic S-matrix  in 1-loop approximation.  

We can rearrange $L_{\theta^{4}}$ as follows  
\ba
 L_{\theta^{4}} =& -\tfrac{1}{2}\eps^{abc}\, \bar\theta\Gamma_{bi}\, \partial_{a}\theta\, \bar\theta\Gamma^{i}\partial_{c}\theta
 -\tfrac{1}{2}\eps^{abc}\bar\theta\Gamma_{bd}\partial_{a}\theta\, \bar\theta\Gamma^{d}\partial_{c}\theta
+\tfrac{1}{2}\bar\theta\Gamma_{i}\partial_{a}\theta\, \bar\theta\Gamma^{i}\partial^{a}\theta
{-\tfrac{1}{2}}\bar\theta\Gamma_{a}\partial_{b}\theta\, \no 
\bar\theta\Gamma^{b}\partial^{a}\theta+{\tfrac{1}{2}}(\bar\theta\slashed{\partial}\theta)^{2}
\\
 =& \tfrac{1}{2} \bar\theta\Gamma^{\star}\Gamma_{i}\Gamma^{ac}\, \partial_{a}\theta\, \bar\theta\Gamma^{i}\partial_{c}\theta
 -\tfrac{1}{2}\eps^{abc}\bar\theta\Gamma_{bd}\partial_{a}\theta\, \bar\theta\Gamma^{d}\partial_{c}\theta
+\tfrac{1}{2}\bar\theta\Gamma_{i}\partial_{a}\theta\, \bar\theta\Gamma^{i}\partial^{a}\theta
{-\tfrac{1}{2}}\bar\theta\Gamma_{a}\partial_{b}\theta\, 
\bar\theta\Gamma^{b}\partial^{a}\theta+{\tfrac{1}{2}}(\bar\theta\slashed{\partial}\theta)^{2}.
\ea
Using  that $\eps^{abc}\Gamma_{cd}=  \Gamma^{\star}(\delta^{a}_{d}\Gamma^{b}-\delta^{b}_{d}\Gamma^{a})$   we get 
\ba
\la{E.23}
 L_{\theta^{4}} =& 
 \tfrac{1}{2} \bar\theta\Gamma^{\star}\Gamma_{i}\Gamma^{ac}\, \partial_{a}\theta\, \bar\theta\Gamma^{i}\partial_{c}\theta
+\tfrac{1}{2}\bar\theta\Gamma_{i}\partial_{a}\theta\, \bar\theta\Gamma^{i}\partial^{a}\theta 
{-\tfrac{1}{2}}\bar\theta(1-\Gamma^{\star})\Gamma_{a}\partial_{b}\theta\, 
\bar\theta\Gamma^{b}\partial^{a}\theta+{\tfrac{1}{2}}\bar\theta(1-\Gamma^{\star})\slashed{\partial}\theta\, \bar\theta\slashed{\partial}\theta\, .
\ea
Fixing the  $\kappa$-symmetry gauge  as in \rf{3.34} 
we conclude that  $L_{X\theta^2}$ in \rf{a55} vanishes. 
Using that $[\Gamma^{\star},\Gamma^{a}]=0$ and $\{\Gamma^{\star}, \Gamma^{i}\}=0$, the terms in the  first  line  in (\ref{E.23})
vanish because of $\bar\theta=\bar\theta P_{-}$,  
 $P_{-}\Gamma_{i}=\Gamma_{i}P_{+}$ and $P_{+}\theta=0$. 
 We thus   end up with the expressions
 in  \rf{5.10},\rf{510},(\ref{511})  (after the rescaling  $\theta\to \frac{1}{\sqrt 2}\theta$).

In the GS string  case \rf{3100},\rf{001} 
the volume part $L_{1}$ is the  same, i.e.  we get again the expansion in  (\ref{E.6}).
For the WZ part we  have (after using that in the static gauge $\del_a X^b = \delta^a_b$)  
\ba
L_{2} =&
 -i\, \eps^{ab}\, \bar\theta\Gamma_{\mu}\Gamma_{2}\, \partial_{a}\theta\, (\partial_{b}X^{\mu}-\tfrac{i}{2}\bar\theta\Gamma^{\mu}\partial_{b}\theta)\\
= 
&-i\, \eps^{ab}\, \bar\theta\Gamma_{b}\Gamma_{2}\, \partial_{a}\theta
-i\, \eps^{ab}\, \bar\theta\Gamma^{i}\Gamma_{2}\, \partial_{a}\theta\, \partial_{b}X^{i}
-\tfrac{1}{2} \eps^{ab}\, \bar\theta\Gamma_{c}\Gamma_{2}\, \partial_{a}\theta\, \bar\theta\Gamma^{c}\partial_{b}\theta
-\tfrac{1}{2} \eps^{ab}\, \bar\theta\Gamma^{i}\Gamma_{2}\, \partial_{a}\theta\, \bar\theta\Gamma^{i}\partial_{b}\theta\no .
\ea
Fixing the $\kappa$-symmetry gauge as in \rf{444} 
 we have again $\bar\theta=\bar\theta P_{-}$, $P_{-}\theta=0$;   since $
[P_{-},\Gamma_{a}]= [P_{-},\Gamma_{2}] = 0, \  P_{\pm}\Gamma^{i}=\Gamma^{i}P_{\mp}$  we 
get 
\ba
L_{2} =& 
-i\, \eps^{ab}\, \bar\theta\Gamma_{b}\Gamma_{2}\, \partial_{a}\theta
-\tfrac{1}{2} \eps^{ab}\, \bar\theta\Gamma_{c}\Gamma_{2}\, \partial_{a}\theta\, \bar\theta\Gamma^{c}\partial_{b}\theta
= i\,  \bar\theta\slashed{\partial}\theta
-\tfrac{1}{2} \bar\theta\, \Gamma^{b}\partial_{c}\theta\, \bar\theta\Gamma^{c}\partial_{b}\theta
+\tfrac{1}{2} (\bar\theta\, \slashed{\partial}\theta)^{2}, \la{a11}
\ea
where we used  that 
$
\eps^{ab}\Gamma_{b}\Gamma_{2} = -\Gamma^{a}\Gamma^{\star}, \  \eps^{ab}\Gamma_{c}\Gamma_{2} = \Gamma^{\star}(\delta^{a}_{c}\Gamma^{b}-\delta^{b}_{c}\Gamma^{a})$. 
Combining this with (\ref{E.6}) (where  $\bar\theta\Gamma^{i}\partial_{a}\theta\, \bar\theta\Gamma^{i}\partial^{a}\theta$ 
 vanishes due to  the $P_{-}$ projection) we find 
\ba
L =&
 2i \bar\theta\slashed{\partial}\theta 
+i (\tfrac{1}{2}\eta^{ab}\partial_{c}X^{i}\partial_{c}X^{i}-\partial^{a}X^{i}\partial^{b}X^{i})
\bar\theta\Gamma_{a}\partial_{b}\theta \lp
+\tfrac{1}{2}\bar\theta\Gamma_{b}\partial_{a}\theta\, \bar\theta\Gamma^{b}\partial^{a}\theta
-{\tfrac{1}{2}}\bar\theta\Gamma_{a}\partial_{b}\theta\, 
\bar\theta\Gamma^{b}\partial^{a}\theta
{-\tfrac{1}{2}\bar\theta\Gamma_{a}\partial_{b}\theta\, 
\bar\theta\Gamma^{a}\partial^{b}\theta}
-\tfrac{1}{2} \bar\theta\, \Gamma^{b}\partial_{c}\theta\, \bar\theta\Gamma^{c}\partial_{b}\theta
+ (\bar\theta\, \slashed{\partial}\theta)^{2}\no \\ 
= &2i \bar\theta\slashed{\partial}\theta 
+i (\tfrac{1}{2}\eta^{ab}\partial_{c}X^{i}\partial_{c}X^{i}-\partial^{a}X^{i}\partial^{b}X^{i})
\bar\theta\Gamma_{a}\partial_{b}\theta 
-\bar\theta\Gamma_{a}\partial_{b}\theta\, 
\bar\theta\Gamma^{b}\partial^{a}\theta
+ (\bar\theta\, \slashed{\partial}\theta)^{2}\ . 
\ea
This  leads to  \rf{5210},\rf{5211} after  the rescaling $\theta\to \frac{1}{\sqrt 2}\theta$.

\section{Some  useful relations} 
\la{app7}

In $D=11$, the Majorana fermions have $32$ real components. 
After $\kappa$-symmetry gauge  fixing we are left  with  $n_{F}=16$ real components.
The number of physical  3d fermionic degrees of freedom is further halved by the Dirac equation $\slashed{\partial} \theta=0$, i.e 
it is given by  $\frac{1}{2}n_{F}=8$. This is same as the number $\wh D = D-3=8$ of scalar $X^i$   bosonic degrees of freedom. 
 We have for the spinor traces 
\be
\tr I = 2n_{F},\qquad  \qquad \tr P_{-} = n_{F}\ .
\ee
Considering a  loop of Majorana fermions, we have an extra $-1/2$ factor    so that effectively we get 
$
\tr P_{-}\to -\tfrac{1}{2}n_{F}=-8.
$
Some other basic traces are 
\ba
\la{5.13}
& \tr(P_{-}\Gamma_{a}\Gamma_{b}) = \eta_{ab}\,\ \tr P_{-},\qquad \qquad 
 \tr(P_{-}\Gamma_{a}\Gamma_{b}\Gamma_{c}\Gamma_{d}) = (\eta_{ab}\eta_{cd}-\eta_{ac}\eta_{bd}+\eta_{ad}\eta_{bc})\ \tr P_{-},\\
& \tr(P_{-}\Gamma_{a_{1}}\dots \Gamma_{a_{n}}\Gamma^{ij}) = 0, \qquad \qquad 
 \tr(P_{-}\Gamma_{a}\Gamma_{b}\Gamma^{ij}\Gamma^{k\ell}) = \eta_{ab}(\delta^{i\ell}\delta^{jk}-\delta^{ik}\delta^{j\ell})\ \tr P_{-}.
\ea
 Using the definition of  $P_{-}$  in \rf{3.34} 
 we have 
\be
\la{5.18}
\tr(P_{-}\Gamma_{a}\Gamma_{b}\Gamma_{c}) =\eps_{abc} \,  \tr P_{-} \,   \ .
\ee
Similarly, 
$
\tr(P_{-}\Gamma_{a}\Gamma_{b}\Gamma_{c}\Gamma^{i_{1}i_{2}}\Gamma^{i_{3}i_{4}}\Gamma^{i_{5}i_{6}})
$
 factorizes into $\tr P_{-}\ \eps_{abc}$ times a combination of $\delta$-symbols. 

We use also that 
\be
\la{5.17}
\epsilon^{abc} \epsilon^{a'b'c'} = -\begin{vmatrix}
\eta^{aa'} & \eta^{ab'} & \eta^{ac'} \\
\eta^{ba'} & \eta^{bb'} & \eta^{bc'} \\
\eta^{ca'} & \eta^{cb'} & \eta^{cc'}
\end{vmatrix}\, , 
\ee
which  implies, in particular, that   $(\eps^{abc}p_{1,a}p_{2,b}p_{3,c})^{2}=\frac{1}{4}stu$.

Some basic  momentum  integrals are 
\ba
\la{A.1}
&\frac{1}{P_{1}^{\alpha_{1}}\dots P_{n}^{\alpha_{n}}} =\frac{\Gamma(\alpha_{1}+\dots+\alpha_{n})}{\Gamma(\alpha_{1})\dots\Gamma(\alpha_{n})} 
\int_{0}^{1}dx_{1}\dots dx_{n} \delta\big(1-\sum_{i=1}^{n} x_{i}\big)\, \frac{x_{1}^{\alpha_{1}-1}\dots x_{n}^{\alpha_{n}-1}}{(\sum_{i=1}^{n}x_{i}P_{i})^{\alpha_{1}+\dots+\alpha_{n}}}, \\
\la{A.2}
&\int\frac{d^{d}p}{(2\pi)^{d}}\frac{1}{(p^{2}+M^{2})^{n}} = \frac{i\, \Gamma(n-\frac{d}{2})}{(4\pi)^{d/2}\Gamma(n)}(M^{2})^{\frac{d}{2}-n}, \\
&\int\frac{d^{d}p}{(2\pi)^{d}}\frac{p^{a}p^{b}}{(p^{2}+M^{2})^{n}} = \frac{i\, \Gamma(n-1-\frac{d}{2})}{(4\pi)^{d/2}\Gamma(n)}(M^{2})^{\frac{d}{2}+1-n}\frac{1}{2}\eta^{ab}, \\
&\int\frac{d^{d}p}{(2\pi)^{d}}\frac{p^{a}p^{b}p^{c}p^{d}}{(p^{2}+M^{2})^{n}} = 
\frac{i\, \Gamma(n-2-\frac{d}{2})}{(4\pi)^{d/2}\Gamma(n)}(M^{2})^{\frac{d}{2}+2-n}\frac{1}{4}(\eta^{ab}\eta^{cd}+\eta^{ac}\eta^{bd}
+\eta^{ad}\eta^{bc}).
\ea

\section{Geometrical  relations for induced metric } 

\subsection{Extrinsic curvature  identities}
\la{app:ext}

The  induced metric on a brane  in flat target space  is  defined as $h_{ab}=\eta_{\mu\nu}\partial_{a}X^{\mu}\partial_{b}X^{\nu}$.
 The tangent space is spanned by $\{\partial_{a}X^{\mu}\}$.
 The normal bundle   has  the basis 
 $\{ (n^{i})^\mu \}$ 
  such that 
\be
n^{i}_{\mu}(n^{j})^{\mu}=\delta^{ij}, \qquad n^{i}_{\mu}\partial_{a}X^{\mu}=0, \qquad i,j=1, \dots, D-d\ . 
\ee
  Let  $\nabla_a$  be  the 
 covariant derivative with respect to the  induced metric  so that  $\nabla_a h_{bc}=0$.
 Then  $\nabla_{a}\partial_{b}X^{\mu}$ is in  the normal  bundle so that 
we can introduce the extrinsic curvature $K$ as  its  coefficients 
\be
\la{B.2}
\nabla_{a}\partial_{b}X^{\mu} = K^{i}_{ab}(n^{i})^{\mu},\qquad  \qquad K_{ab}^{i}=K_{ba}^{i}\ . 
\ee
In  the static gauge $X^{a}=\sigma^{a}$ we have  
$
h_{ab}=\eta_{ab}+\partial_{a}X^i \partial_{b}X^i$
  and the 
 tangent vectors are $\partial_{a}X^{\mu} = (\delta^{b}_{a}, \partial_{a}X^{i})$. The normal basis  vectors can be  chosen as 
\be
\la{B.4}
n^{i}_{\mu} = (-\partial_{a}X^{i}, \delta^{i}_j)+O((\partial X)^{2}) \ .
\ee 
 Then  from  (\ref{B.2})  we  get that\foot{Corrections  to the leading term in $K$ 
 come from the covariant derivative in  (\ref{B.2})  and  also  from 
Gram-Schmidt orthogonalization procedure. The    contributions  from the connection  are  found using that 

$
h^{ab} = \eta^{ab}-\partial^{a}X^{i}\partial^{b}X^{i}+\mc O(X^{4}), \qquad
\Gamma^{c}_{ab} = \tfrac{1}{2}(\partial_{b}h_{ca}+\partial_{a}h_{cb}-\partial_{c}h_{ab})+O(X^{4}) = 
\partial^{c}X^{j}\partial_{ab}X^{j}+\mc O(X^{4}).$}
\be\la{b4}
\nabla_{a}\partial_{b}X^{i} = K^{j}_{ab}(n^{j})^{i} = K^{i}_{ab}+\OO((\partial X)^{2}), \ \ \ \ \ \qquad 
K^{i}_{ab} = \partial_{a}\partial_{b}X^{i}+ \OO((\del X)^3). 
\ee
In the case of the   flat target  space, we have the Gauss-Codazzi relation\footnote{In 
  $d=2$ one has 
$R_{abcd}= \frac{1}{2} ( g_{ac} g_{bd} - g_{ad} g_{bc}) R$ and $R_{ab} = \frac{1}{2} g_{ab} R$, while in 
 $d=3$ where the Weyl tensor vanishes     

$\quad R_{abcd} =    g_{ac} R_{bd}-g_{ad} R_{bc}-g_{bc} R_{ad}+g_{bd} R_{ac}-  \frac{1}{2} R 
(g_{ac} g_{bd}-g_{ad}  g_{bc})$.}
\be
\la{B.7}
R_{abcd} = K^{i}_{ac}K^{i}_{bd}-K^{i}_{ad}K^{i}_{bc}.
\ee
Let us  discuss  some relations for  $K$    and its derivatives  when  expanding them  in
 powers of $X^i$ assumed to  be subject to the free  equations of motion. 
Let us  denote by  $\approx$   the equality up to terms that 
vanish on-shell (i.e. for $\del^2 X=0$) or  are of higher order in $X$.
From (\ref{B.2})   we have  \be
\la{B.8}
K^{i}\equiv  \eta^{ab} K^{i}_{ab} \approx 0.
\ee
Contracting  two indices in  (\ref{B.7}) we get\footnote{We also have  the identity 
  $\nabla^a R_{ab} = \frac{1}{2} \nabla_{b}  R$   which implies
relations between  the derivatives   and products of $K$. }
\be
\la{B.9}
R_{ab} = K^{i} K^{i}_{ab} - K^{i}_{ac}{K^{i}}\indices{_{b}^{c}}\approx -K^{i}_{ac}{K^{i}}\indices{_{b}^{c}}\ . 
\ee
 Using that 
$
0 = \nabla_{a}h_{cd} = \nabla_{a}(\eta_{cd}+\partial_{c}X^{i}\partial_{d}X^{i}) = K^{i}_{ac}\partial_{d}X^{i}+K^{i}_{ad}\partial_{c}X^{i}
$ and 
contracting  this  with  $\eta^{ac}$ and using (\ref{B.8}) gives
\be
\la{B.11}
K^{i}_{ab}\partial^{a}X^{i} \approx 0.
\ee
Considering 
\be
\la{B.12}
\nabla_{a}K^{i}_{bc}-\nabla_{b}K^{i}_{ac} = [\nabla_{a}, \nabla_{b}]\partial_{c}X^{i} = R_{dc ab}  \partial^d X^{i}, 
\ee
 and contracting this with $\eta^{bc}$  and using (\ref{B.8}),(\ref{B.9})  we get 
\be
\la{B.13}
\nabla^{a}K^{i}_{ab} = R_{bc}  \partial^c X^{i} \approx K^{j}_{bd}\,  K^{j}{}_{c}^{\, d}\, \partial^{c}X^{i}.
\ee

\subsection{Expansion of  scalar curvature density}
\la{app:GB} 
Given 
\be\la{C11}   L=\sqrt{-h}\, R^{(d)}  \ , \ \ \ \ \ \ \ \ \   h_{ab}=\eta_{ab} + \hh_{ab}, \qquad \qquad   \hh_{ab}= \del_a X^i \del_b X^i \ , \ee
let us  consider the expansion of 
$L$  in powers of  $\hh_{ab}$.
The linear  term in $\hh~_{ab}$ is  a total derivative: $L_1= \partial_{a}\partial_{b}\hh^{ab}-\Box\, \hh, \ \  \hh= \eta^{ab} \hh_{ab}$.
The quadratic term is 
 \be
 \la{C.1}
 L_2 = -\tfrac{1}{4}\big( \partial^{c}\hh^{ab}\, \partial_{c}\hh_{ab}-2\partial_{a}\hh^{ab}\,\partial^{c}\hh_{bc}+2\partial_{a}\hh\, \partial_{b}\hh^{ab}-\partial^{a}\hh\partial_{a}\hh\big) \ . 
 \ee
 In the special case  of $d=2$   this  can be shown to be a total derivative. 
 Indeed, in 2 dimensions   we have  an identity following from the fact that the  total antisymmetrization of 3 indices vanishes. 
 In particular (here  the repeated indices  are contracted with $\eta_{ab}$ and we use that $[ab'c]=0$) 
  \ba
0 =&  \partial_{c}\hh_{ab}\partial_{c'}\hh_{a'b'} [
\delta_{aa'}\delta_{bb'}\delta_{cc'}-\delta_{ac'}\delta_{bb'}\delta_{ca'}
+\delta_{ac'}\delta_{bc}\delta_{b'a'}-\delta_{ab}\delta_{cc'}\delta_{a'b'}
-\delta_{aa'}\delta_{bc}\delta_{b'c'}+\delta_{ab}\delta_{a'c}\delta_{b'c'}] \lp
=
\partial_{c}\hh_{ab}\partial_{c}\hh_{ab}
-\partial_{c}\hh_{ab}\partial_{a}\hh_{bc}
+\partial_{c}\hh_{ac}\partial_{a}\hh_{a'a'}
-\partial_{c}\hh_{aa}\partial_{c}\hh_{a'a'}
-\partial_{c}\hh_{ac}\partial_{b'}\hh_{ab'}
+\partial_{c}\hh_{aa}\partial_{b'}\hh_{cb'}\lp
= (\partial_{a}\hh_{bc})^{2}
-\partial_{c}\hh_{ab}\partial_{a}\hh_{bc}
+2\partial_{c}\hh_{ac}\partial_{a}\hh
-\partial_{c}\hh\partial_{c}\hh
-\partial_{c}\hh_{ac}\partial_{b'}\hh_{ab'}.
 \ea
 Using this in  the first term in  \rf{C.1}  we get 
\ba
L_2\big|_{d=2} =&  -\tfrac{1}{4}[
\partial_{c}\hh_{ab}\partial_{a}\hh_{bc}
-2\partial_{c}\hh_{ac}\partial_{a}\hh
+\partial_{c}\hh\partial_{c}\hh
+\partial_{c}\hh_{ac}\partial_{b'}\hh_{ab'}
-2\partial_{a}\hh_{ab}\,\partial_{c}\hh_{bc}+2\partial_{a}\hh\, \partial_{b}\hh_{ab}-\partial_{a}\hh\partial_{a}\hh
]\lp
= -\tfrac{1}{4}[
\partial_{c}(\hh_{ab}\partial_{a}\hh_{bc})
-\partial_{a}(\hh_{ab}\,\partial_{c}\hh_{bc})
]  
= \tfrac{1}{4}\partial_{a}[\hh_{ab}\partial_{c}\hh_{bc}-\hh_{bc}\partial_{c}\hh_{ab}] \ . 
\ea
In fact, $L\big|_{d=2}$   is a total   derivative  to  all orders in $\hh_{ab}$   
 (this is  an integrand of  the Euler number invariant  in 2d).

For general $d$  one finds from \rf{C.1}  using the expression for   $\hh_{ab}$ in \rf{C11}  and 
dropping total derivative terms and  terms that vanish on the equation of motion $\partial^2 X^{i}=0$ 
\be\la{C19}
L_2 = -\partial^{a}X^i \partial^b \del^c X^i\  \partial_{b}X^j \partial_{a} \del_c X^j \ . 
\ee
This agrees  with  the expression  used, e.g.,  in   \cite{Dubovsky:2012sh,Conkey:2016qju}.

\newpage

\section{Details of  computation of 2-loop diagrams}
\la{app:NG-2-loop}

Here we present some details  of the computation of the three types of diagrams in \rf{222}  
in the bosonic brane case using
IBP and tensor reduction implemented in
  {\sf FIRE}
 \ci{Smirnov:2008iw}. One   can treat  the  string  $d=2$ and  membrane  $d=3$ cases in parallel
keeping $d$ generic till the final  expansion  stage.

\subsection{Double-bubble diagram}

The  double-bubble diagram gives the following integral  expression (the vertex $V$ is given in \rf{240}) 
\be
\begin{tikzpicture}[thick,baseline=0]
	\begin{feynman}
	\vertex (a1) at (-1,1);
	\vertex (a2) at (-1,-1);
	\vertex (a3) at (5,1);
	\vertex (a4) at (5,-1);
	\vertex (oo1) at (0,0);
	\vertex (oo2) at (2,0);
	\vertex (oo3) at (4,0);
	\diagram* {
	(a1) -- (oo1);
	(a2) -- (oo1);
	(a3) -- (oo3);
	(a4) -- (oo3);
	(oo1) -- [half left] (oo2);
	(oo1) -- [half right] (oo2);
	(oo2) -- [half left] (oo3);
	(oo2) -- [half right] (oo3);
	};
	\end{feynman}

	\draw[->,thin] ($(a1)!0.25!(oo1)+(0,0.15)$) -- ($(a1)!0.75!(oo1)+(0,0.15)$);
	\draw[->,thin] ($(a2)!0.25!(oo1)-(0,0.15)$) -- ($(a2)!0.75!(oo1)-(0,0.15)$);
	\draw[->,thin] ($(a3)!0.25!(oo3)+(0,0.15)$) -- ($(a3)!0.75!(oo3)+(0,0.15)$);
	\draw[->,thin] ($(a4)!0.25!(oo3)-(0,0.15)$) -- ($(a4)!0.75!(oo3)-(0,0.15)$);
	
         \draw[->,thin]  ($ (oo1)!0.25!(oo2) + (0,0.6) $) to [out=15,in=165] ($ (oo1)!0.75!(oo2) + (0,0.6) $);
         \draw[->,thin]  ($ (oo2)!0.25!(oo1) - (0,0.6) $) to [out=180+15,in=-15] ($ (oo2)!0.75!(oo1) - (0,0.6) $);
         \draw[->,thin]  ($ (oo2)!0.25!(oo3) + (0,0.6) $) to [out=15,in=165] ($ (oo2)!0.75!(oo3) + (0,0.6) $);
         \draw[->,thin]  ($ (oo3)!0.25!(oo2) - (0,0.6) $) to [out=180+15,in=-15] ($ (oo3)!0.75!(oo2) - (0,0.6) $);
         	
	\node[left] at (a1) {$(p_{1},i_{1})$};	
	\node[left] at (a2) {$(p_{2},i_{2})$};	
	\node[right] at (a3) {$(p_{3},i_{3})$};	
	\node[right] at (a4) {$(p_{4},i_{4})$};	
	\node at (1-0.3,1.2) {\small $(k_{1}+p_{1}+p_{2}, j_{1})$};
	\node at (1,-1.2) {\small $(k_{1}, j_{2})$};
	\node at (3+0.3,1.2) {\small $(k_{2}+p_{1}+p_{2}, j_{3})$};
	\node at (3,-1.2) {\small $(k_{2}, j_{4})$};
	\node at (-3,0) {${\rm DB}^{i_{1}i_{2}i_{3}i_{4}}_{p_{1},p_{2},p_{3},p_{4}} = $};
\end{tikzpicture}\no 
\ee
\be
 =\tfrac{1}{4i}\int \wt{dk_{1}}\, \wt{dk_{2}}
\frac{
V^{i_{1}i_{2}j_{1}j_{2}}_{p_{1},p_{2},-p_{1}-p_{2}-k_{1}, k_{1}}
V^{j_{1}j_{2}j_{3}j_{4}}_{p_{1}+p_{2}+k_{1}, -k_{1}, -p_{1}-p_{2}-k_{2}, k_{2}}
V^{j_{3}j_{3}i_{3}i_{4}}_{p_{1}+p_{2}+k_{2}, -k_{2}, -p_{3}, -p_{4}}
}{k_{1}^{2}k_{2}^{2}(k_{1}+p_{1}+p_{2})^{2}(k_{2}+p_{1}+p_{2})^{2}}, \la{D2}
\ee
where 
$
\wt{dk} = \frac{d^{d}k}{(2\pi)^{d}}
$. 
Application of IBP and tensor reduction implemented in {\sf FIRE} gives
\ba
&{\rm DB}^{i_{1}i_{2}i_{3}i_{4}}_{p_{1},p_{2},p_{3},p_{4}} =  A^{(2)}_{\DB} \delta^{i_{1}i_{2}}\delta^{i_{3}i_{4}}+ B^{(2)}_{\DB}
\delta^{i_{1}i_{3}}\delta^{i_{2}i_{4}}+C^{(2)}_{\DB}\delta^{i_{1}i_{4}}\delta^{i_{2}i_{3}}\ ,  
\la{D5}  \\ 
A^{(2)}_{\rm DB} =&  \tfrac{s^4} {2048 (-1+d^2)^2} \Big[((-2+d) (24 (-2+{\wh D})-d^2 (-48+{\wh D}^2)+d^3 \
(48-12 {\wh D}+{\wh D}^2)\lp\qquad \qquad \qquad 
-4 d (12-9 {\wh D}+{\wh D}^2)) \
s^2-16 {\wh D}^2 s t-16 {\wh D}^2 t^2)\Big]\, [G_{1,1}]^{2},\no \\
B^{(2)}_{\rm DB} =& -\tfrac{s^5 ((2-2 d+d^2) s+2 t)}{64 (-1+d)^2}\, [G_{1,1}]^{2} , \qquad 
C^{(2)}_{\rm DB} = -\tfrac{s^5 (-2 d s+d^2 s-2 t)}{64 (-1+d)^2}\, [G_{1,1}]^{2} \ ,    \no 
\ea
where $G_{1,1}$ was  given in \rf{3.9}. 
In the string case  we set   $d=2-2\eps$ and $\wh D= D-2$.  Expanding $[G_{1,1}]^{2}$   in $\eps$ we then get 
\be
[G_{1,1}]^{2} =-\tfrac{1}{4\pi^{2}} \Big[ \frac{1}{\eps^{2}}- \frac{2}{\eps}\big(\log\frac{-s}{4\pi} + \gamma_{\rm E}\big)\Big]{1\ov s^2} 
+\text{finite}.
\ee
The contributions \rf{D5}   should be summed over diagrams  related by  crossing.

\subsection{Wine-glass diagram}

For this  diagram we get 
\be
\begin{tikzpicture}[thick,baseline=0,scale=0.8]
	\begin{feynman}
	\vertex (a1) at (-4,1.4);
	\vertex (a2) at (-4,-1.4);
	\vertex (a3) at (4,1.4);
	\vertex (a4) at (4,-1.4);
	\vertex (p1) at (-2,1.4);
	\vertex (p2) at (2,1.4);
	\vertex (p3) at (0,-1.4);
	\diagram* {
	(a1) -- (p1);
	(a2) -- (p3);
	(a3) -- (p2);
	(a4) -- (p3);
	(p1) -- (p2);
	(p1) -- (p3) -- (p2);
	(p1) -- [bend left=45] (p2) ;
	};
	\end{feynman}

	\draw[->,thin] ($(a1)!0.25!(p1)+(0,0.15)$) -- ($(a1)!0.75!(p1)+(0,0.15)$);
	\draw[->,thin] ($(a3)!0.25!(p2)+(0,0.15)$) -- ($(a3)!0.75!(p2)+(0,0.15)$);
	\draw[->,thin] ($(a2)!0.25!(p3)+(0,0.15)$) -- ($(a2)!0.75!(p3)+(0,0.15)$);
	\draw[->,thin] ($(a4)!0.25!(p3)+(0,0.15)$) -- ($(a4)!0.75!(p3)+(0,0.15)$);
	\draw[->,thin] ($(p1)!0.25!(p2)-(0,0.15)$) -- ($(p1)!0.75!(p2)-(0,0.15)$);
	\draw[->,thin] ($(p1)!0.25!(p2)+(0,0.75)$) to [out=20, in=180-20]  ($(p1)!0.75!(p2)+(0,0.75)$);
	\draw[->,thin] ($(p3)!0.25!(p1)-(0.15,0)$) -- ($(p3)!0.75!(p1)-(0.15,0)$);
	\draw[->,thin] ($(p2)!0.25!(p3)+(0.15,0)$) -- ($(p2)!0.75!(p3)+(0.15,0)$);
         	
	\node[left] at (a1) {$(p_{1},i_{1})$};	
	\node[left] at (a2) {$(p_{2},i_{2})$};	
	\node[right] at (a3) {$(p_{3},i_{3})$};	
	\node[right] at (a4) {$(p_{4},i_{4})$};	
	\node at ($(p1)!0.5!(p2)+(0,1.5)$)  {$(k_{1},j_{1})$};
	\node at ($(p1)!0.5!(p2)-(0,0.6)$)  {$(-k_{1}+k_{2},j_{2})$};
	\node at ($(p3)!0.5!(p1)-(1.5,0)$) {$(k_{2}-p_{1},j_{4})$};
	\node at ($(p3)!0.5!(p2)+(1.5,0)$) {$(k_{2}+p_{3},j_{3})$};
	\node at (-6,0) {${\rm W}^{i_{1}i_{2}i_{3}i_{4}}_{p_{1},p_{2},p_{3},p_{4}} = \qquad \qquad\qquad  \ \ \ $};
\end{tikzpicture}\no
\ee
\be
=\tfrac{1}{2i}\int \wt{dk_{1}}\, \wt{dk_{2}}
\frac{
V^{i_{1}j_{1}j_{2}j_{4}}_{p_{1},-k_{1},k_{1}-k_{2},k_{2}-p_{1}}
V^{j_{1}i_{3}j_{3}j_{2}}_{k_{1},p_{3},-k_{2}-p_{3},-k_{1}+k_{2}}
V^{i_{2}j_{4}j_{3}i_{4}}_{p_{2},-k_{2}+p_{1},k_{2}+p_{3},p_{4}}
}{k_{1}^{2}(k_{1}-k_{2})^{2}(k_{2}+p_{3})^{2}(k_{2}-p_{1})^{2}}.\la{D55}
\ee
Application of {\sf FIRE} gives
\ba
&{\rm W}^{i_{1}i_{2}i_{3}i_{4}}_{p_{1},p_{2},p_{3},p_{4}} = A^{(2)}_{\rm W} \delta^{i_{1}i_{2}}\delta^{i_{3}i_{4}}+ B^{(2)}_{\rm W}
\delta^{i_{1}i_{3}}\delta^{i_{2}i_{4}}+C^{(2)}_{\rm W}\delta^{i_{1}i_{4}}\delta^{i_{2}i_{3}}\ , \la{D56}\\
 A^{(2)}_{\rm W} =& Q\,  (-3+d) t^5 (-8 (-1+d) (1+d) (-10+d (7+2 d)) s+(-2+d) (-12+d (1+d) 
(2+d)) {\wh D} s\lp
-8 (-1+d) (1+d) (-8+d (8+d)) t-(-2+d) d (8+(-3+d) 
d) {\wh D} t) ,\no\\
B^{(2)}_{\rm W} =&Q\, (-3+d) t^4 ((-8 (-1+d) (1+d) (2+d)+(-24+d (32+d (8-d (7+3 d)))) 
{\wh D}) s^2\lp
+(-8 (-1+d) (1+d) (2+d)+(-24+d (32+d (8-d (7+3 d)))) 
{\wh D}) s t\lp
+2 (-2+d) (-1+d) (20+d (8+3 d (-4+{\wh D})-5 
{\wh D})-6 {\wh D}) t^2), \no \\
C^{(2)}_{\rm W} =& -Q\, ((-3+d) t^5 ((-8 (-1+d) (1+d) (-10+d (7+2 d))\lp
+(-2+d) (-12+d (1+d) 
(2+d)) {\wh D}) s +2 (-2+d) (-1+d) (-4 (1+d)^2+(6+d+d^2) 
{\wh D}) t)), \no\\
& Q\equiv  \tfrac{1}{144(d^{2}-1)(3d-8)(3d-4)(3d-2)} \, \, G_{1,1,1,1} \ . \no 
\ea
The master integral here  is 
\be
G_{1,1,1,1} = 
\int \wt{dk_{1}}\, \wt{dk_{2}}
\frac{1}{k_{1}^{2}(k_{1}-k_{2})^{2}(k_{2}+p_{3})^{2}(k_{2}-p_{1})^{2}}.
\ee
Integrating over $k_1$ using (\ref{3.9}) we get 
\be
G_{1,1,1,1} = C_{d}
\int \wt{dk}
\frac{1}{(k^{2})^{2-\frac{d}{2}}(k+p_{3})^{2}(k-p_{1})^{2}}, \qquad \qquad C_{d}=\tfrac{i}{(4\pi)^{d/2}}\tfrac{\Gamma(2-\frac{d}{2})[\Gamma(\frac{d}{2}-1)]^{2}}{\Gamma(d-2)}.
\ee
Introducing  Feynman parameters  as in \rf{A.1}   with $x_3=1-x_1-x_2$ we have 
\ba
G_{1,1,1,1} = C_{d}\tfrac{\Gamma(4-\frac{d}{2})}{\Gamma(2-\frac{d}{2})}\int [d^{3}x] \, x_{1}^{1-\frac{d}{2}}\int \wt{dk}
\frac{1}{[x_{1} k^{2}+x_{2}(k+p_{3})^{2}+x_{3}(k-p_{1})^{2}]^{4-\frac{d}{2}}}.
\ea
Shifting   $k$ and using $p_{i}^{2}=0$ we  have 
$x_{1} k^{2}+x_{2}(k+p_{3})^{2}+x_{3}(k-p_{1})^{2} \to k^{2}+2x_{2}x_{3} p_{1}\cdot p_{3}$. 
 Integrating  over $k$  using \rf{A.2}     and also over $x_i$    we get 
\be
G_{1,1,1,1} = -\tfrac{1}{(4\pi)^{d}}\tfrac{\Gamma(4-d)\Gamma(2-\frac{d}{2})[\Gamma(\frac{d}{2}-1)]^{2}[\Gamma(d-3)]^{2}}
{\Gamma(d-2)\Gamma(\frac{3}{2}d-4)}\ (2p_{1}\cdot p_{3})^{d-4} \ .
\ee
Expanded for $d=2-2 \eps$ this   gives 
\be
G_{1,1,1,1} = \tfrac{3}{2(4\pi)^{2}}\Big[\frac{1}{\eps^2} - {2\ov \eps}\Big(\log\frac{-t}{4\pi}  + \gamma_E - \tfrac{1}{2}\Big) +\dots\Big]\, \frac{1}{t^{2}} + {\rm finite} \ . 
\ee
The sum over crossing  contains  6 diagrams where the   labels (1,3) are replaced by  all distinct ordered pairs.

\subsection{Evanescent counterterm diagram}

The  third diagram in \rf{222}   contains one   vertex  from the  1-loop  
 evanescent counterterm \rf{3.14} with the vertex $E$ in \rf{214}\foot{As in the expressions  above we do not explicitly include  normalization $ \mu^{2\eps}$ factor (cf. footnote \rf{M}).}
\be
\begin{tikzpicture}[thick,baseline=0]
	\begin{feynman}
	\vertex (a1) at (-1,1);
	\vertex (a2) at (-1,-1);
	\vertex (a3) at (3,1);
	\vertex (a4) at (3,-1);
	\vertex (oo1) at (0,0);
	\vertex (oo2) at (2,0);
	\diagram* {
	(a1) -- (oo1);
	(a2) -- (oo1);
	(a3) -- (oo2);
	(a4) -- (oo2);
	(oo1) -- [half left] (oo2);
	(oo1) -- [half right] (oo2);
	};
	\end{feynman}

	\draw[->,thin] ($(a1)!0.25!(oo1)+(0,0.15)$) -- ($(a1)!0.75!(oo1)+(0,0.15)$);
	\draw[->,thin] ($(a2)!0.25!(oo1)-(0,0.15)$) -- ($(a2)!0.75!(oo1)-(0,0.15)$);
	\draw[->,thin] ($(a3)!0.25!(oo2)+(0,0.15)$) -- ($(a3)!0.75!(oo2)+(0,0.15)$);
	\draw[->,thin] ($(a4)!0.25!(oo2)-(0,0.15)$) -- ($(a4)!0.75!(oo2)-(0,0.15)$);
	
         \draw[->,thin]  ($ (oo1)!0.25!(oo2) + (0,0.6) $) to [out=15,in=165] ($ (oo1)!0.75!(oo2) + (0,0.6) $);
         \draw[->,thin]  ($ (oo2)!0.25!(oo1) - (0,0.6) $) to [out=180+15,in=-15] ($ (oo2)!0.75!(oo1) - (0,0.6) $);
         
         	\draw[fill=gray, thin] (oo1) circle (0.15); 
	
	\node[left] at (a1) {$(p_{1},i_{1})$};	
	\node[left] at (a2) {$(p_{2},i_{2})$};	
	\node[right] at (a3) {$(p_{3},i_{3})$};	
	\node[right] at (a4) {$(p_{4},i_{4})$};	
	\node at (1,1.2) {\small $(k_{1}+p_{1}+p_{2}, j_{1})$};
	\node at (1,-1.2) {\small $(k_{1}, j_{2})$};
	\node at (-3,0) {${\rm CT}^{i_{1}i_{2}i_{3}i_{4}}_{p_{1},p_{2},p_{3},p_{4}} =\qquad \qquad  $};
\end{tikzpicture}\no 
\ee
\be\qquad \qquad \qquad 
 =-\tfrac{1}{2i} \,\int \wt{dk_{1}}
\ \frac{
E^{i_{1}j_{1}j_{2}i_{2}}_{p_{1},-k_{1}-p_{1}-p_{2},k_{1},p_{2}}\ 
V^{j_{1}i_{3}i_{4}j_{2}}_{k_{1}+p_{1}+p_{2},p_{3},p_{4},-k_{1}}
}{k_{1}^{2}(k_{1}+p_{1}+p_{2})^{2}} \ . \la{D12}
\ee
There is also a similar diagram with the counterterm  vertex on the right, i.e. with  the vertices $E$ and $ V$
interchanged. 
Application of {\sf FIRE} gives for the sum  of these two diagrams  
\ba
&{\rm CT}^{i_{1}i_{2}i_{3}i_{4}}_{p_{1},p_{2},p_{3},p_{4}} = A^{(2)}_{\rm CT} \delta^{i_{1}i_{2}}\delta^{i_{3}i_{4}}+B_{\rm CT}
\delta^{i_{1}i_{3}}\delta^{i_{2}i_{4}}+C_{\rm CT}\delta^{i_{1}i_{4}}\delta^{i_{2}i_{3}}\ ,
\\ 
&A^{(2)}_{\rm CT} =\frac{i \CounterTerm s^3 ((-2+d) (-4+d (-2+{\wh D})) s^2+8 (2+{\wh D}) s 
t+8 (2+{\wh D}) t^2)}{32 (-1+d^2)}  \, G_{1,1} ,\\
&B^{(2)}_{\rm CT} = C^{(2)}_{\rm CT} = -\frac{i (-2+d) \CounterTerm s^5}{8 (-1+d)}  \, G_{1,1} \ , 
\ea
where $\CounterTerm$ is the  coefficient in \rf{3.14} and $G_{1,1}$ is given by \rf{3.9}. 
Summing over crossing one gets for the corresponding  contribution to  the  pole part of the 2-loop amplitude  in the string case 
($d=2-2\eps$, $\wh D=D-2$)
\ba
\la{D.28}
& \mc M^{(2)\, i_{1}i_{2}i_{3}i_{4}}_{{\rm CT}, \frac{1}{\eps}}(p_{1}, p_{2}, p_{3}, p_{4})=\frac{D}{24\pi}\frac{\CounterTerm}{\eps}\, stu\, \big[
s\,\delta^{i_{1}i_{2}}\delta^{i_{3}i_{4}}
+t\,\delta^{i_{1}i_{3}}\delta^{i_{2}i_{4}}
+u\,\delta^{i_{1}i_{4}}\delta^{i_{2}i_{3}}
\big]\ .  
\ea

\subsection{Finite PPS counterterm diagram}\la{ap1}

Similar  result  is found for the   diagram   with the insertion of the vertex  ${\rm V}$  in \rf{400} coming  from 
the  finite  PPS counterterm \rf{13}   required to preserve the  integrability of the bosonic S-matrix   for any $D$. 
   The corresponding contribution to the 2-loop  diagram  is  
\ba
&  \mc M^{(2)\, i_{1}i_{2}i_{3}i_{4}}_{{\rm PPS}}= \frac{ib s^{3}}{32(d^{2}-1)}\Big[
\tfrac{1}{2}((24-2 d-17 d^2+4 (-2+d) (1+d) \wh D) 
s^2+24 s t+24 t^2) \delta^{i_{1}i_{2}} 
\delta^{i_{3}i_{4}}\no  \\  &\qquad \qquad \    +   (2+d) s ((-2+d) s-6 t) 
\delta^{i_{1}i_{3}}\delta^{i_{2}i_{4}}  + 
(2+d) s((4+d) s+6 t)  \delta^{i_{1}i_{4}}\delta^{i_{2}i_{3}}
\Big]\, G_{1,1}(s).\la{D16}
\ea
After summing  over crossing, the expansion around $d=2$ of the  corresponding $A^{(2)}_{\rm PPS} (s,t,u)$ coefficient  of $\delta^{i_{1}i_{2}}$ part reads
\ba
A^{(2)}_{\rm PPS}  = &-\tfrac{7b}{8\pi}s^{2}t(s+t)\,  \big(\tfrac{1}{\eps}+\gamma_{\rm E}-\log(4\pi)\big)
+\tfrac{b}{48\pi}(-78s^{3}t-70s^{2}t^{2}+16st^{3}+8t^{4}+s^{4}(6D-37))\lp
+\tfrac{b}{8\pi}s^{2}(-2s^{2}+st+t^{2})\log(-s)-\tfrac{b}{4\pi}st^{3}\log(-t)+\tfrac{b}{4\pi}s(s+t)^{3}\log(s+t).\la{D17}
\ea
 Setting $t=0$  one finds that it  
 contributes only to the finite part of the 2-loop  bosonic  string  amplitude
according to \rf{6.75}:  
\ba\la{D18}
A^{(2)}_{\rm PPS} (s,0, -s)  
= \tfrac{b}{48\pi}s^{4}(6D-37)  +\tfrac{b}{4\pi}s^{4}\, [\log s-\log(-s)]
 = \tfrac{1}{48\pi} \, b\, (6D-37)\, s^4
+i\tfrac{1}{4}\, b\, s^{4}.
\ea 
A similar counterterm \rf{13}  is required in the  $D=10$ GS  string case 
 \cite{Seibold:2024oyr}  where its coefficient is given by \rf{78}.\foot{One may wonder if there is a supersymmetric partner of \rf{13}   with  two $X$ and two $\theta$   legs  
 (cf. \ci{Yang:1991ua}) that may 
  contribute  extra 2-loop diagram with  one counterterm $X^2\theta^2$   vertex  and  the  fermionic loop. 
  If this happens,   the only change should be a real rational contribution to  the amplitude 
  as the imaginary parts of $A$ and $C$  in \rf{990}  are already cancelled by the contribution \rf{6.75} of the bosonic PPS counterterm.}{

\iffa

\section{Unitarity}

maybe we can just say we checked 2-particle unitarity  at the order we are working and it is ok.

The 2-particle unitarity conditions are, see Eqs.~(5,6,7) of \cite{Conkey:2016qju},
\bea
\la{6.67}
\text{Im}\, A &= \frac{1}{4s}(2|A|^{2}+AB^{*}+A^{*}B+AC^{*}+A^{*}C), \\
\text{Im}\, B &= \frac{1}{4s}(|B|^{2}+|C|^{2}), \\
\text{Im}\, C &= \frac{1}{4s}(B^{*}C+BC^{*}).
\eea
At 2-loops at least  they hold because inelastic process $2\to 4$ happens at higher order
and violates 2-particle unitarity, see comments 
in page 5 of \cite{Conkey:2016qju}. See also Appendix A in \cite{EliasMiro:2019kyf}.

If we write the 2-loop expansion as 
\bea
A &= (A^{(1)}+i\, \wt A^{(1)})\, s^{3}+(A^{(2)}+i\, \wt A^{(2)})\, s^{4}+\dots, \\
C &= (C^{(1)}+i\, \wt C^{(1)})\, s^{3}+(C^{(2)}+i\, \wt C^{(2)})\, s^{4}+\dots, \\
B &= \tfrac{1}{2}s^{2}+(B^{(1)}+i\, \wt B^{(1)})\, s^{3}+(B^{(2)}+i\, \wt B^{(2)})\, s^{4}+\dots, 
\eea
the constraints (\ref{6.67}) give the conditions
\be
\la{6.69}
\wt A^{(1)}=0, \quad \wt A^{(2)} = \tfrac{1}{4}A^{(1)}, \quad \wt B^{(1)}=\tfrac{1}{16}, \quad \wt B^{(2)} = \tfrac{1}{4}B^{(1)},
\quad \wt C^{(1)}=0, \quad \wt C^{(2)} = \tfrac{1}{4}C^{(1)}.
\ee
We have so far (blue for bosonic, red for fermionic at $D=10$, and $K_{i}$ real; $s^{2}$ is tree, $s^{3}$ 1-loop, $s^{4}$ 2-loops)
\ba
\la{6.70}
A &= \Big[\blue{-\frac{D-26}{192\pi}}\red{-\frac{1}{48\pi}}\Big]s^{3}+\frac{i}{4}\, \Big[\blue{-\frac{D-26}{192\pi}}\red{-\frac{1}{48\pi}}\Big]\, s^{4}+K_{1}s^{4}, \\
C &= \Big[\blue{\frac{D-26}{192\pi}}\red{+\frac{1}{48\pi}}\Big]s^{3}+\frac{i}{4}\Big[\blue{\frac{D-26}{192\pi}}\red{+\frac{1}{48\pi}}\Big]\, s^{4}+K_{2}s^{4}, \\
B &= \frac{1}{2}s^{2}+\blue{\frac{i}{16}s^{3}}+K_{3}s^{4}.
\ea
Unitarity conditions are satisfied (at order $s^{4}$) for any $D$ in the bosonic terms. Notice that they fix the two loop imaginary part in $A$ and $C$ to the value we found ($i/(192\pi)s^{4}$),
given the tree and 1-loop values. 
\fi

{\small  \bibliography{BT-Biblio}
\bibliographystyle{JHEP-v2.9}
}
\end{document}